\documentclass{jfm}

\usepackage[english]{babel}
\usepackage[usenames, dvipsnames]{color}
\usepackage{graphicx}
\usepackage{caption}
\usepackage{subcaption}
\usepackage{natbib}
\usepackage{fixltx2e}
\usepackage{textgreek}
\usepackage{textcomp}
\usepackage{hyperref}
\usepackage{multirow}
\usepackage{array}
\usepackage{pbox}
\usepackage{gensymb}
\usepackage{comment} 

\usepackage{xcolor}
\definecolor{cinnamon}{rgb}{0.82, 0.41, 0.12}
\definecolor{pink}{rgb}{0.858, 0.188, 0.478}
\definecolor{black}{rgb}{0.0, 0.0, 0.0}

\graphicspath{{./Figures/}}

\begin{document}

\newtheorem{lemma}{Lemma}
\newtheorem{corollary}{Corollary}

\shorttitle{Particle-laden flow of an elastoviscoplastic fluid} 

\shortauthor{S. Zade, T. J. Shamu, F. Lundell and L. Brandt} 

\title{Finite-size spherical particles in a square duct flow of an elastoviscoplastic fluid: an experimental study}

\author
 {
 Sagar Zade\aff{1}
 \corresp{\email{zade@mech.kth.se}},
Tafadzwa John Shamu\aff{2},
 Fredrik Lundell\aff{1},
 \and
 Luca Brandt\aff{1}
  }

\affiliation
{
\aff{1}
Linn\'e Flow Centre and SeRC (Swedish e-Science Research Centre), \\ KTH Mechanics, SE 100 44 Stockholm, Sweden
\aff{2}
Division of Soil and Rock mechanics, KTH, SE-100 44 Stockholm, Sweden
}

\maketitle

\begin{abstract}

The present experimental study addresses the flow of a Yield Stress Fluid (YSF) with some elasticity (Carbopol gel) in a square duct. The behaviour of two fluids with lower and higher yield stress is investigated in terms of the friction factor and flow velocities at multiple Reynolds numbers $Re^* \in$ (1, 200) and, hence, Bingham numbers $Bi \in$ (0.01, 0.35). Taking advantage of the symmetry planes in a square duct, we 
reconstruct the entire 3-component velocity field from 2-dimensional Particle Image Velocimetry (PIV). A secondary flow consisting of eight vortices is observed to recirculate the fluid from the core towards the wall-center and from the corners back to the core. The extent and intensity of these vortices grows with increasing $Re^*$ or, alternately, as the plug-size decreases. The second objective of this study is to explore the change in flow in the presence of particles. To this end, almost neutrally-buoyant finite-size spherical particles with duct height, $2H$, to particle diameter, $d_p$, ratio of 12 are used at two volume fractions $\phi$ = 5 and 10\%. Particle Tracking Velocimetry (PTV) is used to measure the velocity of these refractive-index-matched spheres in the clear Carbopol gel, and PIV to extract the fluid velocity. Additionally, simple shadowgraphy is also used for qualitatively visualising the development of the particle distribution along the streamwise direction. The particle distribution pattern changes from being concentrated at the four corners, at low flow rates, to being focussed along a diffused ring between the center and the corners, at high flow rates. The presence of particles induces streamwise and wall-normal velocity fluctuations in the fluid phase; however, the primary Reynolds shear stress is still very small compared to turbulent flows. The size of the plug in the particle-laden cases appears to be smaller than the corresponding single phase cases. Similar to Newtonian fluids, the friction factor increases due to the presence of particles, almost independently of the suspending fluid matrix. Interestingly, predictions based on an increased effective suspension viscosity agrees quite well with the experimental friction factor for the concentrations used in this study.  

\end{abstract}
 
\section{Introduction}

\subsection{Applications of Yield Stress Fluids (YSF)}

Gels are usually particulate dispersions that undergo solid-liquid or sol-gel transition i.e. they yield beyond a critical shear stress (called the yield stress $\tau_y$); they have many applications in industry \citep[for e.g. in food products like mayoannaise, consumer products like toothpaste and drugs, building materials like concrete and paint, oil and drilling muds, etc.][]{bird1983rheology}, in biology \citep[for e.g. blood flow][]{picart1998human} and environment \citep[for e.g. clay suspension, debris, snow avalanches, lava flows, etc.][]{ancey2007plasticity, chambon2014experimental}. Before yielding, they have solid like properties e.g. they can sustain shear stress and deform elastically, whereas after yielding they behave like a fluid. Thus, the yield stress $\tau_y$ is a practically useful parameter. To give some examples, it is used to assess the shelf-life of paints, keep particulate fillers from settling in many consumer products, and also dictate whether bubbles remain trapped in cement \citep{singh2008interacting}. The latter is an important factor in the structural integrity of buildings. 

The yield stress $\tau_y$ has its origin in the microstructure of the material, which can dynamically adjust under the action of a flow. Hence, thixotropic (time-dependent shear thinning) behaviour is generally observed for many materials (e.g. clay suspension, colloidal gels) and $\tau_y$ may not exhibit a single invariant value \citep{bonn2009yield}. Apart from this, Yield Stress Fluids (YSF) may possess complex macroscopic properties because of their different elastic, viscous and plastic characteristics \citep{piau2007carbopol}. For a recent comprehensive review describing the properties of YSF, please refer to \cite{bonn2017yield}.

\subsection{Non-dimensional numbers}

The Herschel-Bulkley (HB) constitutive equation (i.e. the relation between stress $\tau$ and strain rate $\dot{\gamma}$) is the archetypical form used for YSF, and is given as $\dot{\gamma} = 0$ when $\tau \leq \tau_y$, and $\tau = \tau_y + \kappa \dot{\gamma}^n$ when $\tau \geq \tau_y$. Here, the power law exponent $n$ is the flow behaviour index i.e. it quantifies the degree of non-Newtonian behaviour, and $\kappa$ is the consistency parameter. For the special case of $n = 1$, known as the Bingham plastic model, $\kappa$ is called the plastic viscosity. 

Numerical simulations of YSF are challenging because of the non-smoothness of the stress at the interface separating the unyielded and yielded regions \citep{wachs2019computational}. Amongst others, (i) regularization methods, where unyielded regions are replaced by regions of very high viscosity by modifying the above constitutive equations by a single smooth and differentiable function, and (ii) augmented Lagrangian methods \citep[refer to][]{glowinski2011numerical} are the main approaches to treat these issues \citep{saramito2017progress, balmforth2014yielding}. In general, the behaviour is further complicated by the presence of elastic effects both before and after yielding requiring the need for elastoviscoplastic (EVP) models \citep{saramito2007new}. Also refer to \cite{fraggedakis2016yielding2} for a comparison of various EVP models and to \cite{izbassarov2018computational} for numerical algorithm describing suspension flow of EVP fluid. 

The generalized Bingham number $Bi$ or equivalently, Oldroyd $Od$ or Herschel-Bulkley number, is the ratio of plastic (yield stress) to viscous (shear stress) effects. The average wall-shear stress $\tau_w$, which is proportional to the streamwise pressure gradient $\Delta{p}/\Delta{x}$, can be used to define $Bi = \tau_y / ((H/2)(\Delta{p}/\Delta{x}))$. Here, $H$ is the half height in case of a square duct or radius in case of a round pipe. Alternately, one can use the nominal shear stress based on the HB model to define $Bi = \tau_y /(\kappa (U/L)^n)$, where $U$ and $L$ are the characteristic velocity and length scale respectively. The former definition is used in this work. Less common is the use of the so called plastic number $Pl = \tau_y / (\tau_y + \kappa (U/L)^n)$ which is the ratio of the yield stress to the total stress, thus ranging from 0 (the material exhibits no plastic effects) to 1 (the material exhibits maximum plasticity). 

The choice of a representative Reynolds number $Re^*$ for non-Newtonian fluids is often motivated by its ability to correctly predict the laminar friction factor $f = \tau_w/(0.5\rho U_{Bulk}^2)$, \textit{Fanning} in our case, according to a relationship $f = 16/Re^*$ \citep[see][]{metzner1955flow} where $\rho$ is the density and $U_{Bulk}$ is the average or bulk velocity of the fluid. Using the Rabinowitsch-Mooney equation, \citet{kozicki1966non} provided a general framework to predict the flow rate as a function of the streamwise pressure gradient for flow of non-Newtonian fluids in ducts of arbitrary cross section, yielding a $Re^*$ consisting of two shape-factors corresponding to the shape of the duct. Later, \citet{liu1998non} improved the accuracy of the above relationship by proposing a three shape-factor approach that is better suited for ducts, where the wall shear rate has contributions from terms other than only the wall-normal streamwise velocity gradient. The above approach has been used in the present study to define a generalized Reynolds number as
\begin{equation}
Re^* = \frac{16}{f} = \frac{8  \rho U_{Bulk}^2}{\tau_y + \kappa {(\frac{2 U_{Bulk} a}{H})}^n {(1+\frac{1-n}{bn})}^nc^{n-1}}
  \label{eqn:Re}
\end{equation}
The constants are $a = $ 1.778, $b =  $ 4.382 and $c = $ 1.067 for a square duct \citep{liu1998non}. Expressions for the $Re^*$ for HB fluids that account for the turbulent regime in pipe-flows can be found in \cite{chilton1998pressure,malin1998turbulent} and \cite{garcia1986comparison} amongst others.

\subsection{YSF in a square duct}

Square duct flows of YSF, mostly of the Bingham type, have been a subject of several numerical studies. Following \cite{taylor1997conduit}, \cite{van1998viscoplastic} performed simulations of laminar duct flow using the Bingham model and proposed a \textit{master curve} relating the flow rate to the pressure drop; a useful information in design of extrusion geometries. \citet{saramito2001adaptive} used the augmented Lagrangian method for a Bingham model fluid flow in a square duct at varying $Bi$ to compute the critical value of $Bi$ above which the flow stops, known as the stopping criteria \citep[also see][]{mosolov1965variational}. The authors accurately identified the yield surfaces separating the unyielded plug region in the core and the sheared regions around it, while also capturing the correct concavity of the unyielded stagnant zones near the corners. \citet{huilgol2005application} extended the above method to a HB fluid model and found that for a fixed $Bi$ (or $Od$ in their case), the plug region increases as the power law exponent in the HB model decreases. 

\citet{letelier2017elasto} introduced elasticity in their simulations of a Bingham fluid for an \textit{approximate} square duct geometry and observed that for a fixed $Bi$, the flow rate increases with increasing elasticity (quantified using a Weissenberg number $Wi$, which is the ratio of elastic stresses in the form of normal stress difference to viscous stress due to shear forces). In a later work by the same authors, by including higher order terms in $Wi$, secondary flow in the form of streamwise vortices is seen \citep{letelier2018elastoviscoplastic}. Increasing the $Bi$ at the same $Wi$ reduced the intensity of the secondary flow, and displaced the vortices further away from the center of the duct. Similar observations are also reported in this work, the novelty being the HB nature of the experimental fluid against the Bingham numerical model along with a higher $Bi$ in our case compared to their simulations, as will be described in the following sections. Often a Deborah number $De$ is defined which is the ratio of the relaxation time of the fluid to the characteristic time of the flow, and under certain circumstances, it is equivalent to the $Wi$ \citep{robpooleDeWi}. Since both $Wi$ and $Re^*$ increase with the flow rate for a given elastocviscoplastic fluid, an elasticity number $El$ is defined as $Wi/Re^*$ to quantify the effects of elastic forces compared to inertial forces. 

Many researchers since \citet{ericksen1956overdetermination} have investigated the existence and strength of the aforementioned streamwise vortices in laminar rectangular duct flows. This secondary flow originates from viscoelastic forces in non-circular pipes, specifically the second normal stress difference \citep{dodson1974non}, and its strength increases with the elasticity \citep{debbaut1999secondary}. Thus, no secondary flow would be observed in a purely viscous or viscoplastic fluid or for certain special relationship between the second normal stress difference and the fluid viscosity \citep{xue1995numerical}. Despite being very weak (around two orders in magnitude lower than the mean axial velocity), the presence of such secondary velocity field may have substantial effects on the rate of heat-transfer \citep{kostic1994turbulent, gao1996heat}. Its weak nature also presents challenges in measuring it experimentally \citep{schroeder2011rheo}. \citet{yue2008general} provided a general criteria to predict the direction of these secondary flows.

\subsection{Other aspects of flowing YSF}

There have been numerous studies concerning laminar to turbulent transition of YSF, mostly in the classic pipe geometry, and a few criteria able to predict transition have been proposed \citep{hanks1963laminar, guzel2009predicting}. In general, it is observed that the presence of a yield stress acts to stabilize the flow, thereby delaying transition \citep{peixinho2005laminar}. Moreover, transition exists over a much narrower velocity range compared to Newtonian fluids \citep{park1989pipe}. Similar to other shear-thinning fluids, the streamwise velocity profile develops an asymmetry in the transitional regime \citep{escudier1996pipe, escudier2005observations}, which is associated with the existence of a robust structure in the form of two weak counter-rotating longitudinal vortices \citep{esmael2008transitional}. In turbulence, where the Reynolds stress is sufficient to break the plug, the behaviour of the YSF resembles a shear-thinning fluid \citep{guzel2009observation} characterized by a reduced wall-normal turbulence intensity. Recently, \cite{rosti2018turbulent} simulated the turbulent channel flow of an EVP fluid and observed that in the turbulent regime the friction factor decreases with increasing $Bi$, until the flow becomes laminar at high $Bi$.


\subsection{Studies in particle laden flow}

Dispersed phases like particles, drops or bubbles may appear as desired or undesired components in the final product made using YSF. Hence, their interaction with YSF is of practical importance and this paper partly addresses this problem. 

In the case of particle laden flows, sedimentation of a single spherical particle in a YSF is the most common study on account of its obvious importance in gaining fundamental understanding and as a precursor for multi-particle problems. A heavy isolated particle can settle inside a YSF only when the buoyancy forces exceeds the resistance offered by the yield stress. Accordingly, a dimensionless Yield-gravity parameter $Y = \tau_y / ((\rho_p - \rho_f) g d_p)$ can be defined such that the particle moves only when $Y$ is lower than a threshold critical value $\approx$ 0.2 while noting that the exact value is still a topic of investigation  \citep{chhabra2006bubbles}. It has been observed that the disturbance field created by a particle decays faster in a YSF, compared to a Newtonian fluid, implying that two particles will feel each other at a farther distance in the latter type of fluid \citep{firouznia2018interaction}. Creeping flow of an ideal yield stress fluid (Bingham or Herschel-Bulkely) around a sphere is reversible and display fore-aft symmetry \citep{putz2010creeping, beaulne1997creeping}. 

\subsection{Need for careful preparation of the solution}

In contrast to the above observation, when \cite{putz2008settling} performed PIV measurements of the flow field around a spherical particle sedimenting in a Carbopol solution at low $Re < 1$, they observed a breaking of the symmetry that is observed in Newtonian fluids. In particular, the characteristics of the flow regions around the falling sphere were associated with the rheological properties of the fluid, thus calling for numerical models to include both elastic \citep{fraggedakis2016yielding} and hysteresis/thixotropic effects to accurately reproduce experimental observations as well as for experiments to ensure careful preparation of the fluid. On the other hand, \cite{tabuteau2007drag} could obtain reproducible results in experiments for the drag on a settling sphere in agreement with the simulations of \cite{beaulne1997creeping}. The authors attributed this to the careful preparation (homogenizing for ten days!) of the yield stress fluid. 

Indeed, in the recent work by \cite{dinkgreve2018carbopol}, it was clearly shown that the discrepancy in the \textit{simple yield stress} behaviour of Carbopol observed in previous studies was most likely due to lack of optimal mixing; long stirring breaks the polymers into smaller fragments which exhibit Brownian motion and these small thermal particles cause a depletion interaction. The ensuing attractive forces can then lead to the formation of a percolated network of the larger unbroken Carbopol microgel particles. This network is sensitive to the shear, which leads to the observed thixotropy. At low shear rates, transient shear banding \citep{divoux2010transient} can also cause apparent hysteresis due to lack of sufficient measurement time. The measurement time required for reaching a steady state approaches unrealistically large values in the presence of wall-slip, which occurs when a smooth rotating surface is used during rheometry \citep{poumaere2014unsteady}.

\subsection{Particle migration in non-Newtonian fluids}

Since the Carbopol gel used in this study exhibits small but measurable elastic effects, the normal stresses and ensuing secondary flows are expected to affect the particle migration and their equilibrium distribution. Hence, a review of particle migration inside a duct flow of viscoelastic fluid is appropriate. 

Cross-stream particle migration in viscoelastic fluids is quite different to their migration in a Newtonian fluid \citep{d2015particle}. In the absence of inertia, which is typical of microfluidic applications (e.g. cell focussing and separation), depending on the initial particle position fluid elasticity drives the particle towards the channel center-line or the closest wall and from there towards the nearest corner in case of a duct geometry \citep{villone2013particle}. This can be understood by observing the distribution of the first normal stress difference, which has local minima at the center and the corners of the duct cross section \citep{yang2011sheathless}. The shear-thinning effects reduce the first normal stress difference \citep{li2015dynamics} thus, augmenting particle migration towards the closest wall \citep{d2012single}. When second normal stress differences are present (e.g. in a duct), the particle migration velocity, proportional to $De (d_p/(2H))^2$, may be higher or lower than the ensuing secondary flow velocity, proportional to $De^4$ \citep{yue2008general}. Thus, at larger $De$ and smaller particle confinement $d_p/(2H)$, the secondary flow may overcome the migration velocity and the particle may also find a stable position near the center of the streamwise vortices that are characteristic of the secondary flow \citep{villone2013particle}. 

Under the influence of non-negligible inertia and elasticity, \cite{yang2011sheathless} experimentally managed to focus all the particles towards the duct center-line. On the contrary, \cite{del2013particle} could achieve the same feat without inertia and attributed their observation to purely viscoelastic forces sans shear-thinning. Indeed, as shown by \cite{seo2014lateral}, the complex interplay between elasticity, inertia, shear-thinning and confinement can lead to multiple stable and unstable equilibrium configurations. \cite{lim2014inertio} realized that particles in microfluidic flows can be focussed at a high throughput rate even at very high $Re \approx$ 10000 i.e. high inertia, provided that the elastic normal stresses also increase proportionally i.e. the important criteria governing particle focussing is the elasticity number $El$. Later, \cite{li2015dynamics} investigated the role of these forces in focussing a particle at the center of the duct as well as the associated particle-induced fluid transport.

In the Stokes regime i.e. negligible particle inertia, flow of particle suspension in a YSF inside a tube has been recently simulated by \cite{siqueira2019pressure} using a regularised viscosity function, with the diffusive flux model \citep{phillips1992constitutive} being used to describe the shear-induced particle migration. Particles were found to concentrate at the boundary of the plug and with increasing particle concentration, the maxima shifted towards the tube-wall. No particles were found inside the plug due to the high but finite viscosity gradient (due to the regularisation) that slowly pushed them radially outwards. A similar method was used to model the particle migration for a tube flow of a Bingham fluid in \cite{lavrenteva2016shear}. The maximum particle concentration was found at the interface between yielded and unyielded region. On the other hand, in the simulations of \cite{hormozi2017dispersion} to model particle-laden flow in a fracture, particles were found to concentrate in the plug.






\subsection{Importance of the present study}

Most of the above experimental studies concerning particle migration are microfluidic in their treatment i.e. they have a very low fluid and particle inertia. Also, they deal with a very low particle concentration ($\Phi \leq$ 0.1\%) i.e. negligible inter-particle interactions, that is typical of flow-focusing experiments, and to our knowledge there have not been many experimental studies devoted to multi-particle dynamics with the exception of the impressive observations made by \cite{gauthier1971particle} and \cite{tehrani1996experimental} a while ago. A noteworthy contribution of this study lies in extending previous scientific investigations beyond the above regimes using a novel set-up with a large flow cross section where the flow can be resolved at the particle scale, a feature that is very difficult to capture in microfluidic devices. In addition, the presence of yield stress in the suspending fluid and its interaction with the finite-size dispersed phase is expected to complement studies using the continuum based approaches e.g. \cite{hormozi2017dispersion}. 

\subsection{Outline}

In the following sections, we start by describing the experimental set-up and the PIV-PTV measurement techniques. The fluid rheology and particle properties are also mentioned. Later, in the results section, we first describe the single phase flow of two YSF, with high, HYS, and low, LYS, yield stress, at varying $Re^*$ and $Bi$. This is then followed by the results concerning the particle-laden cases. Finally, in the discussion section, we try to interpret the above results by supporting our main observations with simple visual images obtained from shadowgraphy experiments, which are described therein. This is followed by a summary of the main conclusions. 

\section{Experimental technique}

\subsection{Experimental set-up}

\begin{figure}
\centering
\begin{subfigure}{.45\textwidth}
  \centering
  \includegraphics[height=0.75\linewidth]{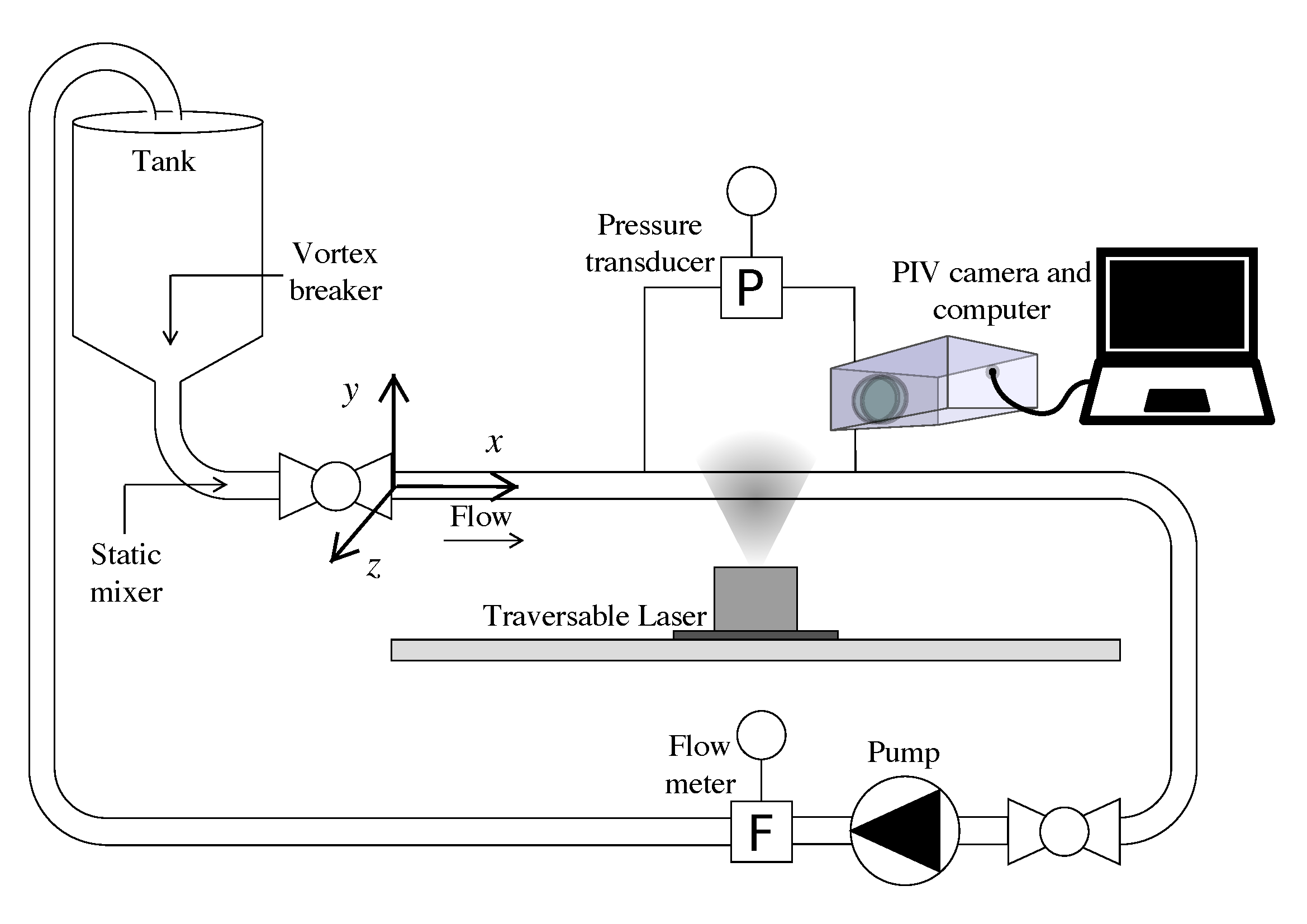}
  \caption{Schematic of the flow-loop}
  \label{fig:Set-up schematic}
\end{subfigure}%
\begin{subfigure}{.45\textwidth}
  \centering
  \includegraphics[height=0.75\linewidth]{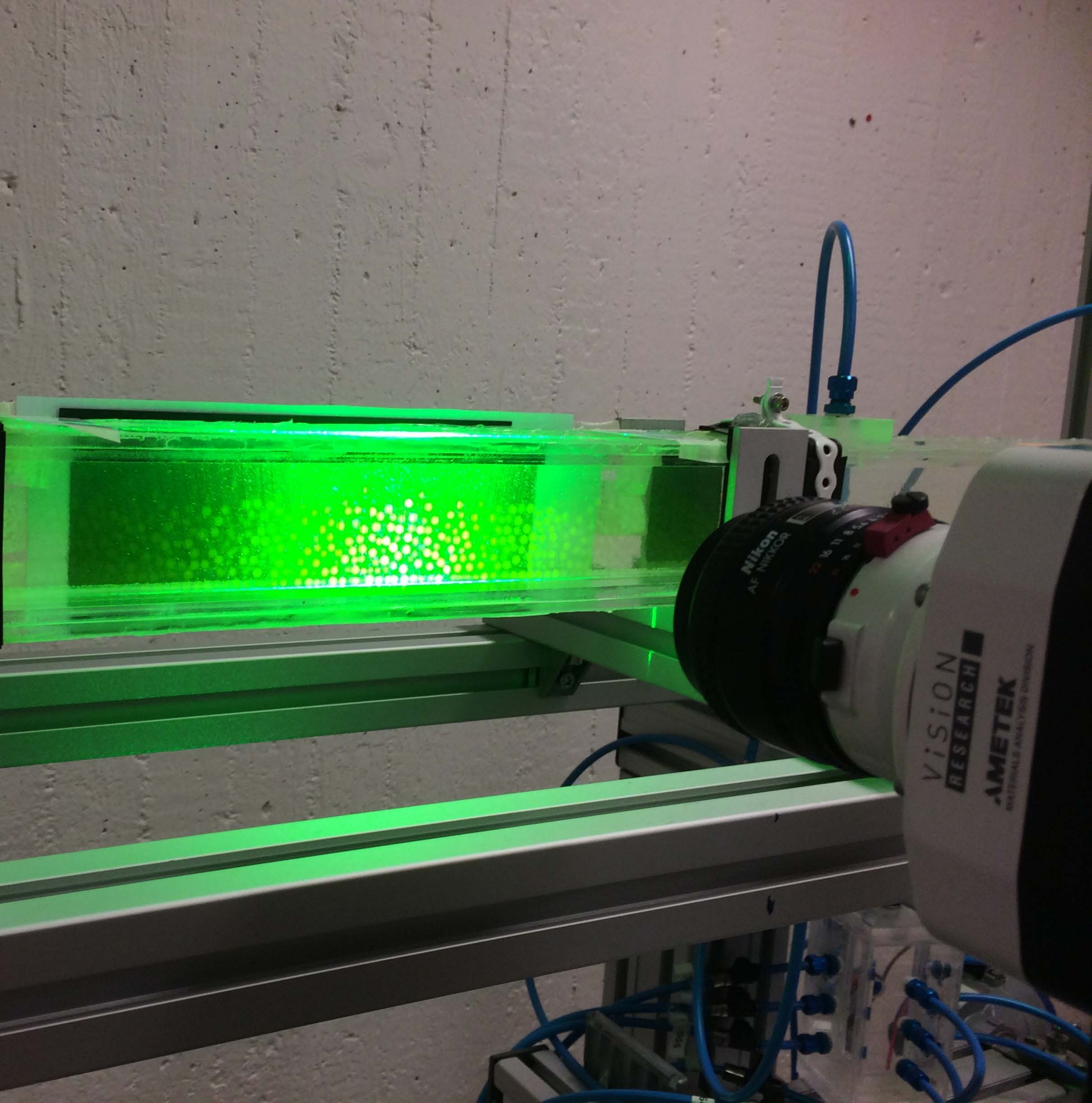}
  \caption{Section where PIV is performed}
  \label{fig:PIV set-up}
\end{subfigure}
\caption{Experimental set-up. Reproduced from \cite{zade2018experimental}} 
\label{fig:Set-up}
\end{figure}

The experimental set-up used in this study has been previously used in \cite{zade2018experimental} for investigating particle laden turbulent flows of Newtonian and non-Newtonian flows up to $\phi$ = 20\%, and is shown in figure \ref{fig:Set-up}. Only the most important features are described here and the interested reader is referred to the previous publications for additional details. The flow loop consists of a 5 m long transparent square duct with 50 mm x 50 mm cross section (refer to figure \ref{fig:Set-up schematic}). The particles are introduced in the conical tank and are circulated through the loop for a long enough time before measurements, so that they homogeneously disperse in the carrier fluid. A static mixer (Vortab Company, CA, USA) is installed close to the inlet of the duct to neutralize any swirling motions that may arise from the long 90$^\circ$ bend at the exit of the tank. The temperature is maintained at nearly 20\textdegree C by means of an immersed-coil heat-exchanger in the tank. A very gentle disc pump (Model: 2015-8-2HHD Close coupled, Discflo Corporations, CA, USA) has been chosen, similar to \cite{draad1998laminar}, to minimise the mechanical breakage of the particles and avoid unwanted pulsations in the flow. 

An electromagnetic flowmeter (Krohne Optiflux 1000 with IFC 300 signal converter, Krohne Messtechnik GmbH, Germany) is used to measure the volume flow rate of the particle-fluid mixture. The pressure drop is measured at a streamwise distance of nearly $140H$ from the inlet across a length of $54H$ using a differential pressure transducer (0 - 1 kPa, Model: FKC11, Fuji Electric France, S.A.S.). Data acquisition from the camera, flow meter and pressure transducer is performed using a National Instruments NI-6215 DAQ card using Labview\textsuperscript{TM} software. The entry flow problem for YSF has been studied by \cite{ookawara2000unified}, amongst others, who proposed that the spatial development of the streamwise velocity at the edge of the plug is a suitable indicator for a fully developed flow. The development length was deduced in terms of a modified Reynolds number accounting for the plug radius, and for the maximum $Re^*$ encountered in our case, this development length would be less than 30$H$. Thus, at our measurement location $\approx$ 150$H$, the flow can be considered to be fully developed for the single phase cases. For the particle laden cases, the evolution of the particle concentration is qualitatively observed using shadowgraphy techniques, described later in the discussion section, and it is observed that an equilibrium concentration is established upstream of the measurement location.

\subsection{Fluid rheology}

An aqueous solution of Carbomer powder supplied as Carbopol$^{\textregistered}$ NF 980 (Lubrizol Corporation, USA), a commercial thickener in anhydrous powder form consisting of cross-linked polyacrylic acid resins, and neutralized with Sodium Hydroxide (NaOH), is used as the model yield stress fluid. It is chosen due to its high transparency, small thixotropy and material stability (very slow ageing). Under neutral pH the resin is anionic i.e. the many side chains carry a negative charge and have the ability to absorb and retain water and swell so enormously that concentrations of even below 0.1\% mass are sufficient for the particles to form a percolated network and form a yield stress fluid \citep{piau2007carbopol}. 

For each of our Carbopol batches, a weighed amount of Carbopol powder (0.25\% by weight) was first dispersed in 24 kg of water at room temperature ($\sim$20\textdegree C). A high shear mixer (Silverson AX5, Silverson Machines, Inc., USA) was used for dispersing  the powder at a maximum of $\sim$1200 rpm for up to $\sim$30 minutes. The dispersion was then left stationary for $\sim$30 minutes to allow for air bubbles to evacuate. Then, a 18 wt./v \% of NaOH solution at 2.3 times the mass of Carbomer powder is used for neutralising the dispersion. The NaOH was gradually added with a pipette whilst stirring the solution gently at low rotational speeds ($\sim$135 rpm) using a helical mixer (RW 20 DZM-P4, IKA-Labortechnik, IKA-Werke GmbH \& Co. KG, Germany). At the end of the neutralisation process, the pH was noted to be ideally within the required range (6.5--7.0) so as to ensure maximum swelling and hence, a high yield stress. The neutralized solution was further mixed in a large cement mixer (1402 HR, AL-KO, Germany) at a low rotational speed for $\sim$5 additional hours to ensure complete homogenisation.

The whole preparation process lead to a concentrated solution with a high yield stress. Experiments had to be performed with a thinner (less viscous) fluid in order to facilitate pumping. Thus, the concentrated solution was gradually diluted in the flow loop by mixing with water and recirculating for $\sim$2 hours before measurements. Since the shear rates in the flow loop are not extremely high, solution degradation and the resulting change in properties can be considered insignificant. Repeatability of the properties of final solution was ascertained by (i) measuring the flow curves in a rheometer and (ii) by monitoring the pressure drop at a fixed flow rate. It was found that by systematically following the above mixing protocol, solutions with nearly the same pressure drop exhibiting the same flow curve were obtained repeatedly. In order to minimise the occurrence of bubbles arising due to dissolved gases in water, tap water was heated so as to remove these dissolved gases and then cooled before being used for preparing the Carbopol solution. Air bubbles entrained during introduction of the hydrogel particles were removed in the conical tank that is open to the atmosphere.

\begin{figure}
\centering
 \includegraphics[height=0.4\linewidth]{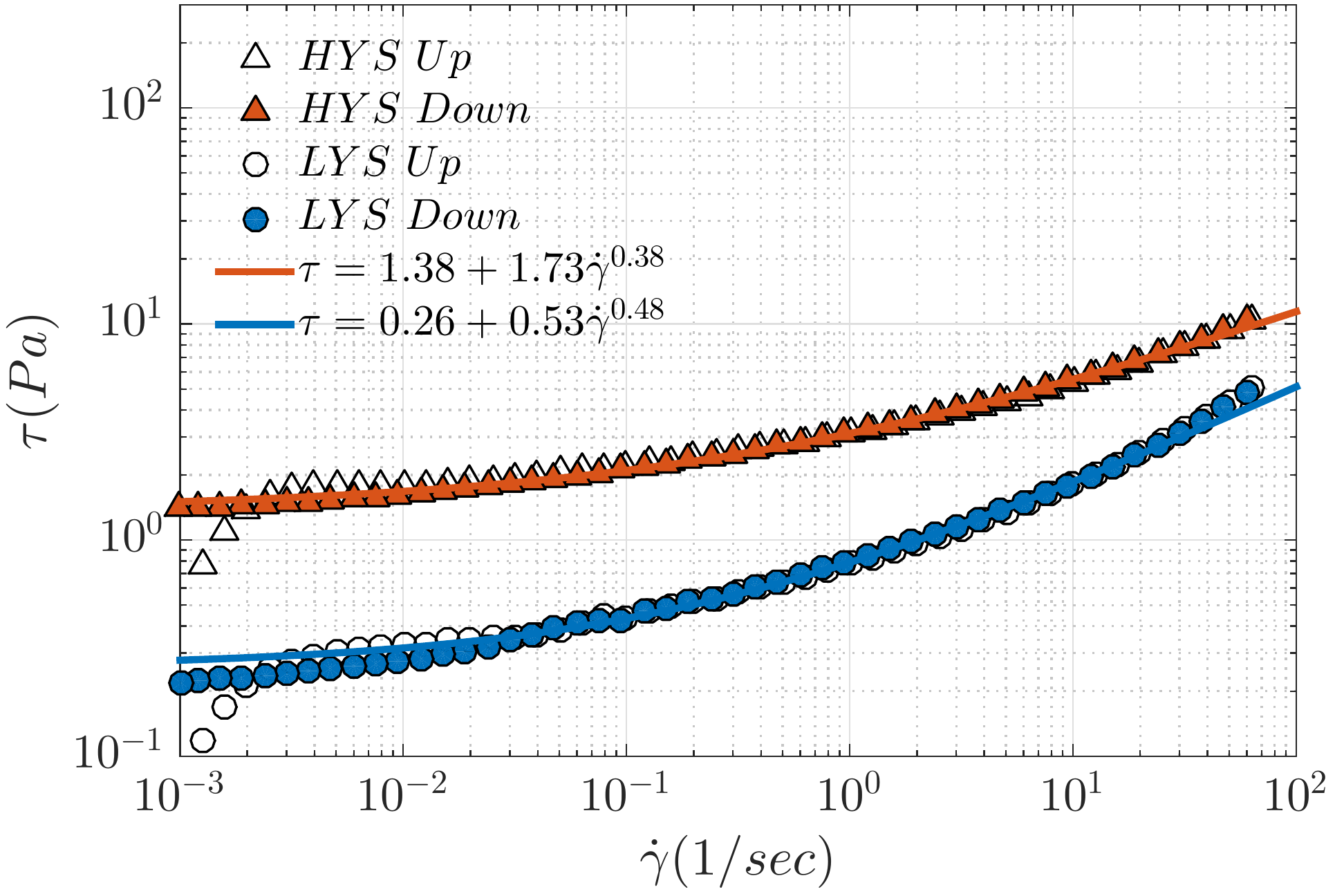}
\caption{Flow curves (symbols) and the HB model fit (solid lines) for the low LYS and high HYS yield stress fluids. The open and closed symbols denote measurements during ascending and descending shear rates respectively.}
\label{fig:Flow_curves}
\end{figure}

The rheological tests focused on the determination of  the steady-state flow curve, i.e. the relation between shear stress and shear rate in simple shear, were performed using a TA AR-2000ex rheometer (TA Instruments, Inc., USA) with a vane and cup geometry. A pre-shear of 300 sec$^{-1}$ was applied for 10 sec followed by a rest period of 40 seconds to establish a reproducible state. Flow sweeps in both ascending and descending controlled shear-rate were then carried out in a range of shear rates (0.001 to 80 1/sec with 10 points per decade) at room temperature to determine the flow curves and to check for thixotropy. This range of shear rates includes the maximum shear rate in the experiments. For each measurement point, the shear rate was held constant for 30 sec. Figure \ref{fig:Flow_curves} shows the flow curve for the two fluids used in this study. Elastic effects on start-up \citep{dinkgreve2017everything} are visible for a few points at low shear rates during the \textit{up} sweep i.e. the accumulated strain in perhaps insufficient for the material to flow. Hence, the \textit{down} sweep curve is used for fitting the Herschel-Bulkley model, whose parameters are mentioned in the legend of figure \ref{fig:Flow_curves}. The repeatability of the flow curves was ensured by performing the above tests on three separate solutions prepared independently of each other, both before and after adding particles, and an average curve is used. The working fluids display a shear-thinning behaviour after yielding i.e. they are yield-pseudo plastic fluids. The dynamic yield stress determined above (static yield stress is determined using creep tests) is low compared to other wall-bounded flow studies of YSF (see \cite{guzel2009observation} and \cite{escudier1996pipe} amongst others) but for such flows, the non-dimensional $Bi$ is the relevant criteria to assess the importance of plasticity on the flow. This point about the relevance of $Bi$ will be illustrated further by means of velocity profiles that exhibit a solid plug.

\begin{figure}
\centering
 \includegraphics[height=0.4\linewidth]{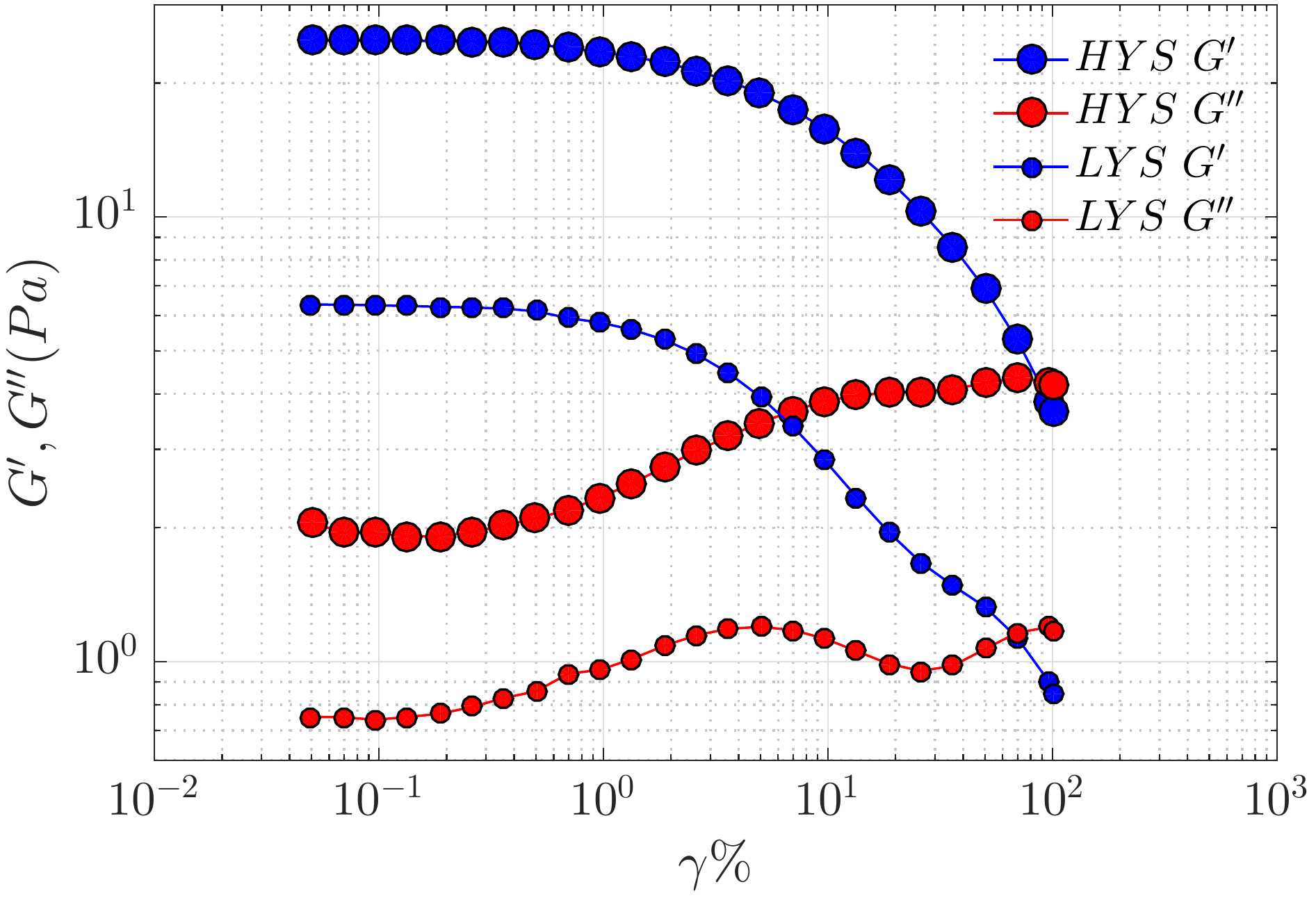}
\caption{Oscillatory measurements: Amplitude of the strain, $\gamma$, sweeps at a fixed angular frequency $\omega$ = 1 Hz for the two fluids used in this work.}
\label{fig:Linear_viscoelastic_Gp_Gpp}
\end{figure}

\begin{figure}
\centering
\begin{subfigure}{0.5\textwidth}
  \centering
  \includegraphics[height=0.65\linewidth]{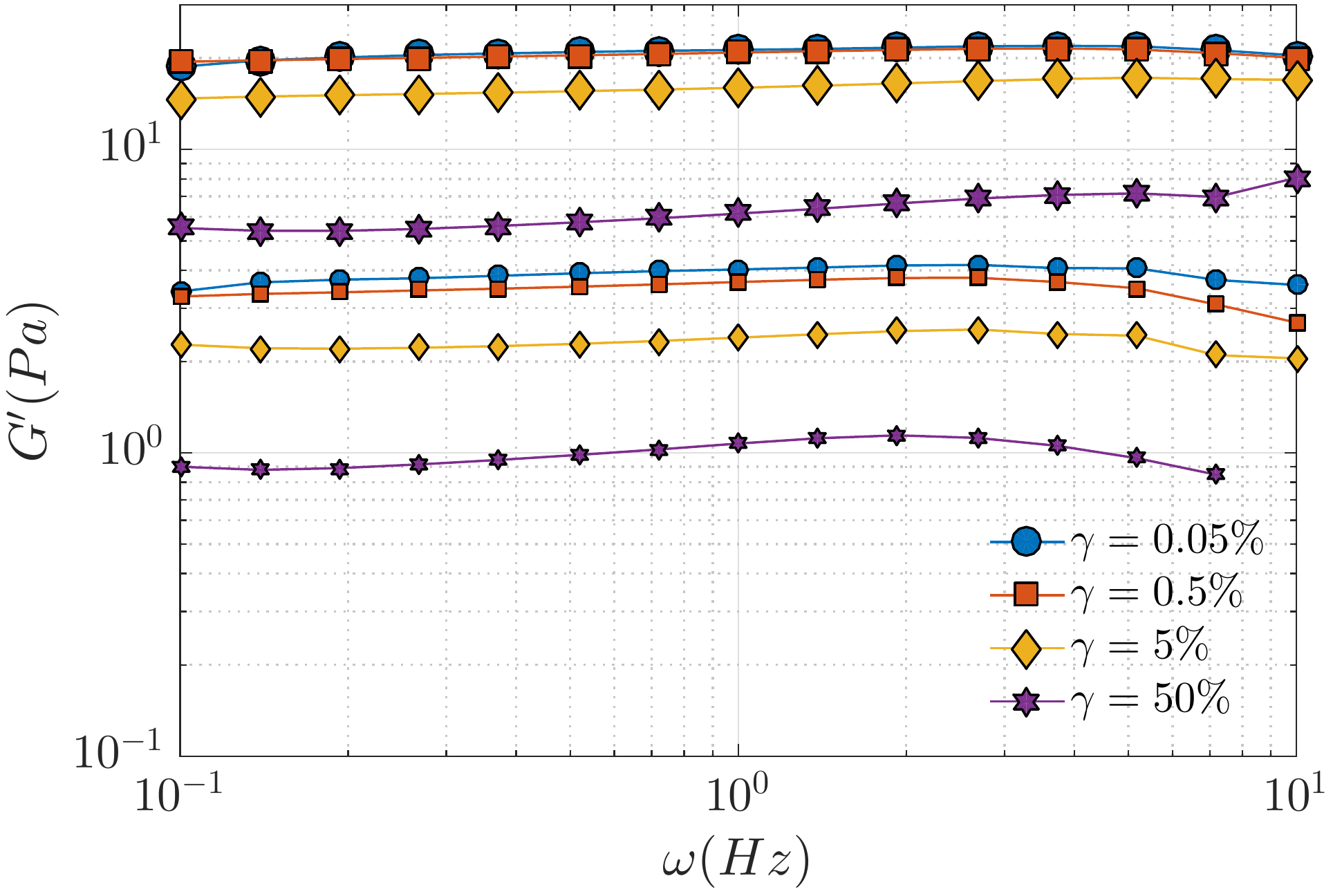}
  \caption{$G'$}
  \label{fig:Gp}
\end{subfigure}%
\begin{subfigure}{0.5\textwidth}
  \centering
  \includegraphics[height=0.65\linewidth]{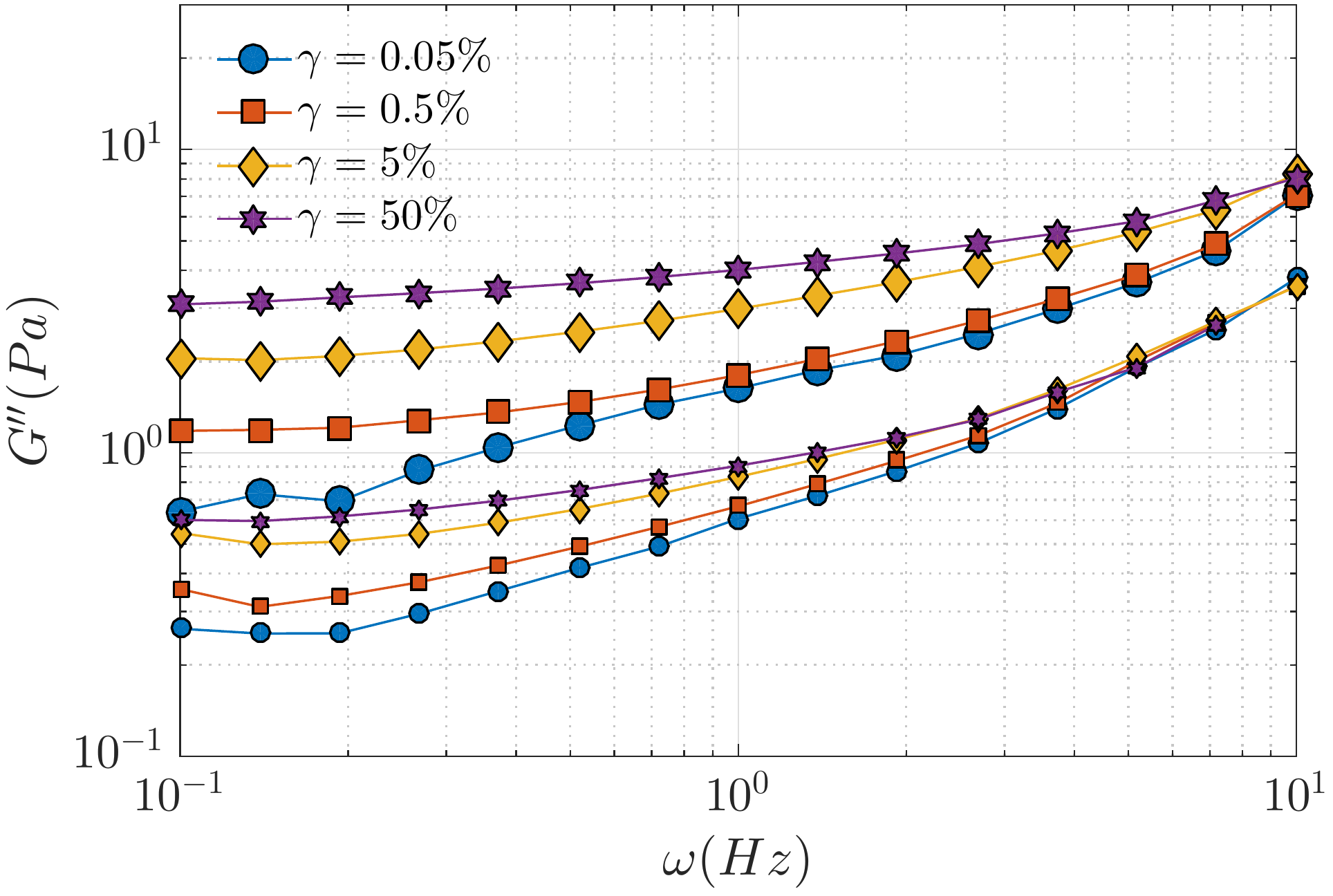}
  \caption{$G''$}
  \label{fig:Gpp}
\end{subfigure}
\caption{Oscillatory measurements: Frequency $\omega$ sweeps for determining (a) the elastic modulus $G'$ and the viscous modulus $G''$ for several strain amplitudes. The larger and smaller symbols refer to the HYS and LYS cases, respectively. }
\label{fig:Gp_and_Gpp}
\end{figure}

The viscoelastic behaviour near the yield limit is measured by oscillatory tests using a splined bob and cub attachment in a Kinexus pro+ rheometer (Malvern Panalytical, UK). The linear viscoelastic region that exists at small strain $\gamma$ amplitudes can be identified by a nearly constant elastic $G'$ and viscous $G''$ moduli over a wide range of strain amplitudes. This is shown in figure \ref{fig:Linear_viscoelastic_Gp_Gpp} where oscillatory measurements are performed at a frequency $\omega$ of 1 Hz. The point of cross over between $G'$ and $G''$ for each of the two fluids at 1 Hz, $H\tau_y\sim$ 4 Pa and $L\tau_y\sim$ 1 Pa, is of similar order of magnitude as the yield stress, albeit slightly larger. Figure \ref{fig:Gp_and_Gpp} shows the variation of $G'$ and $G''$ with the frequency $\omega$ at multiple strain amplitudes $\gamma$, starting from the linear and extending towards the non-linear viscoelastic regime associated with yielding. In the linear regime i.e. at lower $\gamma$, the elastic moduli $G'$ (refer to figure \ref{fig:Gp}) is approximately flat and around one order of magnitude greater than the viscous moduli $G''$ (refer to figure \ref{fig:Gpp}). $G'$ and $G''$ are nearly independent of $\gamma$ in this linear regime ($\gamma \leq$ 0.5\%), as expected. As mentioned in \cite{r2019rheological}, at these low amplitudes, the microgels deform elastically but, they do not move significantly relative to each other and hence, have a low internal dissipation \cite[also see ][]{piau2007carbopol}. As the deformation amplitude increases, the microstructure starts breaking, causing elastohydrodynamic friction forces to rise and dissipation to increase leading to a higher loss modulus. Finally, at higher $\gamma$, outside the purview of small amplitude oscillatory shear, the elastic components decrease and the viscous components increase. Similar observations have been reported in \cite{firouznia2018interaction, gutowski2012scaling, r2019rheological, piau2007carbopol} amongst others. The viscoelastic moduli $G'$ and $G''$ are relevant at small shear rates. In order to quantify the viscoelastic effects at higher shear rates, other steady state viscometric functions namely, the first and second normal stress differences: $N_1$ and $N_2$ needs to be measured. These measurements are fraught with difficulties in YSF and for $N_1$ both positive \citep{tehrani1996experimental, peixinho2005laminar, piau2007carbopol} and negative values \citep{janmey2007negative} are reported. In the recent work of \cite{de2019yield}, the authors investigated common simple i.e. non-thixotropic YSF and found that $N_1 >$ 0 and $N_2 <$ 0 with $N_1 > N_2$; they increase quadratically with the shear stress, both below and above the yield stress. The interested reader is referred to the important work by \cite{mahaut2008yield}, who investigated the influence of varying particle size and volume fraction on the viscoelastic moduli and effective yield stress of different types of yield stress fluids (emulsions, gels and colloidal suspensions) and found a simple expression describing these changes. For further information about the change in the HB parameter due to the particle phase, and measurements concerning particle migration and velocity field during rheological measurements, the work of \cite{ovarlez2015flows} is of particular importance. These authors found that suspensions in YSF have the same HB index as the interstitial fluid.

\subsection{Particle properties}

As already noted in our previous work: \cite{zade2018experimental}, the particles are commercially procured super-absorbent (polyacrylamide based) hydrogel spheres which are delivered in dry form. After grading them into different sizes using sieves, one fairly mono-dispersed fraction is used in these experiments. These dry particles are mixed with tap water, that has a very small amount of fluorescent Rhodamine dye (a few ppm) dissolved in it, and then left submerged for around one day. Ultimately, they grow to an equilibrium size of : 4.2$\pm$0.8 mm (3 times standard deviation) thus, yielding a duct height to particle diameter ratio $2H/d_p$ of $\sim$12. The tiny amount of Rhodamine is absorbed by each particle, and makes them glow in the laser sheet, so that they can be detected in the PIV images as will be shown later. The particle size was determined both by a digital imaging system and from the PIV images of particles in the flow. A small Gaussian like spread in the particle diameter was observed. The fact that a Gaussian like particle size distribution has small effect on the flow statistics has already been shown in \citet{FORNARI201854}. 

The density ratio of the particle to tap water $\rho_{p}/\rho_{f}$ is found to be equal to 1.0035$\pm$0.0003. Since the final concentration of Carbomer in the working fluid is less than 0.1\%, the change in density of the fluid from tap water is assumed to be insignificant. Since the typical flow velocities are relatively small ($U_{Bulk} < 0.3$ m/s) in the present flow configuration, the corresponding dynamical forces are low under these conditions and the hydrogel particles did not exhibit any visible deformation (as also confirmed from the time-resolved movies of the particles in flow). Also, as pointed out by \cite{gondret2002bouncing}, for the low impact Stokes number, collisions between particles or between a particle and the wall are dominated by viscous effects, and particles do not rebound i.e. the effective coefficient of restitution will be very small. Based on the above arguments, the particles can be considered to behave as rigid spheres. For details about the density measurements and calculations pertaining to the restitution coefficient, the reader is referred to our previous work \cite{zade2019buoyant}.

\subsection{Velocity measurement technique}

\begin{figure}
\centering
\begin{subfigure}{.33\textwidth}
  \centering
  \includegraphics[height=0.9\linewidth]{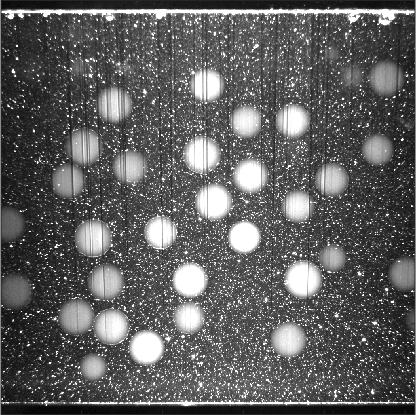}
  \caption{Raw image}
  \label{fig:PIVraw}
\end{subfigure}%
\begin{subfigure}{.33\textwidth}
  \centering
  \includegraphics[height=0.9\linewidth]{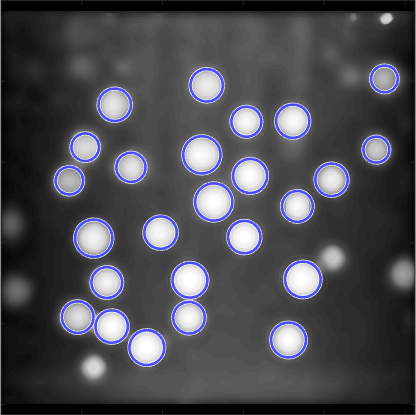}
  \caption{Enhanced image for PTV}
  \label{fig:PTVprocessed}
\end{subfigure}%
\begin{subfigure}{.33\textwidth}
  \centering
  \includegraphics[height=0.9\linewidth]{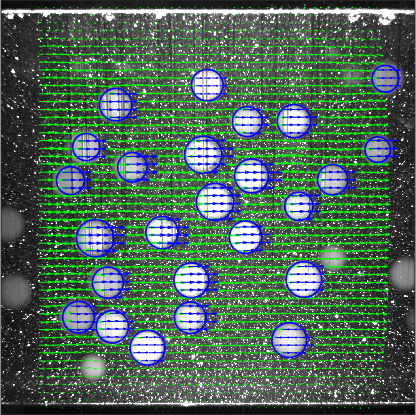}
  \caption{PIV + PTV}
  \label{fig:PIVPTV}
\end{subfigure}%
\caption{Images involved for calculating the fluid and particle velocity field. The raw image (a) is used for PIV analysis. The enhanced image (b) is used to detect circles and subsequent PTV analysis. The combined PIV (for fluid phase denoted by green arrows) and PTV (for particle phase denoted by blue arrows) velocity field is shown in (c). The detected particles are also shown in figure (b). The above images correspond to LYS fluid, $\phi$ = 10\%, $z/H$ = 0.9 and $Re^*$ = 156.}
\end{figure}

The coordinate system is sketched in figure \ref{fig:Set-up schematic} with $x$ in the streamwise, $y$ in the wall-normal and $z$ in the span-wise directions. The velocity field is measured using 2D Particle Image Velocimetry (2D-PIV) in multiple span-wise planes ($z/H$ = 0 to 1) to measure the in-homogenous velocity field in the square duct. The PIV set-up consists of a continuous wave laser (wavelength = 532 nm, power = 2 W) and a high-speed camera (Phantom Miro 120, Vision Research, NJ, USA) as shown in figure \ref{fig:PIV set-up}. The thickness of the laser light-sheet is 1 mm. The following paragraphs are briefer reproductions of the PIV-PTV algorithm already described in \cite{zade2019buoyant}.

PIV image pairs are captured at a resolution of approximately 60 mm/1024 pixels. For each flow rate, a frame rate (acquisition frequency) was chosen so that the maximum pixel displacement, based on the mean velocity, did not exceed a quarter of the size of the final interrogation window IW \citep{raffel2013particle}. Images were processed using an in-house, three-step, FFT-based, cross-correlation algorithm \citep{kawata2014velocity}. The degree of overlap is around 47\% and can be estimated from the fact that the corresponding final resolution is 1 mm x 1 mm per IW. Between 200--400 image pairs have been observed to be sufficient to ensure statistically converged results.

Figure \ref{fig:PIVraw} depicts one image from a typical PIV sequence for particle-laden flow. Raw images captured during the experiment were saved in groups of two different intensity levels. The first group of images, a typical example being figure \ref{fig:PIVraw}, were used for regular PIV processing according to the algorithm mentioned above to find the fluid velocity field. The same first group of images were later contrast-enhanced in the post-processing step and constituted the second group of images, an example being figure \ref{fig:PTVprocessed}. These second group of images were used for detecting the particles. Again, the details of the image post-processing and the PTV algorithm is explained at length in \cite{zade2019buoyant}.

The velocity of both the fluid and particle phases is defined on a spatially fixed Eulerian grid. Consequently, a mask matrix is defined, which assumes the value 1 if the point lies inside the particle and 0, if it lies outside the particle. The particle velocity is determined using PTV at its center, which is assigned to the grid points inside the particle (mask = 1). The velocity field of the particle-phase is, now, available at the same grid points as that of the fluid and the ensemble averaging, reported later, are phase averaged statistics. Figure \ref{fig:PIVPTV} shows the combined fluid (PIV) and particle (PTV) velocity field. It may be observed that the intensity of the laser sheet is weaker away from the centre of the images and particles are not accurately detected in this region. Hence, PIV and PTV results are extracted from a reduced area neglecting this poorly illuminated region, as shown in figure \ref{fig:PIVPTV}.

\section{Results}

The results are described first in terms of  pressure drop. The velocity field for single phase flow is described later, before the velocity and concentration profiles for the particle-laden cases are presented.

\subsection{Pressure drop}

\begin{figure}
\centering
\begin{subfigure}{0.5\textwidth}
  \centering
  \includegraphics[height=0.65\linewidth]{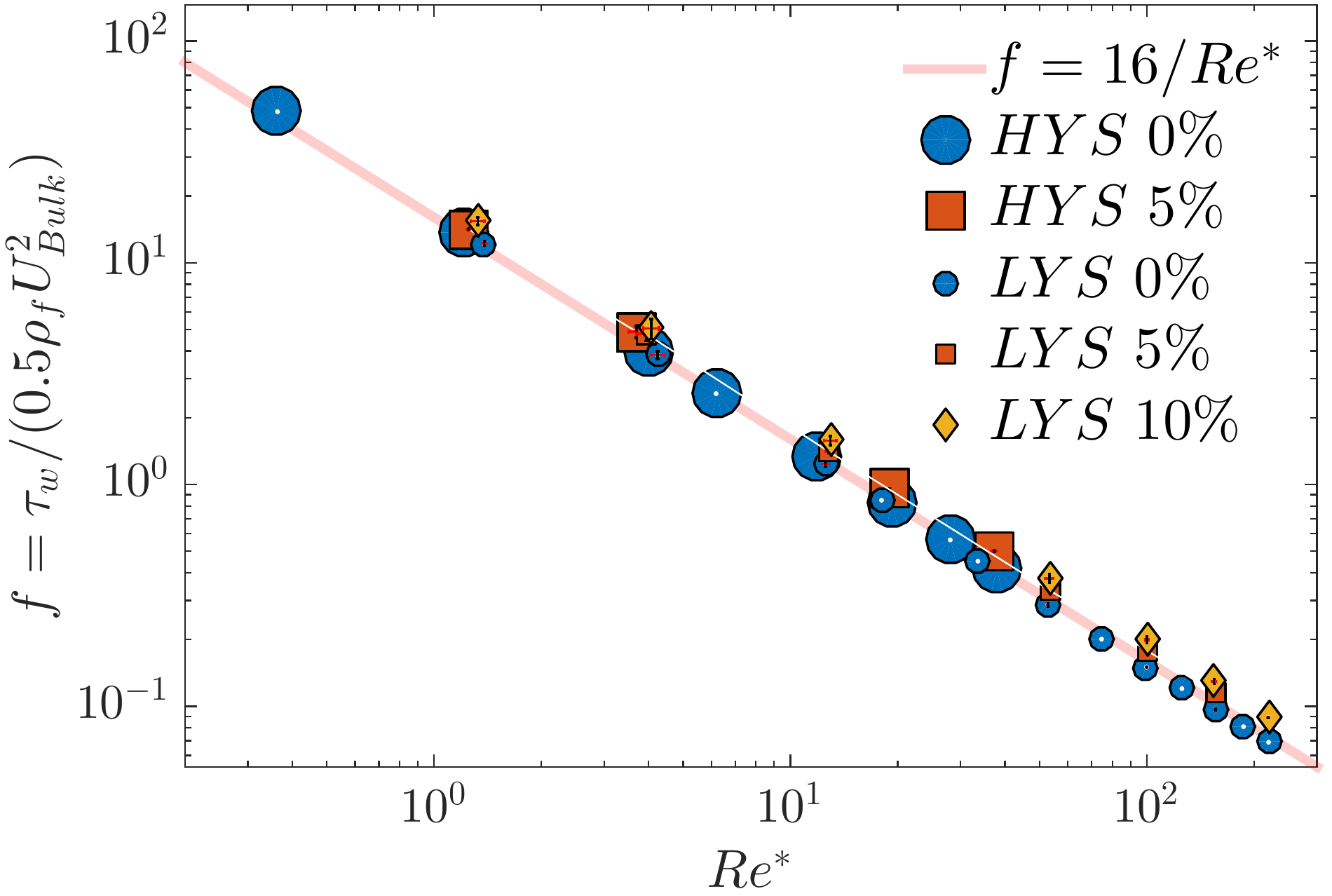}
  \caption{}
  \label{fig:Re_vs_f}
\end{subfigure}%
\begin{subfigure}{0.5\textwidth}
  \centering
  \includegraphics[height=0.65\linewidth]{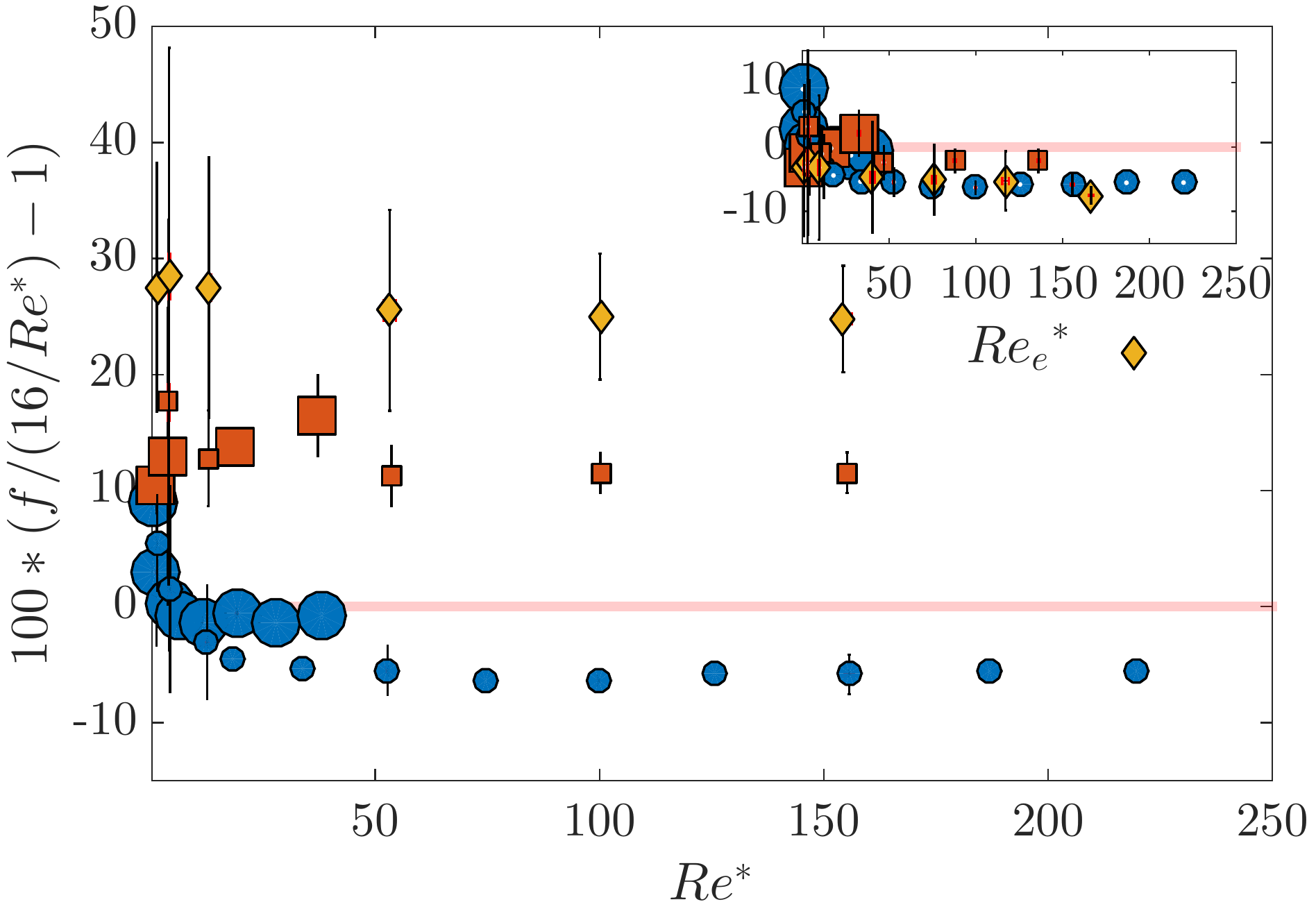}
  \caption{}
  \label{fig:Re_vs_f_diff}
\end{subfigure}
\caption{(a) Variation of friction factor $f$ with $Re^*$ for both single phase and particle laden cases. (b) The percentage deviation of the measured $f$ compared to the $f$ expected for single phase laminar flow. The inset in figure (b) shows the same plot but with a Reynolds number ${Re_e}^*$ accounting for the additional suspension viscosity due to particles.}
\label{fig:Re_vs_f_all}
\end{figure}

Figure \ref{fig:Re_vs_f} displays the friction factor $f$ as a function of the Reynolds number $Re^*$ for all the cases investigated in this study. The single-phase cases seem to agree reasonably well with the $16/Re^*$ curve, which is the analytical solution for Poiseuille flow of a Newtonian fluid. This is noteworthy considering the fact that a complex expression is used to define  $Re^*$, see equation (\ref{eqn:Re}). There are measurable changes due to the addition of particles and to better appreciate them, the data in the figure are re-plotted as a percentage change compared to the analytical solution $16/Re^*$ in figure \ref{fig:Re_vs_f_diff}. The deviation for single phase flow is higher at the lowest $Re^*$ for HYS fluid but, drops to values very close to the analytical solution for higher $Re^*$. At particle volume fraction $\phi$ = 5\%, within the extent of the error-bars, the increase in pressure drop is independent of the suspending fluid and the $Re^*$. The departure in pressure drop from the single phase case is more pronounced for $\phi$ = 10\%. Note that for $\phi$ = 10\%, experiments are conducted only in LYS.

The increase in the viscosity of a suspension due to the addition of particles can be semi-empirically described using the Eilers fit \citep{stickel2005fluid} as follows 
\begin{equation}
  \frac{\mu_e}{\mu} = \Big(1 + \frac{5}{4}\frac{\phi}{1-\phi/0.65}\Big)^2.
  \label{eqn:Eilers fit}
\end{equation}
This effective suspension viscosity $\mu_e$ is only a function of the nominal particle concentration $\phi$. In the limit of low particle inertia and a uniform particle distribution, it may be used to predict the change in the friction factor. Accordingly, in the inset of figure \ref{fig:Re_vs_f_diff}, a new effective Reynolds number ${Re_e}^*$ is defined using $\mu_e$ such that ${Re_e}^* = Re^* \mu / \mu_e$ and the experimentally measured friction factor $f$ at $Re^*$ is compared to the expected friction factor $16/{Re_e}^*$ for an effective fluid. A good collapse of all the data points within $\pm$5\% is observed using this approach. Thus, the friction factor seems to be well described using an effective viscosity formulation for the volume fractions used in this study. In our previous work \citep{zade2018experimental}, we have observed that the friction factor in a turbulent flow is a function of both the particle volume fraction and size. Also, from the results in \cite{zade2019turbulence}, we learnt that the predictions from Eilers fit diverge at higher volume fractions $\phi \ge$ 15\% for Newtonian turbulent flows. Thus, the good collapse observed in this study could be a consequence of the low Reynolds number and low volume fractions considered for these two elastoviscoplastic fluids. Note that the Reynolds number at the particle scale is reasonably small ($Re_{p,\dot{\gamma_{Wall}}} = \rho \dot{\gamma_{Wall}} {(d_p/2)}^2/ \mu_{Wall} \le$ 2) and the particle distribution inside the duct is non-uniform as will be shown later. Nevertheless, the suitability of using an effective suspension viscosity for describing the pressure drop in a complex suspending fluid matrix is impressive. Different particle sizes, volume fractions, Reynolds numbers and types of suspending fluids needs to be investigated to check the limitations of this approximation.

A mean secondary flow is also observed in the duct, as will be shown in the coming sections. However, considering the weak amplitude of this flow, it is assumed to have a negligible effect on the friction factor, as also previously observed by \cite{gao1993steady}. Another point to note is that the Reynolds number, if defined using the viscosity at the wall as $Re_{\mu_{Wall}} = \rho U_{Bulk} (2H)/\mu_{Wall}$, is very similar in magnitude (within 10\%) to the $Re^*$ used in this study. The viscosity at the wall $\mu_{Wall}$, also used in the previous paragraph to estimate the particle scale Reynolds number, can be calculated using the flow curves in figure \ref{fig:Flow_curves} and the average shear stress at the wall,  estimated from the streamwise pressure gradient. 

\subsection{Velocity statistics for single phase flow}

\begin{figure}
\centering
\begin{subfigure}{.50\textwidth}
  \centering
  \includegraphics[height=0.8\linewidth]{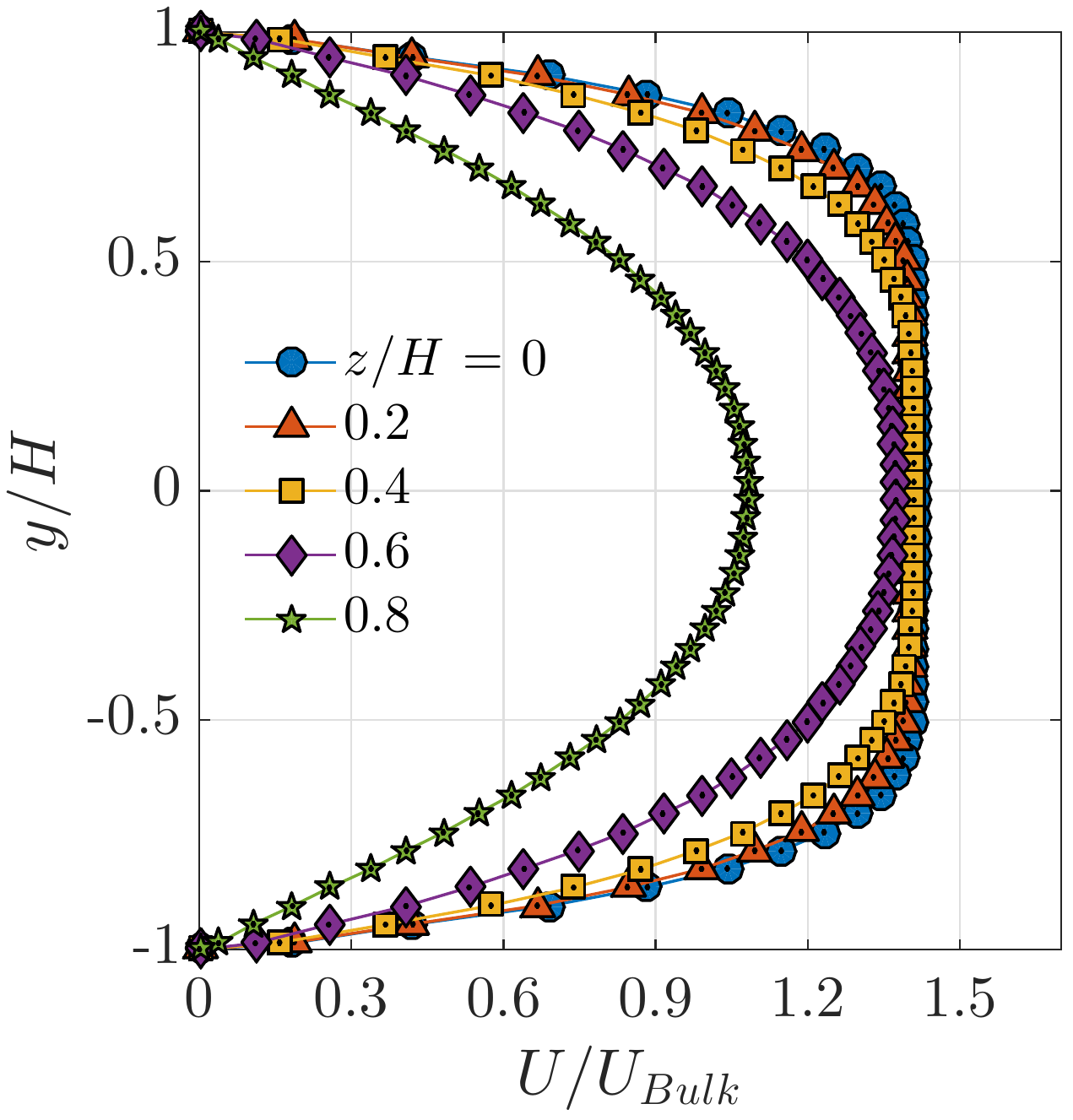}
  \caption{}
  \label{fig:Thickest_8lpm_U_by_Ubulk}
\end{subfigure}%
\begin{subfigure}{.50\textwidth}
  \centering
  \includegraphics[height=0.8\linewidth]{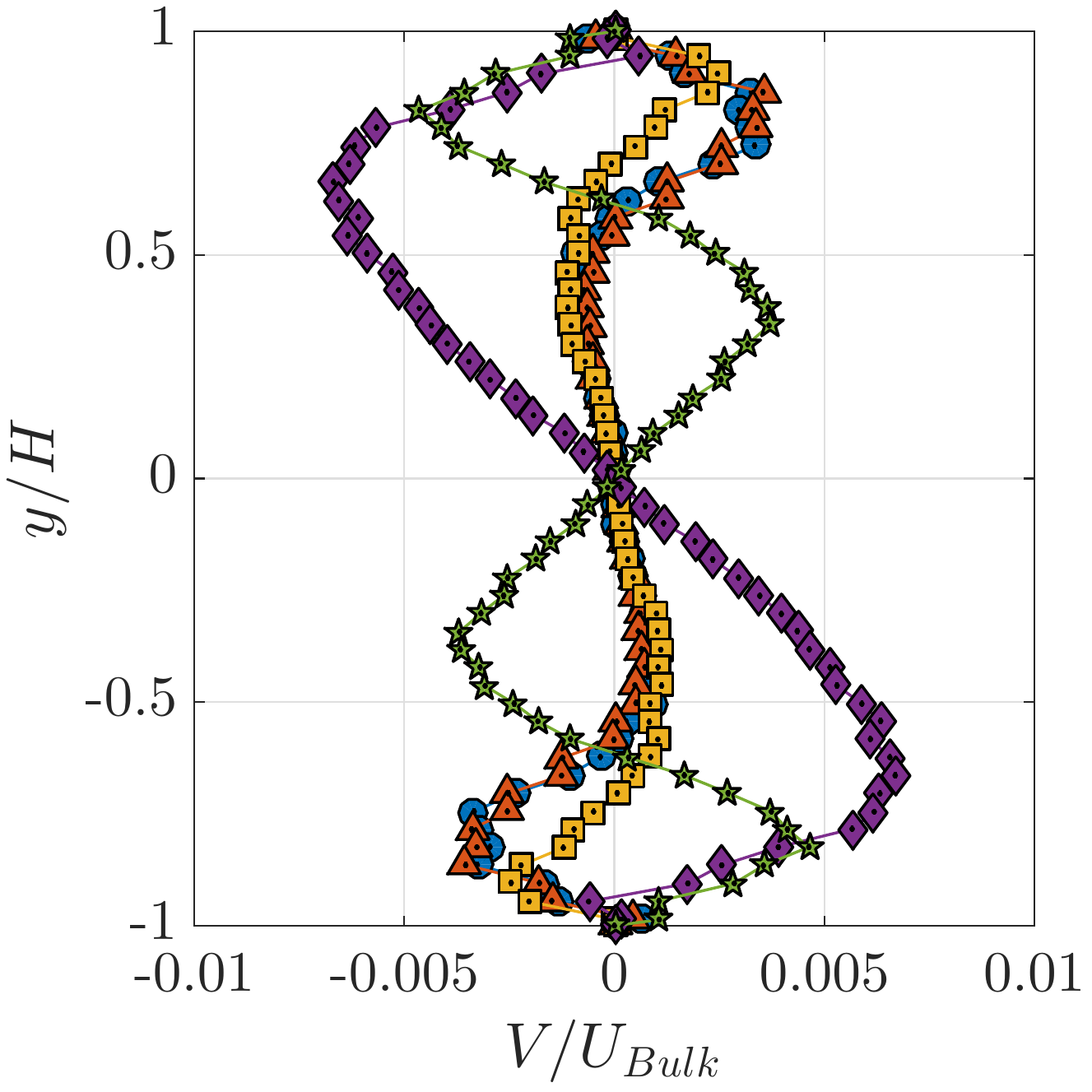}
  \caption{}
  \label{fig:Thickest_8lpm_V_by_Ubulk}
\end{subfigure}
\caption{(a) Streamwise and (b) Wall normal velocity profiles in different span-wise planes for high YSF (HYS). Flow rate (Q) = 8 liters/minute (lpm), $Re^*$ = 4, $Bi$ = 0.26.}
\label{fig:Thickest 8lpm U and V}
\end{figure}

\begin{figure}
\centering
\begin{subfigure}{.50\textwidth}
  \centering
  \includegraphics[height=0.8\linewidth]{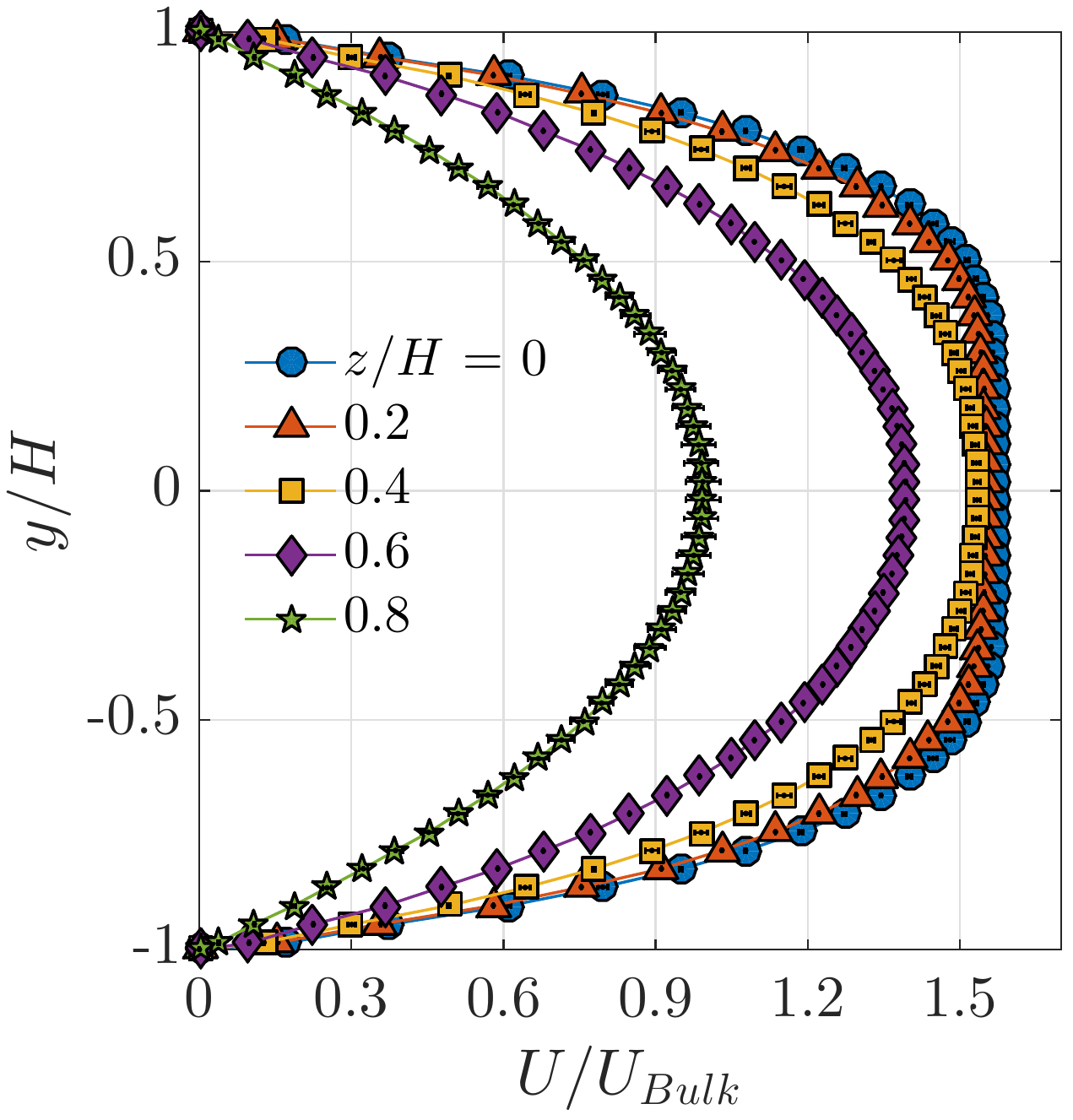}
  \caption{}
  \label{fig:Thick_8lpm_U_by_Ubulk}
\end{subfigure}%
\begin{subfigure}{.50\textwidth}
  \centering
  \includegraphics[height=0.8\linewidth]{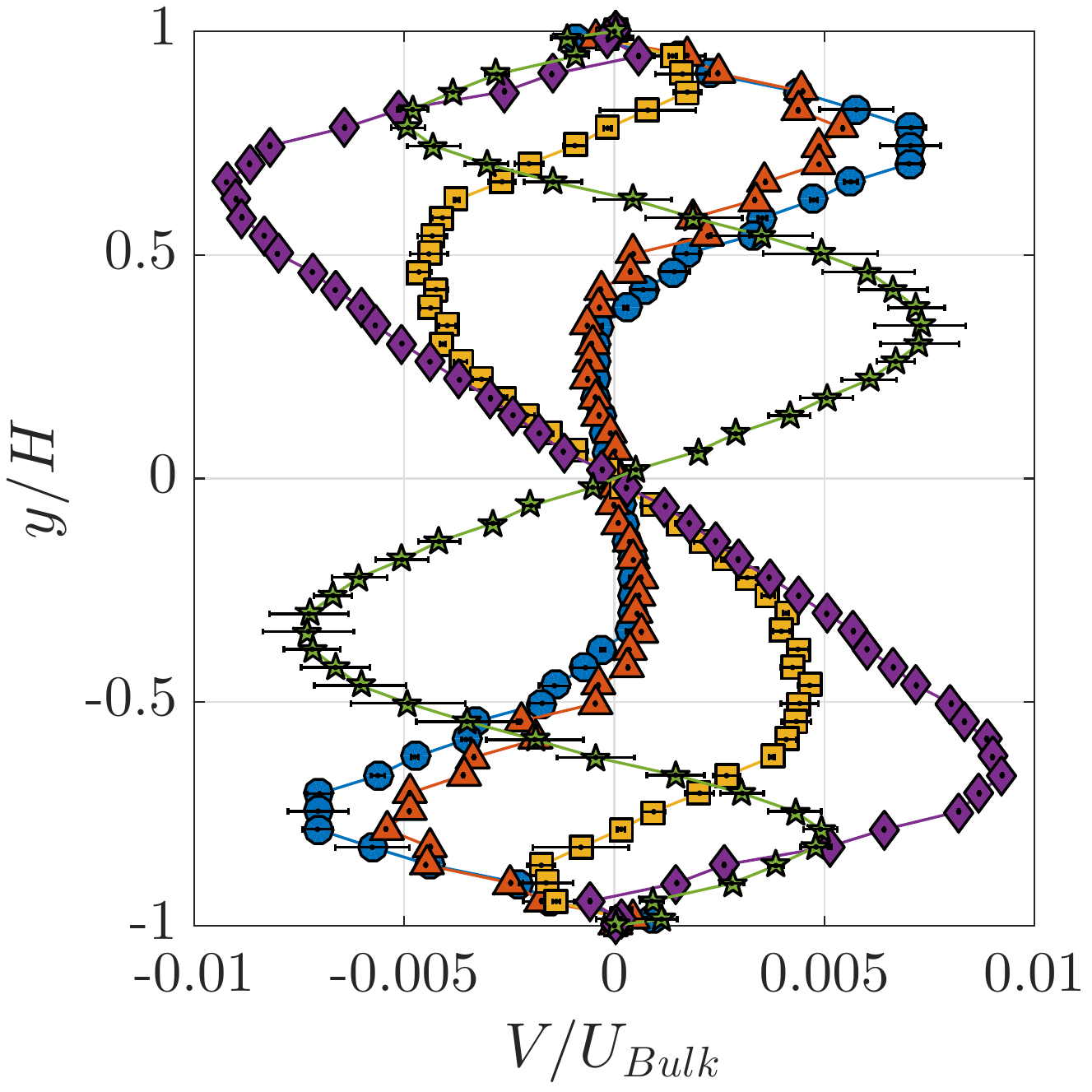}
  \caption{}
  \label{fig:Thick_8lpm_V_by_Ubulk}
\end{subfigure}
\caption{Caption similar to figure \ref{fig:Thickest 8lpm U and V} but this time for the low YSF (LYS) at the same flow rate Q = 8 lpm. $Re^*$ = 12.5, $Bi$ = 0.15.}
\label{fig:Thick 8lpm U and V}
\end{figure}

For the single phase flow, PIV measurements were performed in multiple span-wise planes. An example of results of such measurements  is shown in figures \ref{fig:Thickest 8lpm U and V} and \ref{fig:Thick 8lpm U and V} for HYS and LYS, respectively, for the same flow rate. The streamwise velocity profiles, normalized by the bulk velocity, display a flat region of constant velocity in the center-plane $z/H$ = 0. This region of zero velocity gradient is representative of the solid plug and it shrinks while moving to a plane closer to the side wall $z/H$ = 1. At the same flow rate, the plug is larger for the thicker fluid HYS due to the high yield stress (compare figure \ref{fig:Thickest_8lpm_U_by_Ubulk} and \ref{fig:Thick_8lpm_U_by_Ubulk}). 

Figures \ref{fig:Thickest_8lpm_V_by_Ubulk} and \ref{fig:Thick_8lpm_V_by_Ubulk} show the variation of the wall-normal velocity profiles for the two fluids. Similar symbols represent similar planes. These velocities are a direct measure of the secondary flow and their maximum strength is close to 1\% of the bulk velocity. In the center-plane $z/H$ = 0, the region of zero velocity gradient for the streamwise velocity has also a nearly zero wall-normal velocity i.e. there is no secondary flow inside the plug. The most striking feature by comparing the two figures is that the intensity of the wall-normal velocities increases for the thinner fluid LYS.  The plug region is smaller for LYS as deduced from the corresponding streamwise velocity profiles, allowing for a larger fluidised region and hence, larger and stronger secondary flows. The action of the plug in damping the secondary flow was also observed in the simulations by \cite{letelier2018elastoviscoplastic}.

\begin{figure}
\centering
 \includegraphics[height=0.4\linewidth]{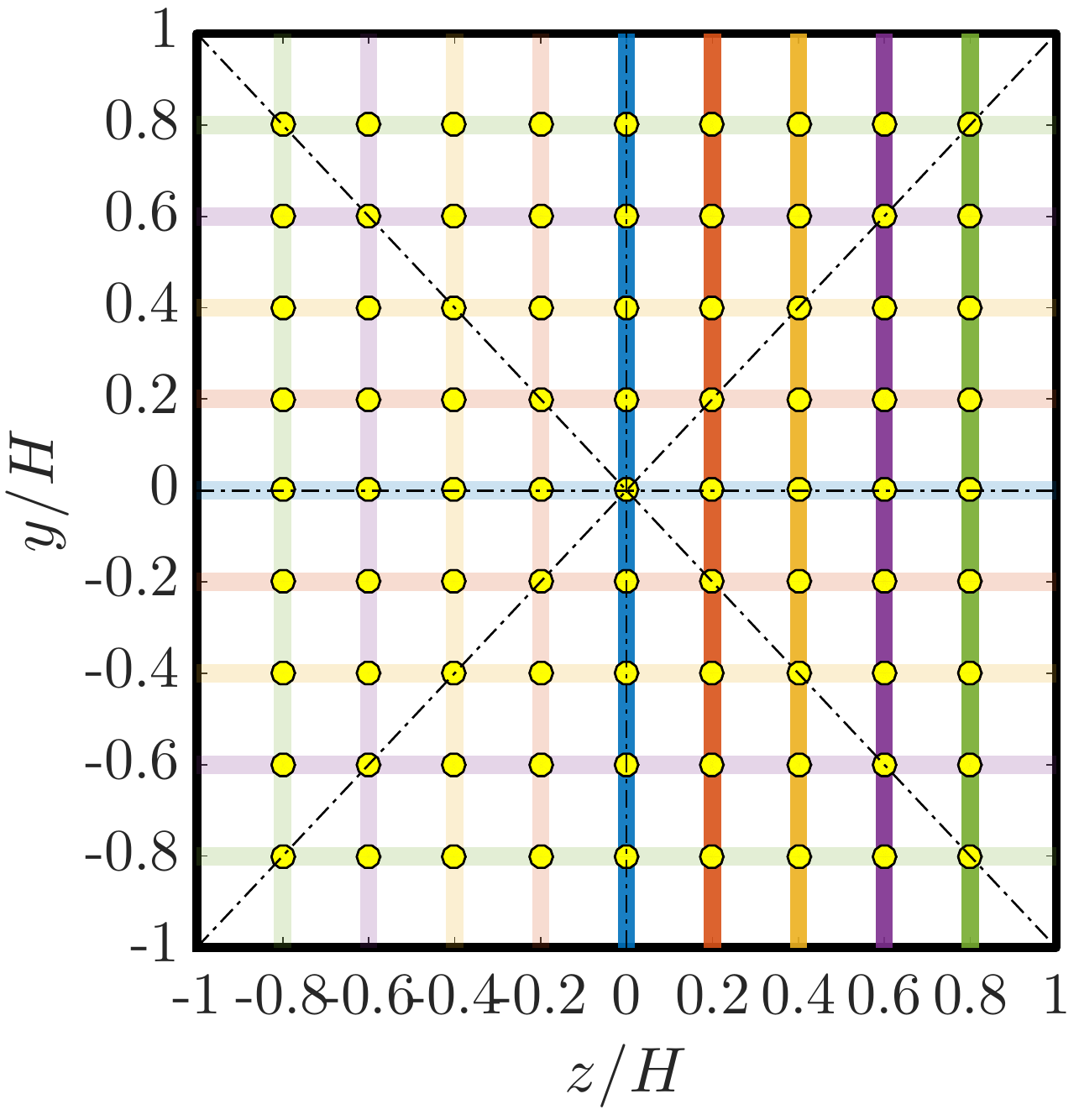}
\caption{Schematic of the symmetry lines in a square duct (dash-dotted lines). The span-wise planes where PIV is performed are represented with a darker colour. The mirror images of these planes about the symmetry lines are represented by lighter version of the same colours. The intersection points of these planes correspond to locations where all three velocity components can be evaluated.}
\label{fig:Symmetry}
\end{figure}

A rectangular duct flow is symmetric about the orthogonal axes $z-z$ and $y-y$. In the special case of the square duct, the flow is also symmetric about its diagonals. Thus, measurements of two velocity components in a limited number of span-wise planes (like in the 2D-PIV used here) can be used to calculate all the three velocity components at other locations in the duct. This approach is illustrated in the schematic of figure \ref{fig:Symmetry}. Thus, 2D PIV measurements in $N$ planes (in figure \ref{fig:Symmetry}, $N$ = 5 denoted by darker lines) can yield the 3D velocity at  $(2N-1)^2$ locations.

\begin{figure}
\centering
\begin{subfigure}{.50\textwidth}
  \centering
  \includegraphics[height=0.9\linewidth]{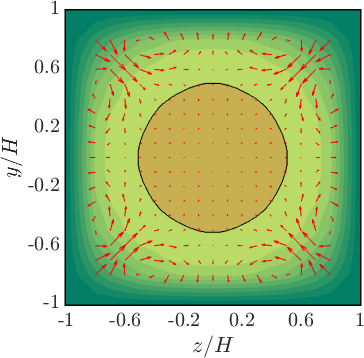}
  \caption{$Re^*$ = 1.2, $Bi$ = 0.30}
  \label{fig:Sec_flow_4lpm_SP}
\end{subfigure}%
\begin{subfigure}{.50\textwidth}
  \centering
  \includegraphics[height=0.9\linewidth]{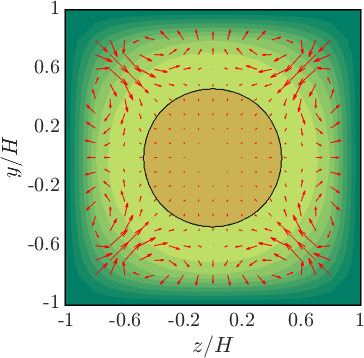}
  \caption{$Re^*$ = 4, $Bi$ = 0.26}
  \label{fig:Sec_flow_8lpm_SP}
\end{subfigure}
\begin{subfigure}{.50\textwidth}
  \centering
  \includegraphics[height=0.9\linewidth]{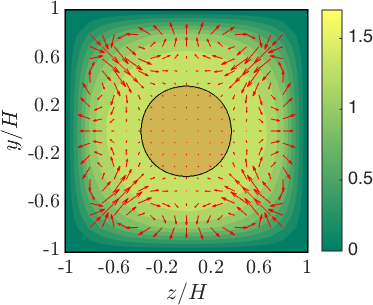}
  \caption{$Re^*$ = 19, $Bi$ = 0.20}
  \label{fig:Sec_flow_20lpm_SP}
\end{subfigure}
\caption{Variation of mean flow with increasing $Re^*$ for the high YSF (HYS). The colormap corresponds to the mean streamwise velocity normalized by the bulk velocity. The arrows represent the direction and strength of the secondary flow. The plug region at the centre is delimited by the solid line.}
\label{fig:Sec_flow_SP}
\end{figure}

The above technique is applied using $N$ = 9 planes for three different flow rates or $Re^*$ of the thicker fluid HYS as shown in figure \ref{fig:Sec_flow_SP}. The secondary flow can now be clearly visualised: it carries the high momentum fluid from the yielded regions around the core to the centre of the wall and low momentum fluid from the corners towards the core. The intensity of this secondary flow increases with increasing $Re^*$, coupled with the reduction of the plug size.

The extent of the plug at the duct core is determined as follows: a two dimensional polynomial of $6^{th}$ order is fitted to the streamwise velocity data (coefficient of determination $R^2 \ge$ 0.99). A no-slip condition is assumed at the boundaries. Higher order polynomials did not show any significant improvement in the quality of the fit. Using this fit, the streamwise velocity field was smoothly mapped on a plane of higher resolution (400$\times$400 points). The norm of the velocity gradient $\sqrt{(\partial {U} /\partial {y})^2 + \partial {U}/\partial {z})^2}$, using a simple forward difference scheme, is evaluated at each of these points and the plug is designated at those points where the norm is less than 1\% of its maximum value (occurs at the wall-center). This procedure naturally suggests the existence of small unyielded plug-like regions also at the corners where the velocity gradients are small. Despite the likely presence of these stagnant zones, as found in many previous numerical studies \citep{saramito2001adaptive}, they are not shown here due to their proximity to the corners where PIV measurements are not performed. Nevertheless, their size is assumed to be very small for the $Bi$ used in this study. The shape of the moving plug in the core region appears to be near-circular, more so for higher $Re^*$ or, equivalently, lower $Bi$.

\begin{figure}
\centering
 \includegraphics[height=0.4\linewidth]{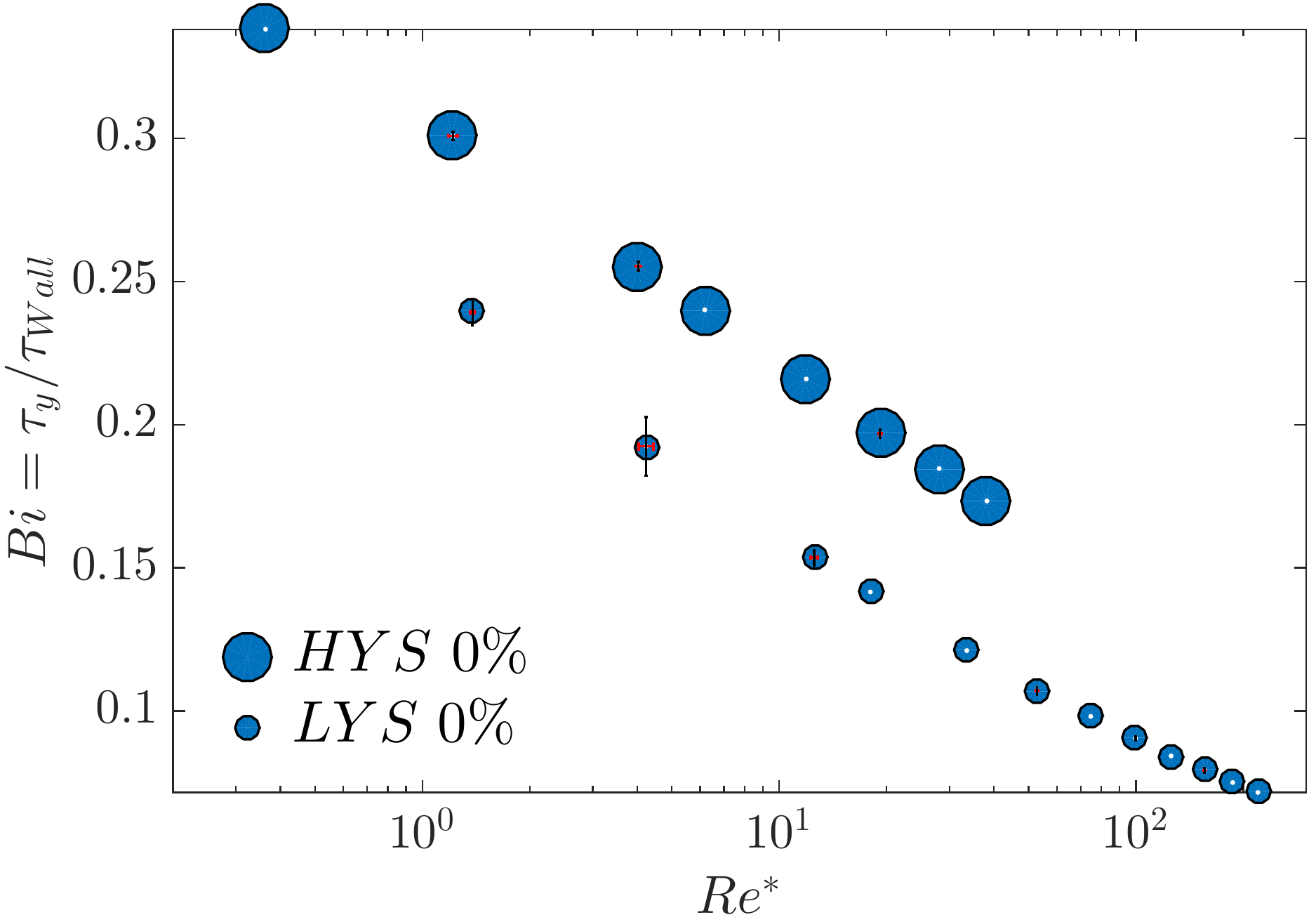}
\caption{Change of Bingham number $Bi$ as a function of the Reynolds number $Re^*$ for both the high HYS and low LYS yield stress fluids (single phase with no particles).}
\label{fig:Re_vs_Bi}
\end{figure}

Figure \ref{fig:Re_vs_Bi} shows the variation of $Bi$ with $Re^*$. For the same $Re^*$, the thicker fluid HYS has a larger $Bi$. As mentioned previously in the introduction, if a Bingham number $Bi$ based on the HB parameters is defined as $Bi_{HB} = \tau_y /(\kappa (U/L)^n)$, it ranges from 0.36 to 0.72 for HYS and from 0.14 to 0.43 for LYS i.e. around twice the value of the $Bi$ shown in figure \ref{fig:Re_vs_Bi}. From \cite{saramito2001adaptive}, it can be expected that with increasing $Bi$, both the moving plug at the center and the stagnant plug at the corners grow. The growing plug in the core would become non-circular before finally merging with the plugs at the corner at some critical $Bi$, whose value for a strictly Bingham fluid was found out to be $4/(2+\sqrt(\pi)) \approx$ 1.06 \citep{mosolov1965variational}. At this stage, the flow stops.

\begin{figure}
\centering
\begin{subfigure}{.5\textwidth}
\centering
 \includegraphics[height=0.8\linewidth]{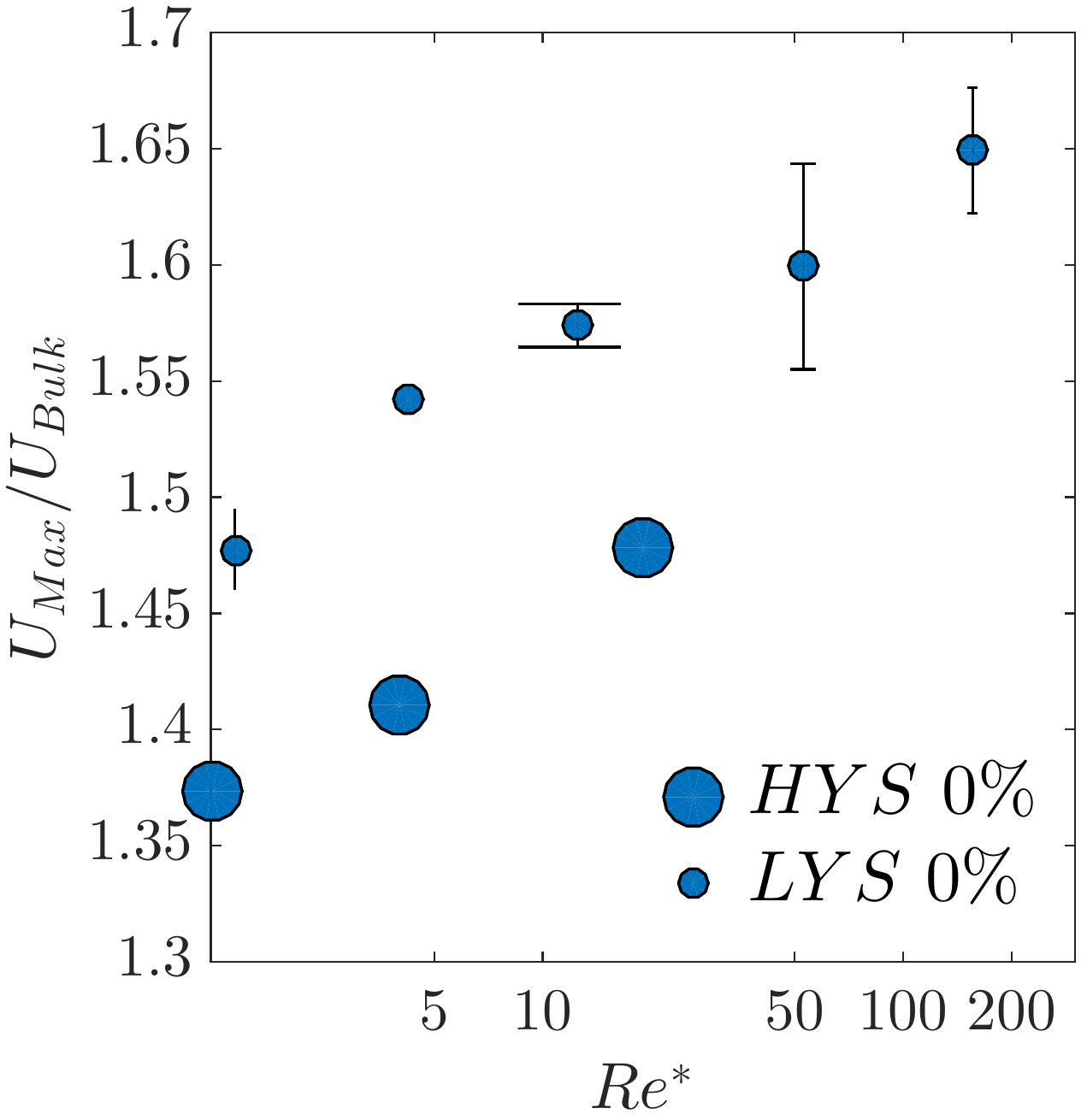}
   \caption{}
  \label{fig:Umax_Ubulk_vs_Re}
\end{subfigure}%
\begin{subfigure}{.5\textwidth}
\centering
 \includegraphics[height=0.8\linewidth]{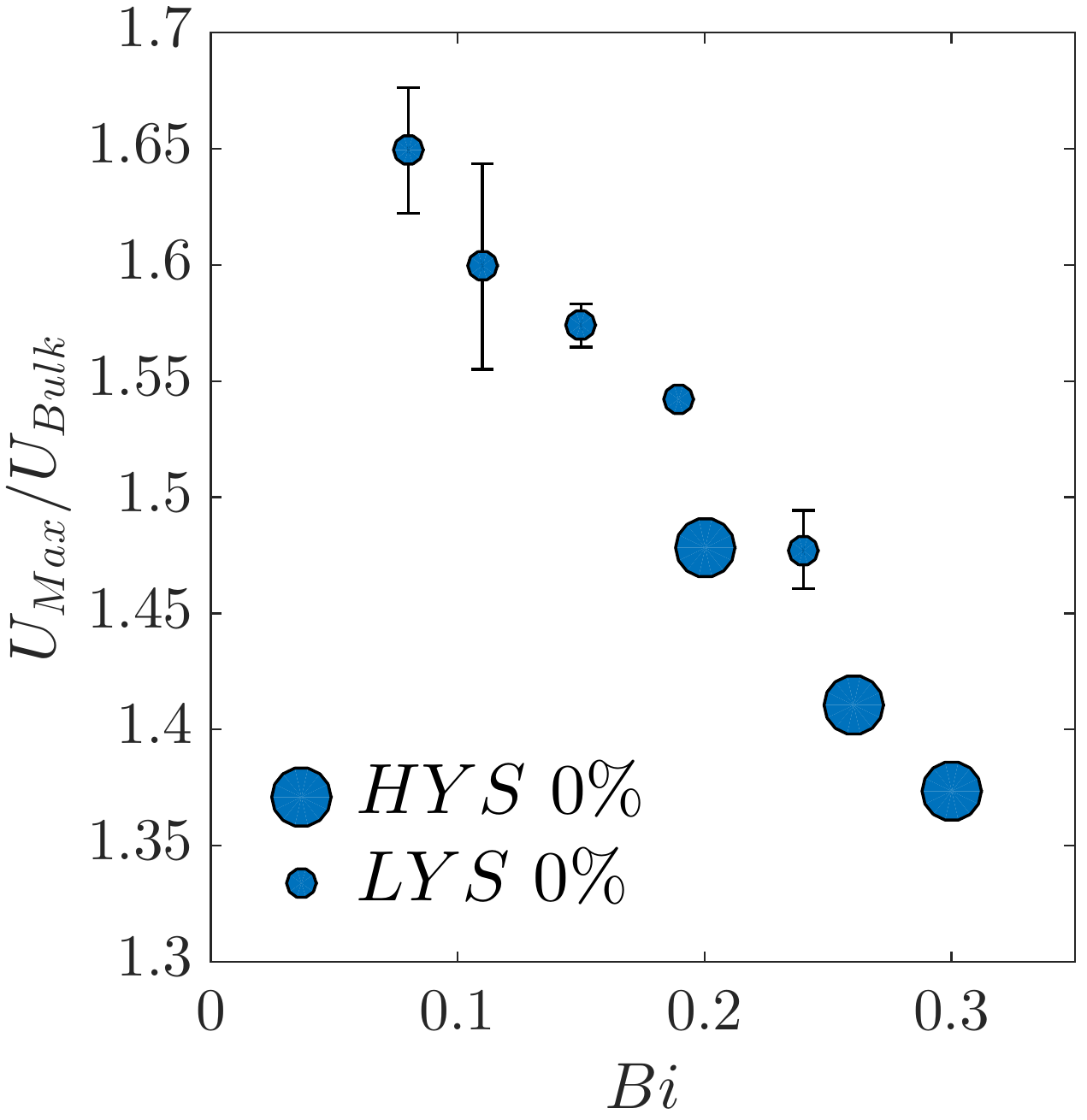}
   \caption{}
  \label{fig:Umax_Ubulk_vs_Bi}
\end{subfigure}%
\caption{Variation of maximum to bulk streamwise velocity ratio as a function of the (a) Reynolds number $Re^*$ and (b) Bingham number $Bi$ for both the high HYS and low LYS yield stress fluids (single phase with no particles). The maximum velocity is measured using PIV and the bulk velocity is measured from the flow meter.}
\label{fig:Umax_Ubulk}
\end{figure}

Before moving to the results for particle laden cases, figure \ref{fig:Umax_Ubulk} shows the change in the ratio of the maximum to bulk velocity for different $Re^*$ and  $Bi$. By virtue of a higher yield stress, at the same $Re^*$, HYS has a larger plug than LYS and hence, a flatter velocity profile i.e. a smaller $U_{Max}/U_{Bulk}$ (see figure \ref{fig:Umax_Ubulk_vs_Re}). Figure \ref{fig:Umax_Ubulk_vs_Bi} shows that, at the same $Bi$, HYS has a smaller $U_{Max}/U_{Bulk}$ i.e. a larger plug region. This is in agreement with \cite{huilgol2005application}, who observed that the plug region is larger for a fluid with a lower power law exponent $n$, for fixed Bingham number $Bi$. From figure \ref{fig:Flow_curves}, it is deduced that LYS has a higher $n$ than HYS but a smaller plug, as indicated by the higher $U_{Max}/U_{Bulk}$. 

\subsection{Particle-laden flow field}

\begin{figure}
\centering
\underline{$\phi$ = 5\%}
\begin{subfigure}{.33\textwidth}
$Q$ = 4 lpm
  \centering
  \includegraphics[height=0.9\linewidth]{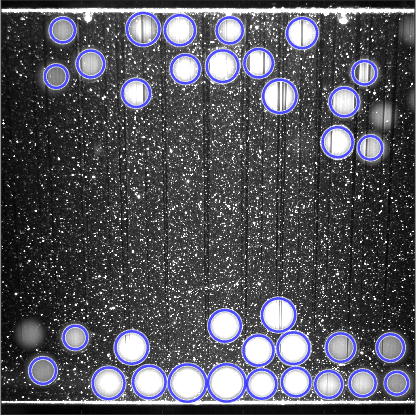}
  \caption{$Re^*$ = 4.2}
  \label{PIV_5p_4lpm}
\end{subfigure}%
\begin{subfigure}{.33\textwidth}
$Q$ = 20 lpm
  \centering
  \includegraphics[height=0.9\linewidth]{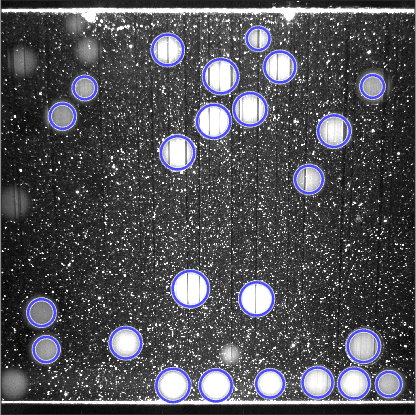}
  \caption{$Re^*$ = 53}
  \label{PIV_5p_20lpm}
\end{subfigure}%
\begin{subfigure}{.33\textwidth}
$Q$ = 40 lpm
  \centering
  \includegraphics[height=0.9\linewidth]{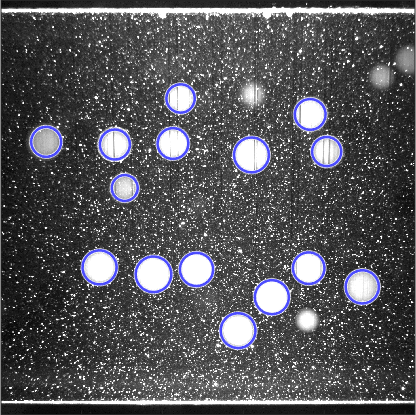}
  \caption{$Re^*$ = 156}
  \label{PIV_5p_40lpm}
\end{subfigure}%
\caption{Instantaneous snapshots of the particle distribution for increasing $Re^*$ (or flow rate $Q$) in a plane close to the wall ($z/H$ = 0.9). The above representative images correspond to the LYS fluid for $\phi$ = 5\%. The corresponding flow rates of the mixture in liters per minute (lpm) are mentioned on the top.}
\label{fig:Particle concentration images}
\end{figure}

The greatest advantage of using particles with the same refractive index as the suspending fluid is the ability to assess the spatio-temporal particle concentration even at high volume fractions. The different panels of figures \ref{fig:Particle concentration images} depict the instantaneous distribution of particles for $\phi$ = 5\% in a plane close to the side-wall of the duct ($z/H$ = 0.9) for increasing $Re^*$. 
A plane closer to the wall is chosen since most of the particles are seen to migrate towards this region. 
At low flow rates, most of the particles migrate towards the top and bottom of this plane i.e. to the corners. Additional particles migrating towards the corners surround the existing particles, thus forming layers. This is especially visible at $\phi$ = 10\% and will be shown in the next figure. A distinct change in their distribution is observed as the flow rate is increased: particles migrate away from the corners towards the centre of the near-wall plane. Since the total number of particles in this near-wall plane decreases with increasing flow rate, particles also migrate away from the side-wall, towards the duct core. Particles detected using the algorithm described earlier are outlined with a blue colour. The difference in size of the particles is not due to the variability in their physical diameter but to the fact that only the portion of the particle intercepting the laser-sheet is illuminated.

\begin{figure}
\centering
\underline{$\phi$ = 5\%}
\begin{subfigure}{.33\textwidth}
$Q$ = 4 lpm
  \centering
  \includegraphics[height=1\linewidth]{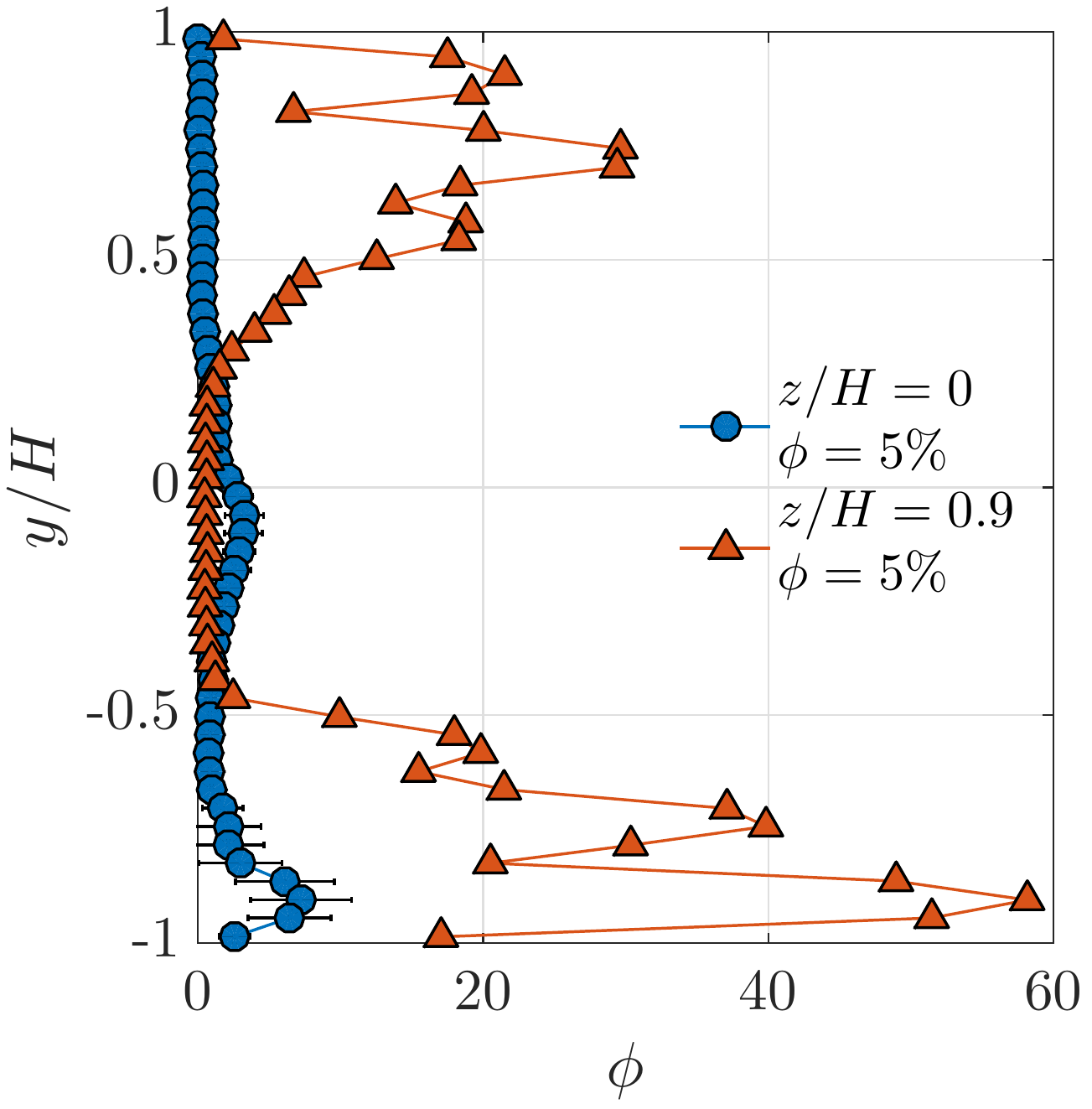}
  \caption{$Re^*$ = 4.2}
  \label{fig:phi_5p_4lpm}
\end{subfigure}%
\begin{subfigure}{.33\textwidth}
$Q$ = 20 lpm
  \centering
  \includegraphics[height=1\linewidth]{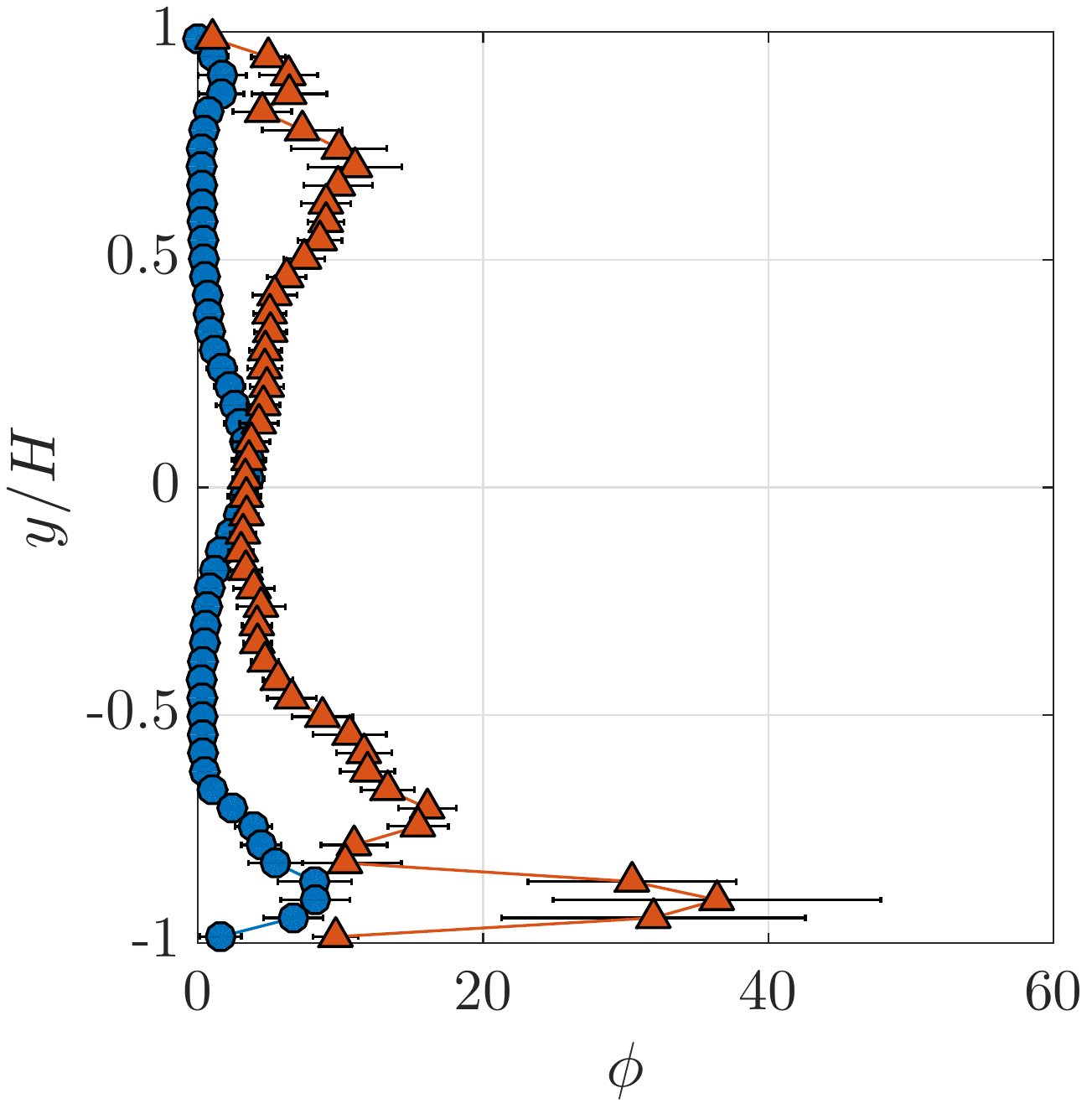}
  \caption{$Re^*$ = 53}
  \label{fig:phi_5p_20lpm}
\end{subfigure}%
\begin{subfigure}{.33\textwidth}
$Q$ = 40 lpm
  \centering
  \includegraphics[height=1\linewidth]{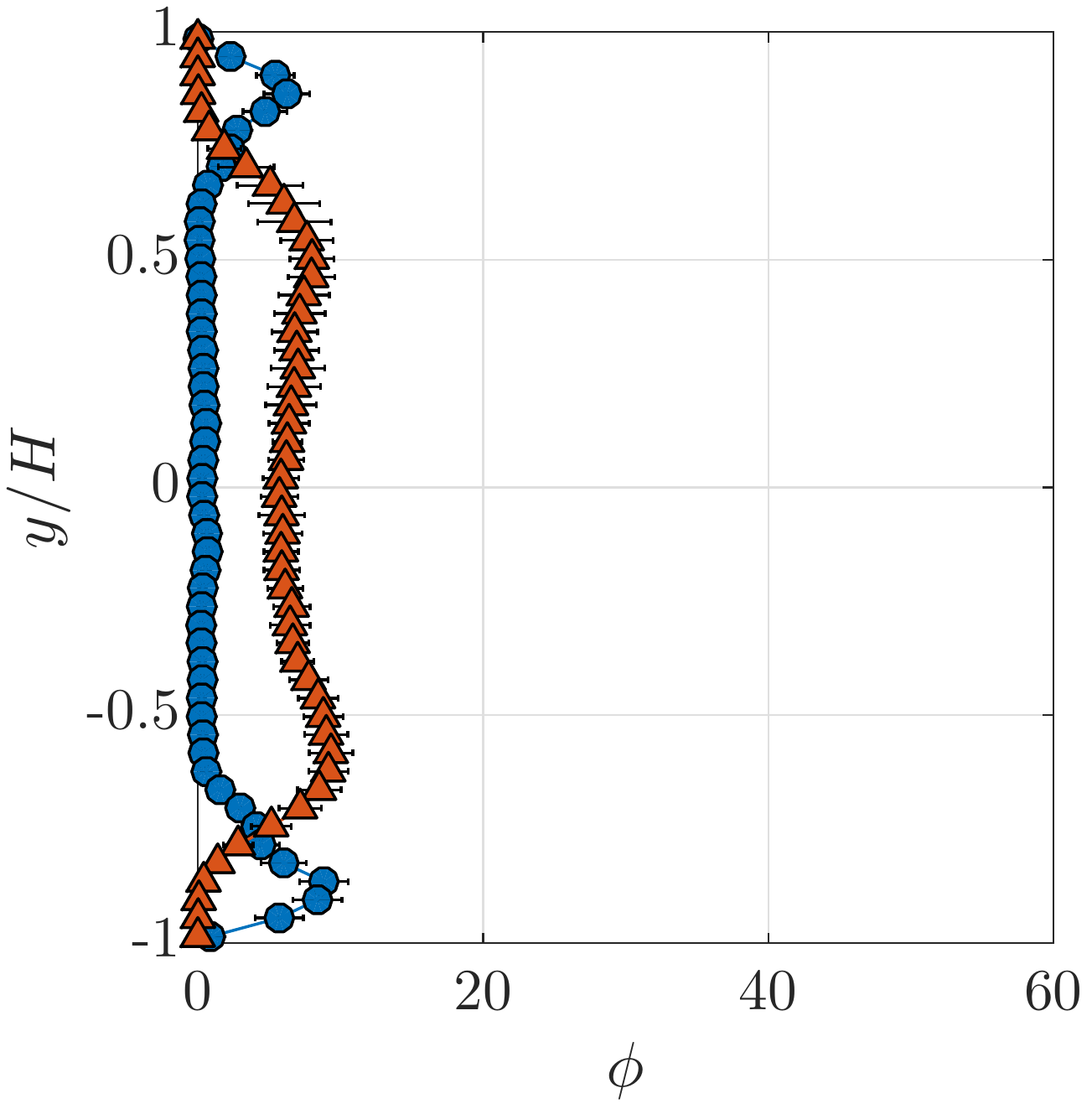}
  \caption{$Re^*$ = 156}
  \label{fig:phi_5p_40lpm}
\end{subfigure}%

\underline{$\phi$ = 10\%}
\begin{subfigure}{.33\textwidth}
  \centering
  \includegraphics[height=1\linewidth]{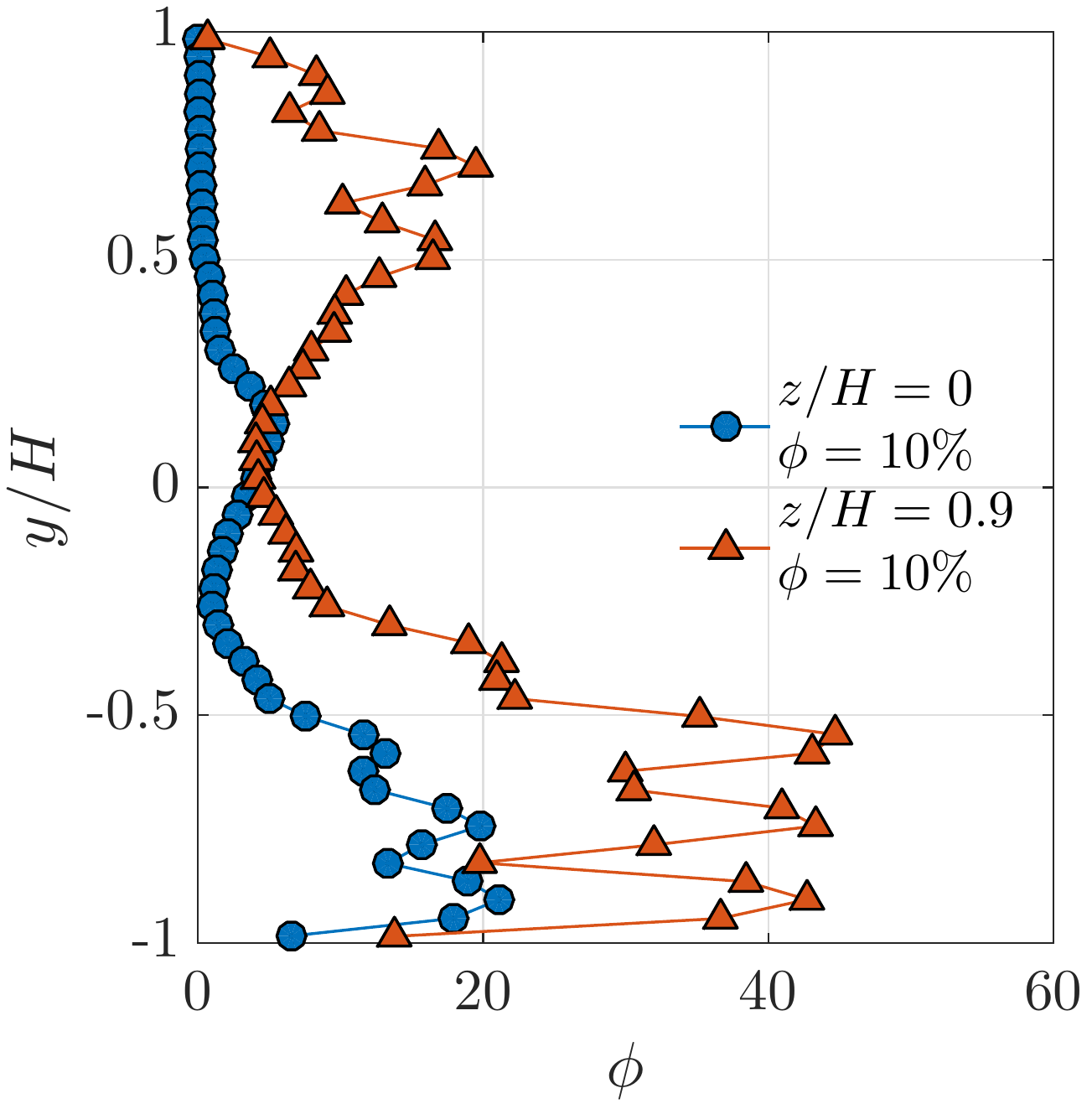}
  \caption{$Re^*$ = 4.2}
  \label{fig:phi_10p_4lpm}
\end{subfigure}%
\begin{subfigure}{.33\textwidth}
  \centering
  \includegraphics[height=1\linewidth]{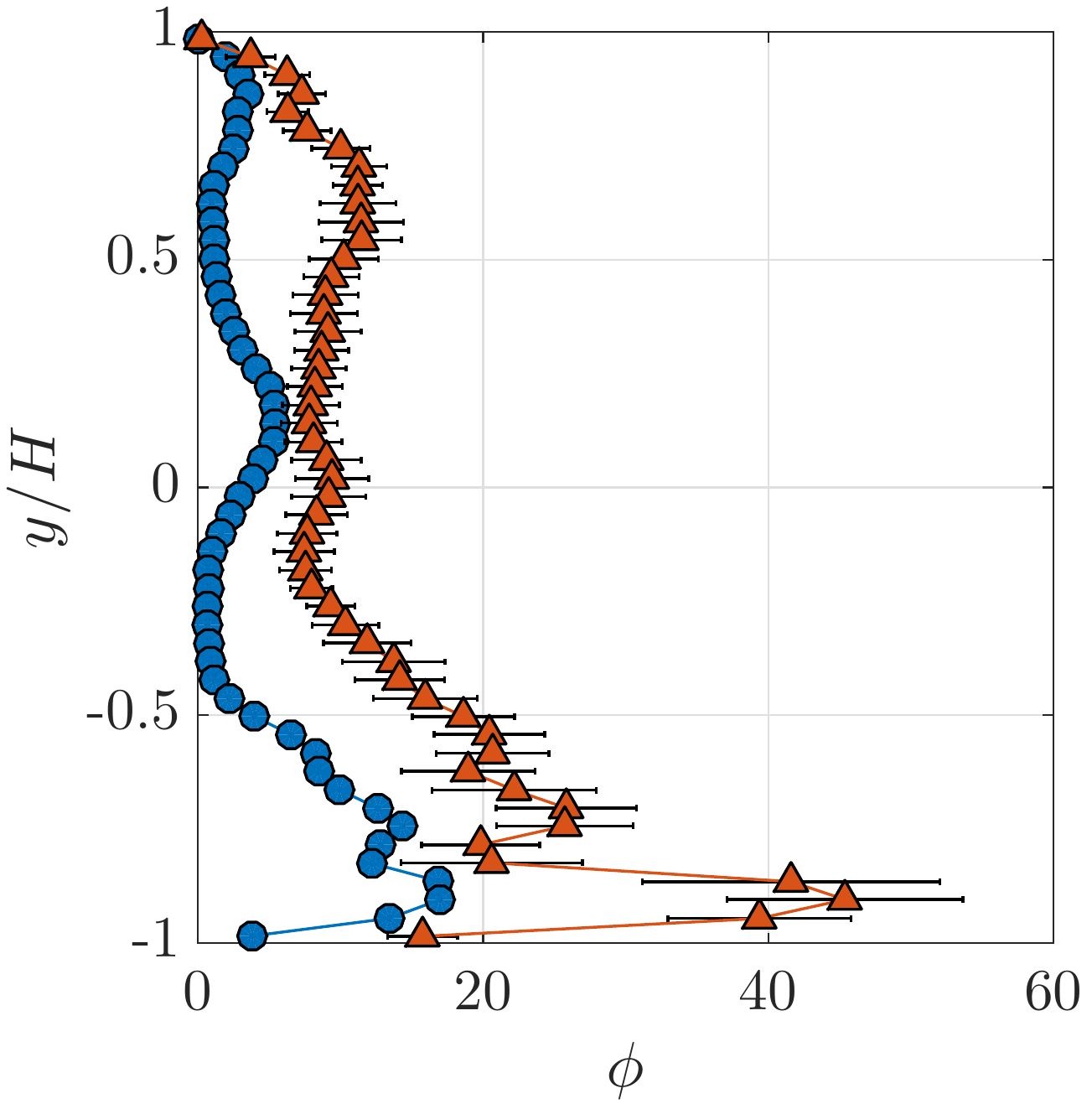}
  \caption{$Re^*$ = 53}
  \label{fig:phi_10p_20lpm}
\end{subfigure}%
\begin{subfigure}{.33\textwidth}
  \centering
  \includegraphics[height=1\linewidth]{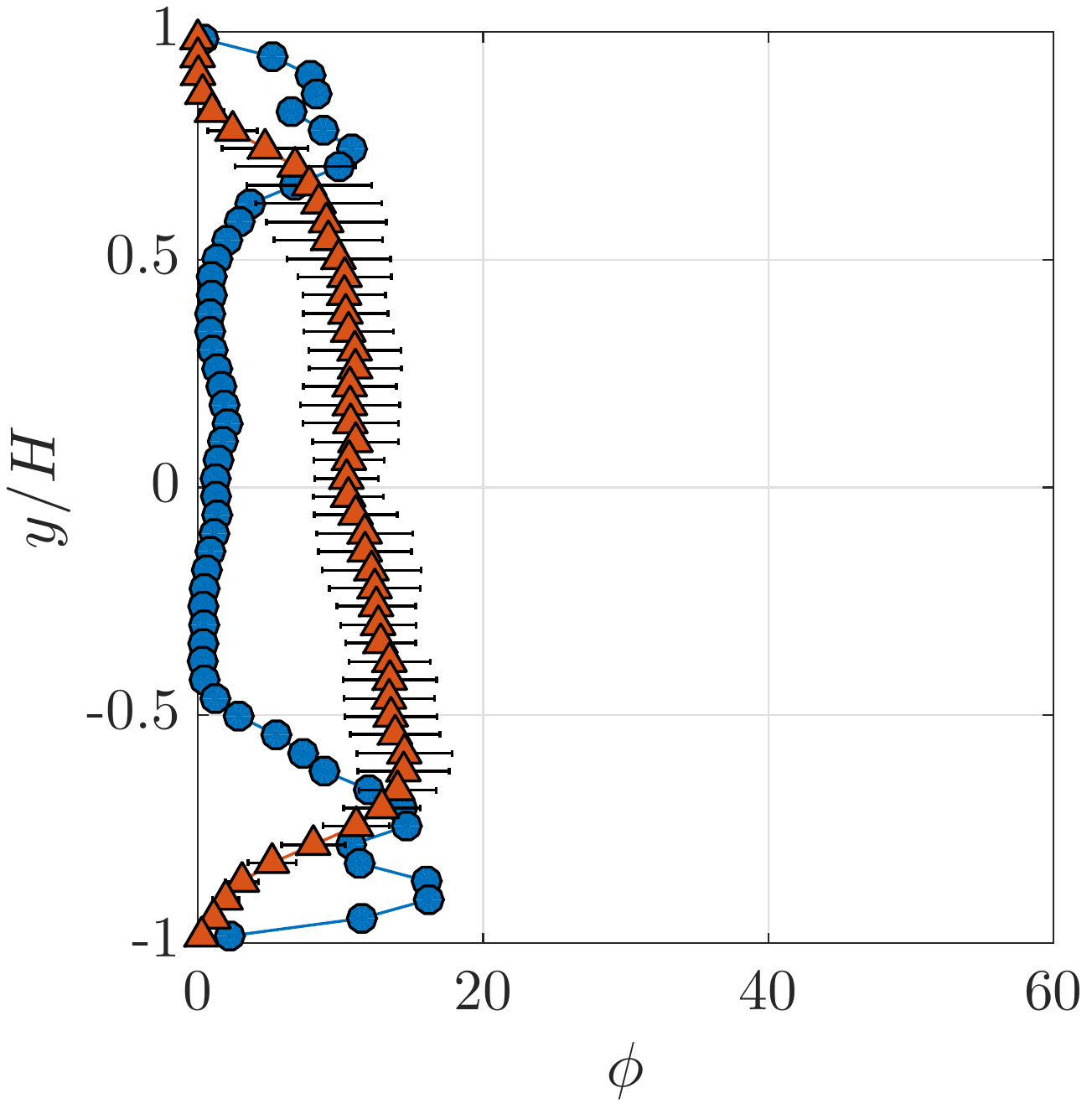}
  \caption{$Re^*$ = 156}
  \label{fig:phi_10p_40lpm}
\end{subfigure}%
\caption{Particle concentration profiles at two span-wise planes: $z/H$ = 0 (center-plane) and $z/H$ = 0.9 (plane close to the side-wall). The top row (a--c) displays results for  $\phi$ = 5\% and the bottom row (d--f) data for the flows with $\phi$ = 10\%. The legend in (a) is applicable to (b--c) and the legend in (d) is applicable to (e--f).}
\label{fig:Particle concentration profiles}
\end{figure}

Several instantaneous snapshots as those reported in figure \ref{fig:Particle concentration images} are processed and the average particle concentration profiles shown in figure \ref{fig:Particle concentration profiles}. The first row (figures \ref{fig:phi_5p_4lpm}--\ref{fig:phi_5p_40lpm}) corresponds to $\phi$ = 5\% with the two symbols corresponding to two span-wise planes: $z/H$ = 0 and 0.9 as mentioned in the caption. 
The high concentration near the corners at the low flow rate (see the plane $z/H$ = 0.9), as seen in the instantaneous snapshots, is quantitatively confirmed in figure \ref{fig:phi_5p_4lpm}. Particle layering is also observed in the form of multiple local maxima, whose value decreases away from the walls. The wall-normal size of each peak closely matches the diameter of the particle. At the highest flow rate, particles move away from the corners and migrate towards the centre of the side walls as seen from the nearly flat concentration in this region in figure \ref{fig:phi_5p_40lpm}. 

Gravitational effects are apparent from the asymmetry in the concentration distribution; more particles at the bottom than the top. However, this asymmetry is reduced at higher flow rates when the particle settling velocity reduces in comparison with the flow velocity. In the same figures \ref{fig:phi_5p_4lpm}--\ref{fig:phi_5p_40lpm}, it is apparent that the concentration in the centre-plane ($z/H$ = 0) is comparatively lower than close to the side wall. A small peak in concentration is observed very close to the core, inside the plug, at lower flow rates, which vanishes completely at the highest flow rates, where the plug is very small. At these high flow rates, concentration peaks are seen at the top and bottom of the centre plane. At high flow rates, the overall picture suggests the existence of a ring like distribution with negligible concentration at the core and corners. On the other hand, at low flow rates, particles are seen to find an equilibrium position predominantly at the corners and in the centre of the relatively larger plug. These observations will be confirmed using shadowgraphy and explained using the secondary flow, the extent of the plug region and inter-particle collisions in the discussion section.

The second row of the same figure (panels \ref{fig:phi_10p_4lpm}--\ref{fig:phi_10p_40lpm}) displays data from the flow at volume fraction $\phi$ = 10\%; it offers the same picture as above but at a higher concentration. Gravitational settling effects are more pronounced here. In this regard, an interesting point to note is that when the flow is stopped by switching off the driving pump, particles come to rest in a very short time (fraction of the time needed to stop when flowing in tap water at the same flow rate) due to the high viscosity of the suspending liquid. If the pump is not switched on, the particles remain at the same position for a period of days, indicating that the yield stress of the fluid is higher than the buoyancy forces of individual particles. The Yield-gravity parameter $Y = \tau_y / ((\rho_p - \rho_f) g d_p)$ for the fluid particle system of this study is around 4, which is substantially greater than the critical value for a particle to move (between 0.04--0.2) \citep{chhabra2006bubbles}. Thus, the sedimentation observed in figure \ref{fig:Particle concentration profiles} is due to particles falling down through the suspending material fluidised under shear during flow as described in \cite{ovarlez2012shear}. Also, particles migrating towards the four corners form packed layers and the resulting inter-particle contact inhibits particle rotation in this dense region. Thus, particles slide when surrounded by a sufficient number of neighbouring particles.

\begin{figure}
\centering
\underline{$\phi$ = 5\%}
\begin{subfigure}{.33\textwidth}
$Q$ = 4 lpm
  \centering
  \includegraphics[height=1\linewidth]{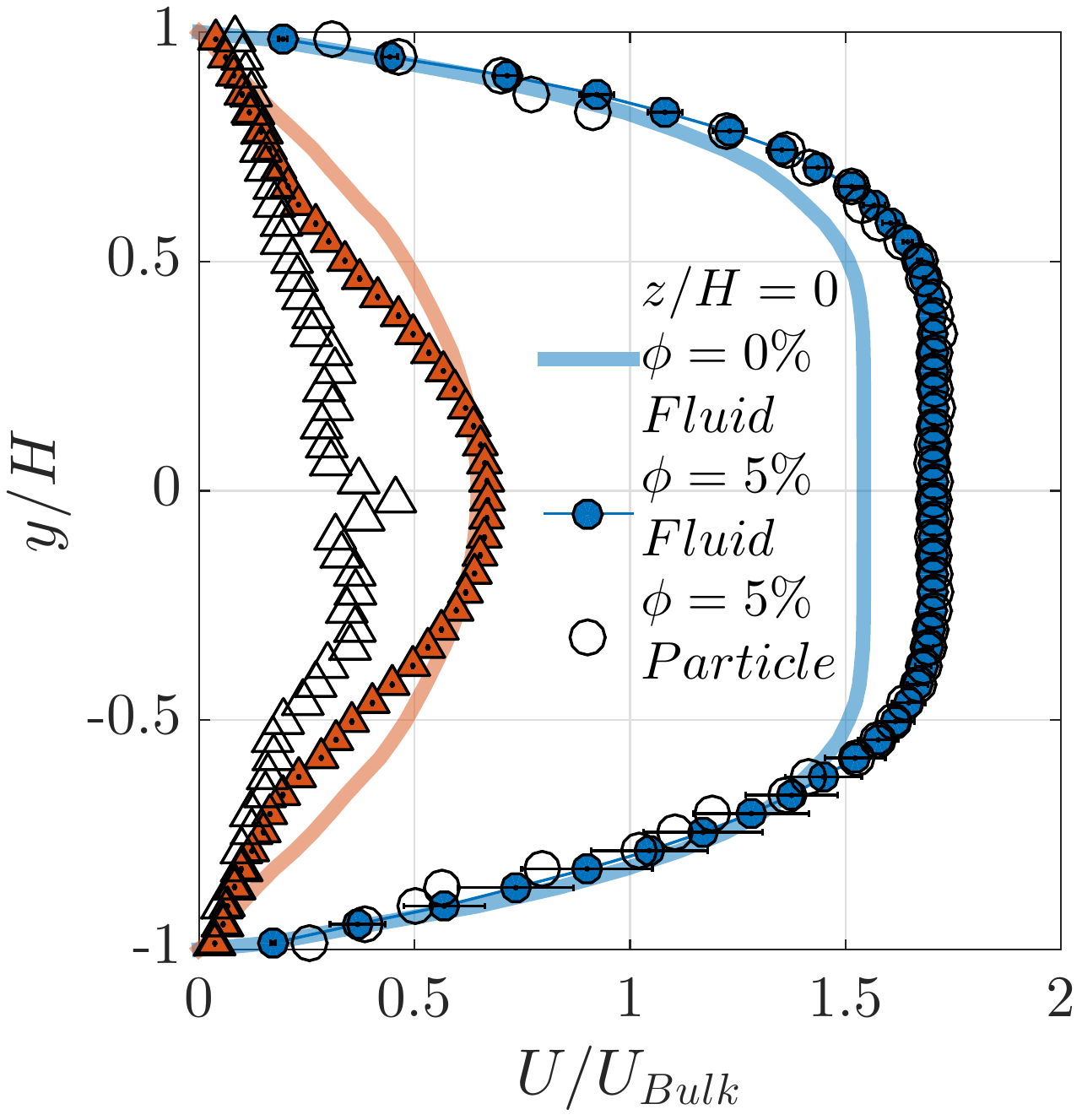}
  \caption{$Re^*$ = 4.2}
  \label{fig:U_5p_4lpm}
\end{subfigure}%
\begin{subfigure}{.33\textwidth}
$Q$ = 20 lpm
  \centering
  \includegraphics[height=1\linewidth]{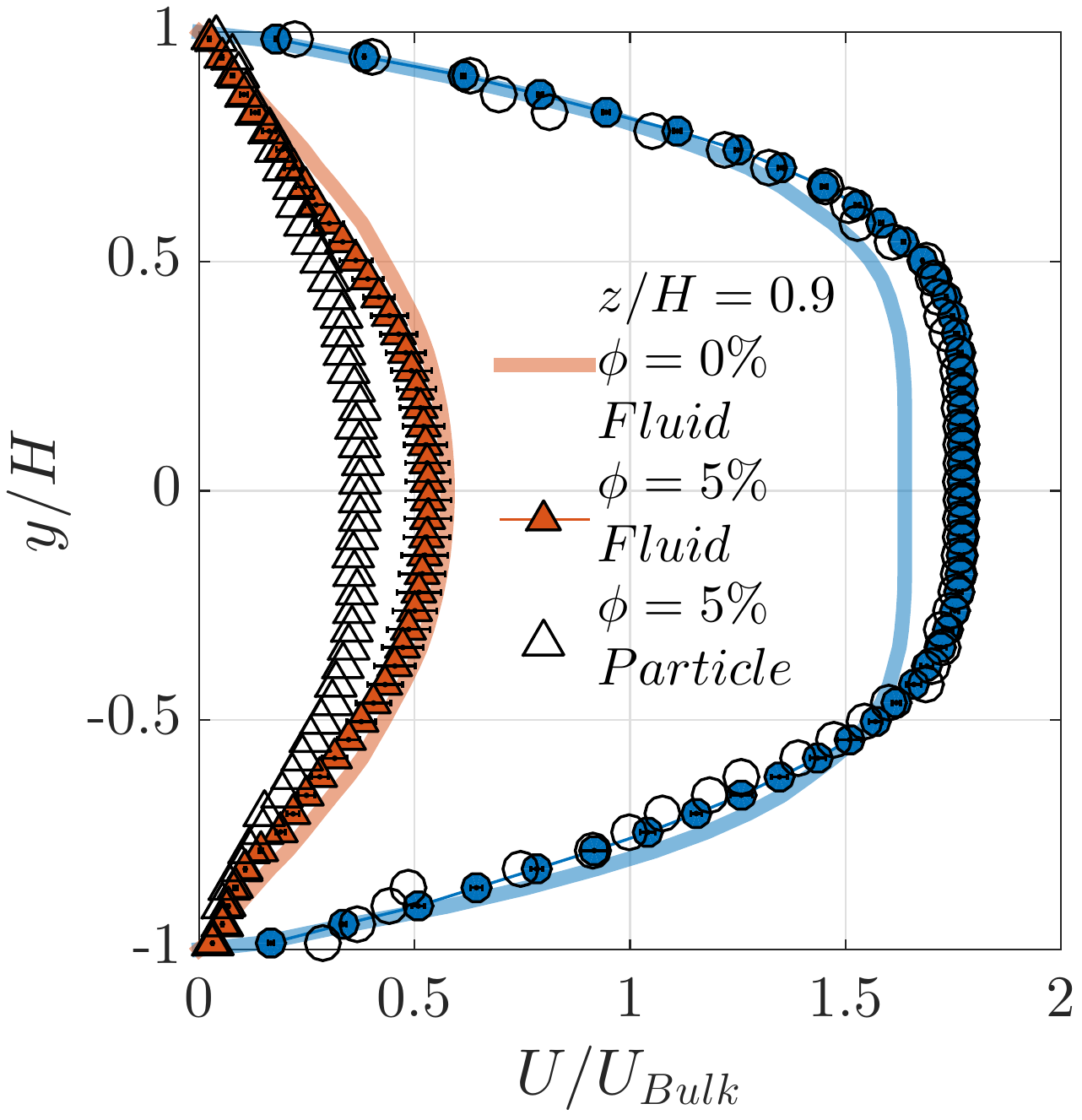}
  \caption{$Re^*$ = 53}
  \label{fig:U_5p_20lpm}
\end{subfigure}%
\begin{subfigure}{.33\textwidth}
$Q$ = 40 lpm
  \centering
  \includegraphics[height=1\linewidth]{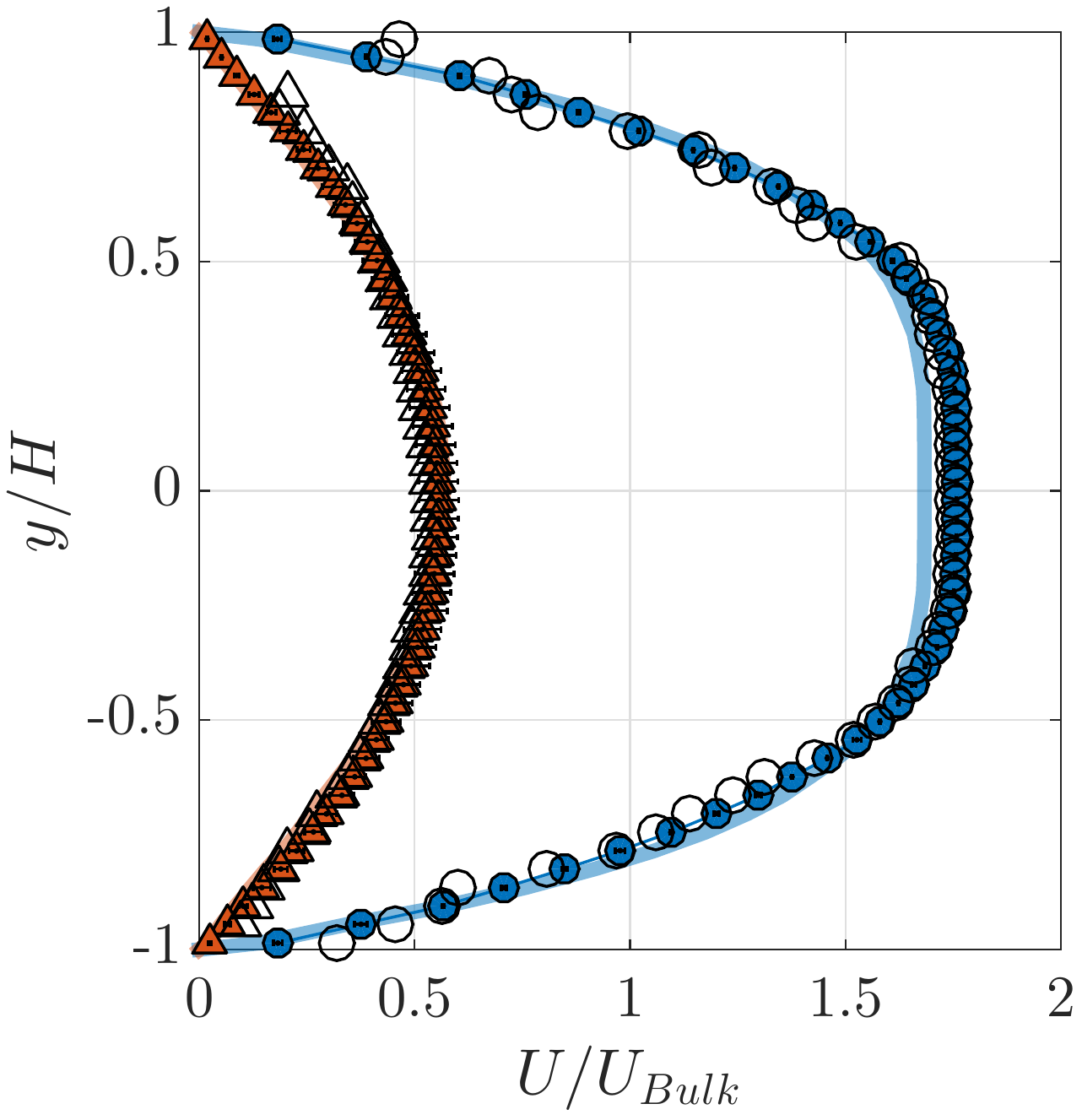}
  \caption{$Re^*$ = 156}
  \label{fig:U_5p_40lpm}
\end{subfigure}%

\underline{$\phi$ = 10\%}
\begin{subfigure}{.33\textwidth}
  \centering
  \includegraphics[height=1\linewidth]{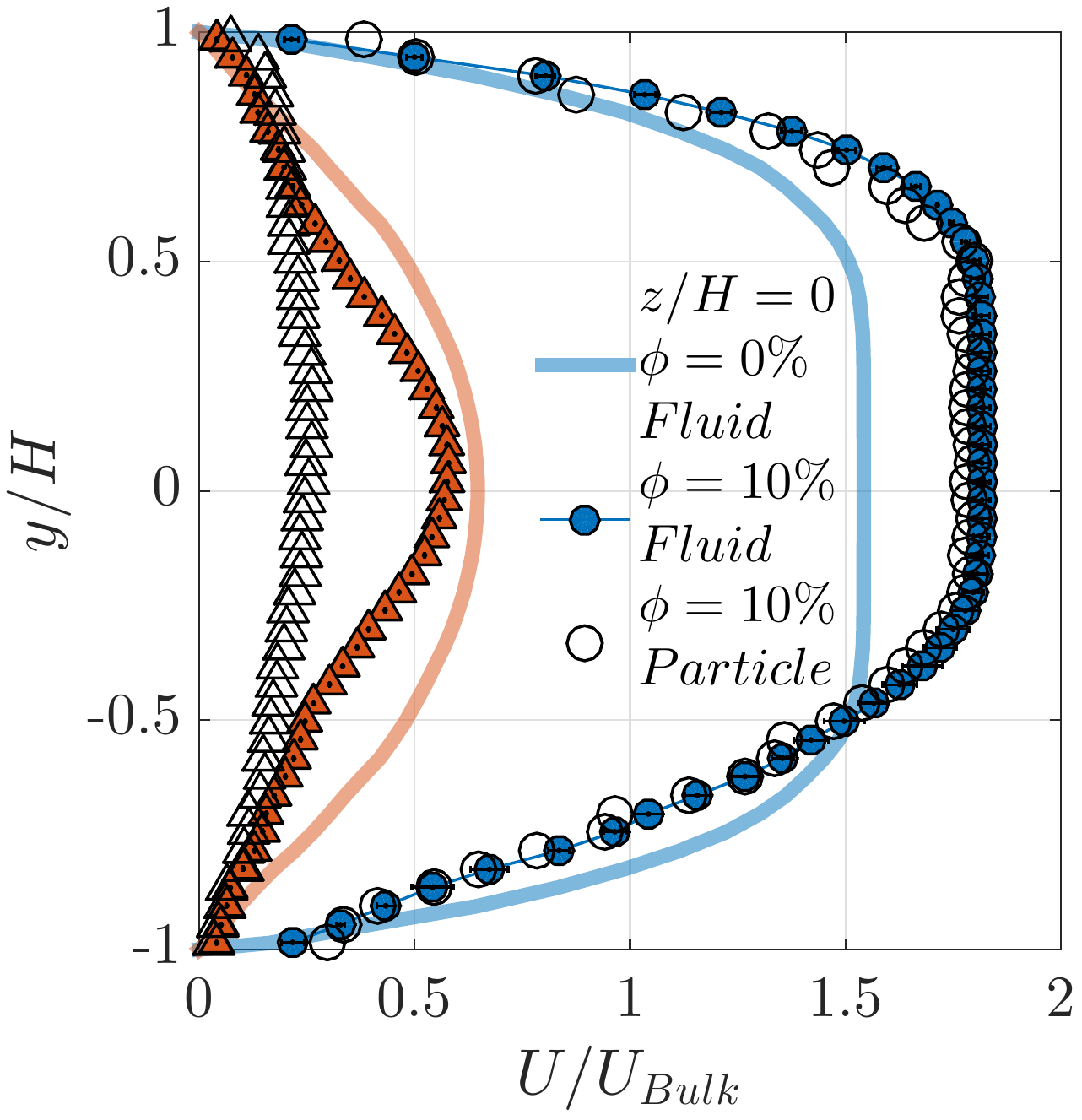}
  \caption{$Re^*$ = 4.2}
  \label{fig:U_10p_4lpm}
\end{subfigure}%
\begin{subfigure}{.33\textwidth}
  \centering
  \includegraphics[height=1\linewidth]{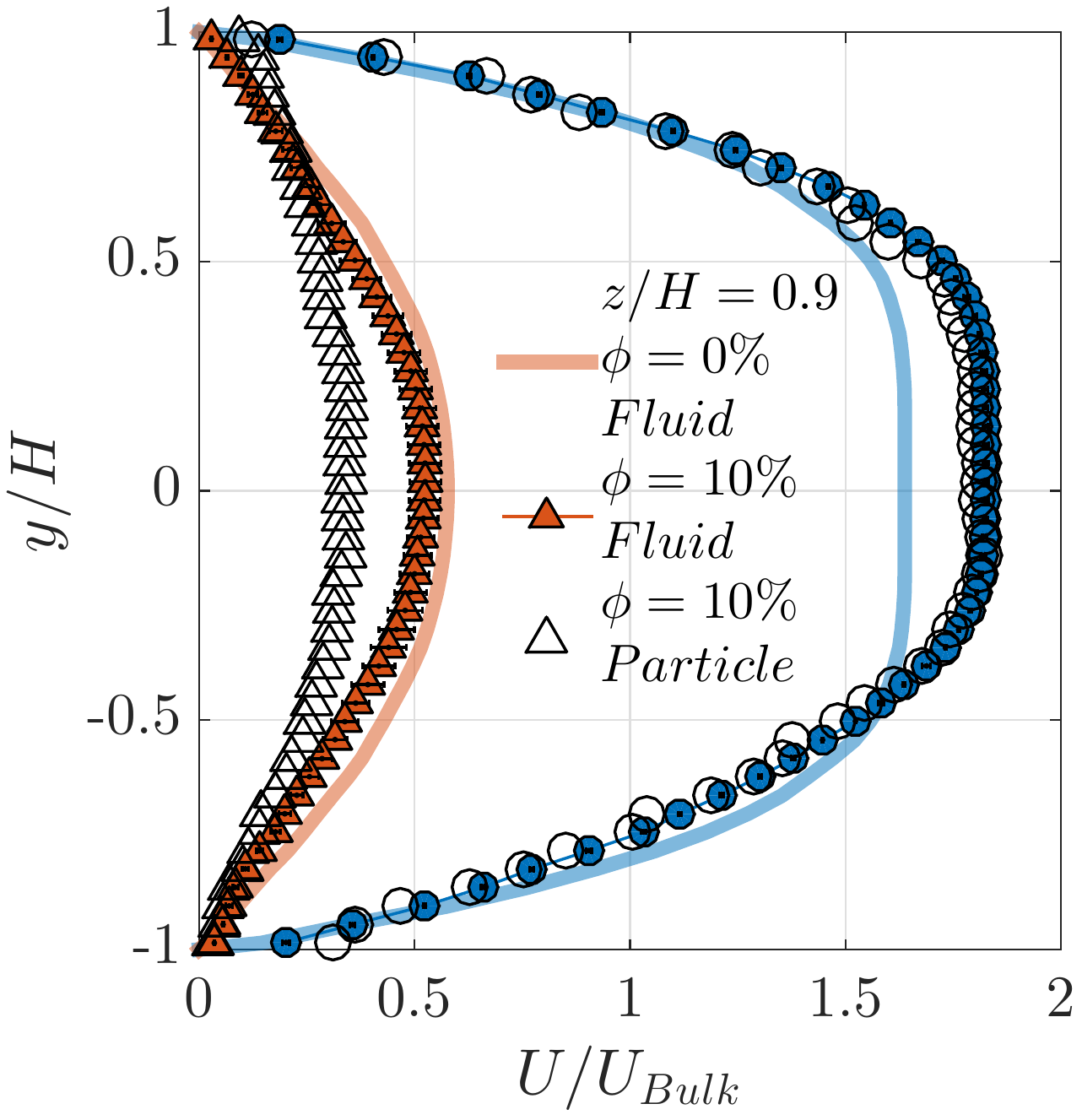}
  \caption{$Re^*$ = 53}
  \label{fig:U_10p_20lpm}
\end{subfigure}%
\begin{subfigure}{.33\textwidth}
  \centering
  \includegraphics[height=1\linewidth]{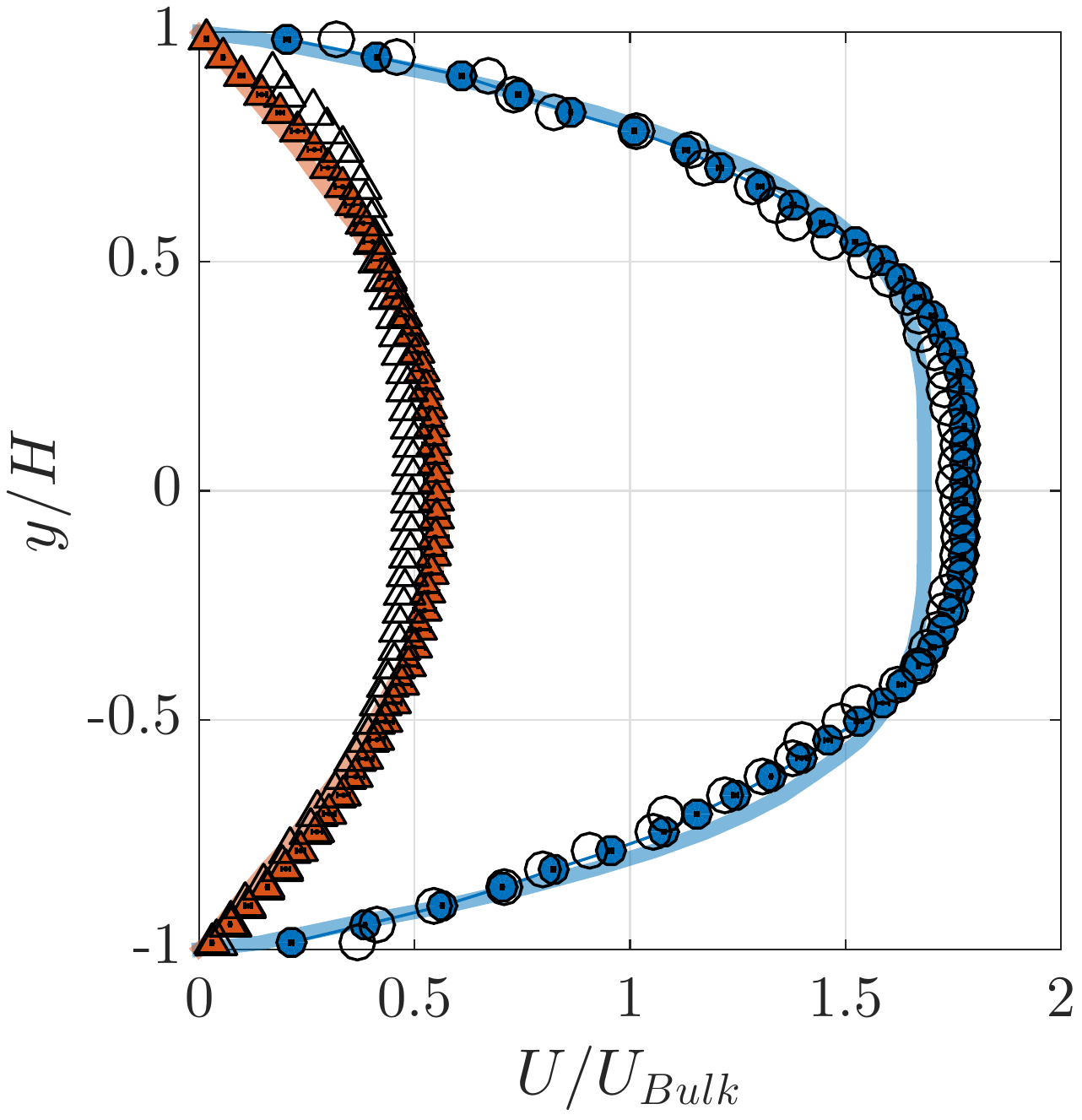}
  \caption{$Re^*$ = 156}
  \label{fig:U_10p_40lpm}
\end{subfigure}%
\caption{Change of mean streamwise velocity profiles due to introduction of particles at two span-wise planes: $z/H$ = 0 (center-plane) and $z/H$ = 0.9 (plane close to the side-wall). The top row (a--c) corresponds to $\phi$ = 5\% and the bottom row (d--f) corresponds to $\phi$ = 10\%. The filled and hollow symbols represent the fluid and particle velocity respectively, and the solid lines represent the fluid velocity when no particles are present i.e. for $\phi$ = 0\%. Figures (a--c) share the same legends mentioned in (a) and (b), and (d--f) share the same legends mentioned in (d) and (e).}
\label{fig:Umean particles}
\end{figure}

The flow-dependent non-uniform particle distribution causes a change in the mean streamwise velocity as shown in figure \ref{fig:Umean particles}. As in the previous figure, we display the particle and fluid velocities in two span-wise planes with the top row corresponding to $\phi$ = 5\% and the bottom row to $\phi$ = 10\%. The single phase reference case at the same $Re^*$ is also shown for comparison. 

The most striking changes happen for the low flow rates (figures \ref{fig:U_5p_4lpm} and \ref{fig:U_10p_4lpm}), when the particles occupy the corners. In these regions of high particle concentration, the fluid velocity is reduced as compared to its single phase counterpart (see $z/H$ = 0.9 plane in figures \ref{fig:U_5p_4lpm} and \ref{fig:U_10p_4lpm}). To make-up for the velocity deficit in the near-wall planes, the velocity in the plug region increases (see $z/H$ = 0 plane in figures \ref{fig:U_5p_4lpm} and \ref{fig:U_10p_4lpm}) and as a consequence, the size of the plug reduces. Again, sedimentation also affects the mean velocity profiles, more for the higher $\phi$. The particles move with approximatively the velocity of the fluid in the centre plane but lag significantly the fluid phase in the near-wall plane, especially near the wall bisectors. It should be recalled that, at these low flow rates, most of the particles are concentrated at the corners and there are very few particles near the middle of the side walls, as seen from the concentration profiles in figures \ref{fig:phi_5p_4lpm} and \ref{fig:phi_10p_4lpm}; we speculate that these slowly moving particles near the corners occasionally move towards the centre of the side walls thus contributing to the apparent lag in the particle velocity. 

With increasing the flow rate, the particles redistribute in to a porous ring-like region between the corners and the core, and, as a result,  the mean velocity distribution has smaller deviations from the corresponding single phase cases. Indeed, the plug size reduces with increasing particle concentration. At the highest flow rate considered, the plug in the single phase is already very small and hence, the reduction is not significant. A small but finite slip is also seen near the top and bottom wall of the centre plane due to the finite size of the spherical particles. This slip, due to finite size effects, is more pronounced at the higher wall shear that occurs in turbulent flows \citep{zade2018experimental}.

\begin{figure}
\centering
\underline{$\phi$ = 5\%}
\begin{subfigure}{.33\textwidth}
$Q$ = 4 lpm
  \centering
  \includegraphics[height=1\linewidth]{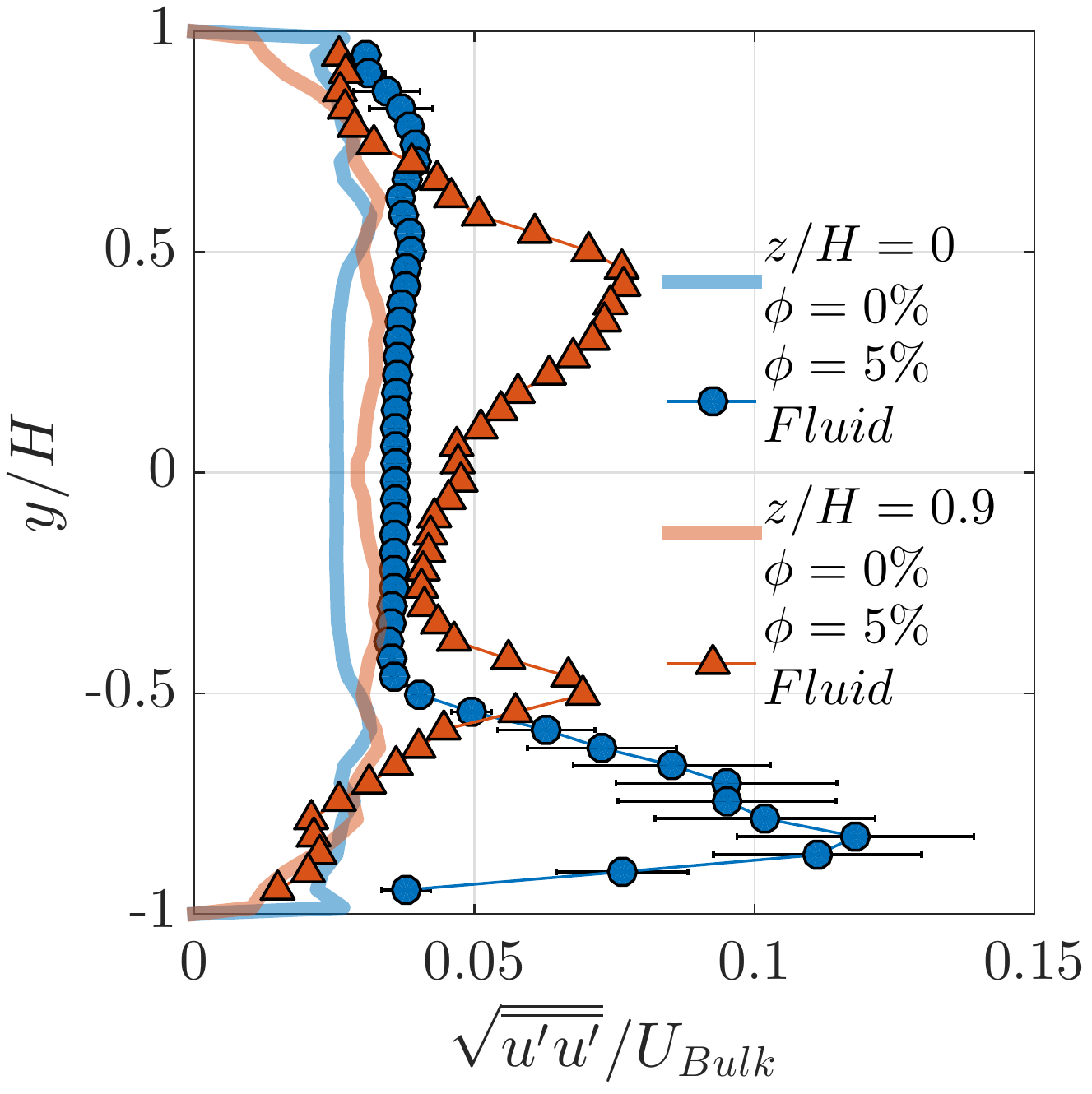}
  \caption{$Re^*$ = 4.2}
  \label{fig:uu_5p_4lpm}
\end{subfigure}%
\begin{subfigure}{.33\textwidth}
$Q$ = 20 lpm
  \centering
  \includegraphics[height=1\linewidth]{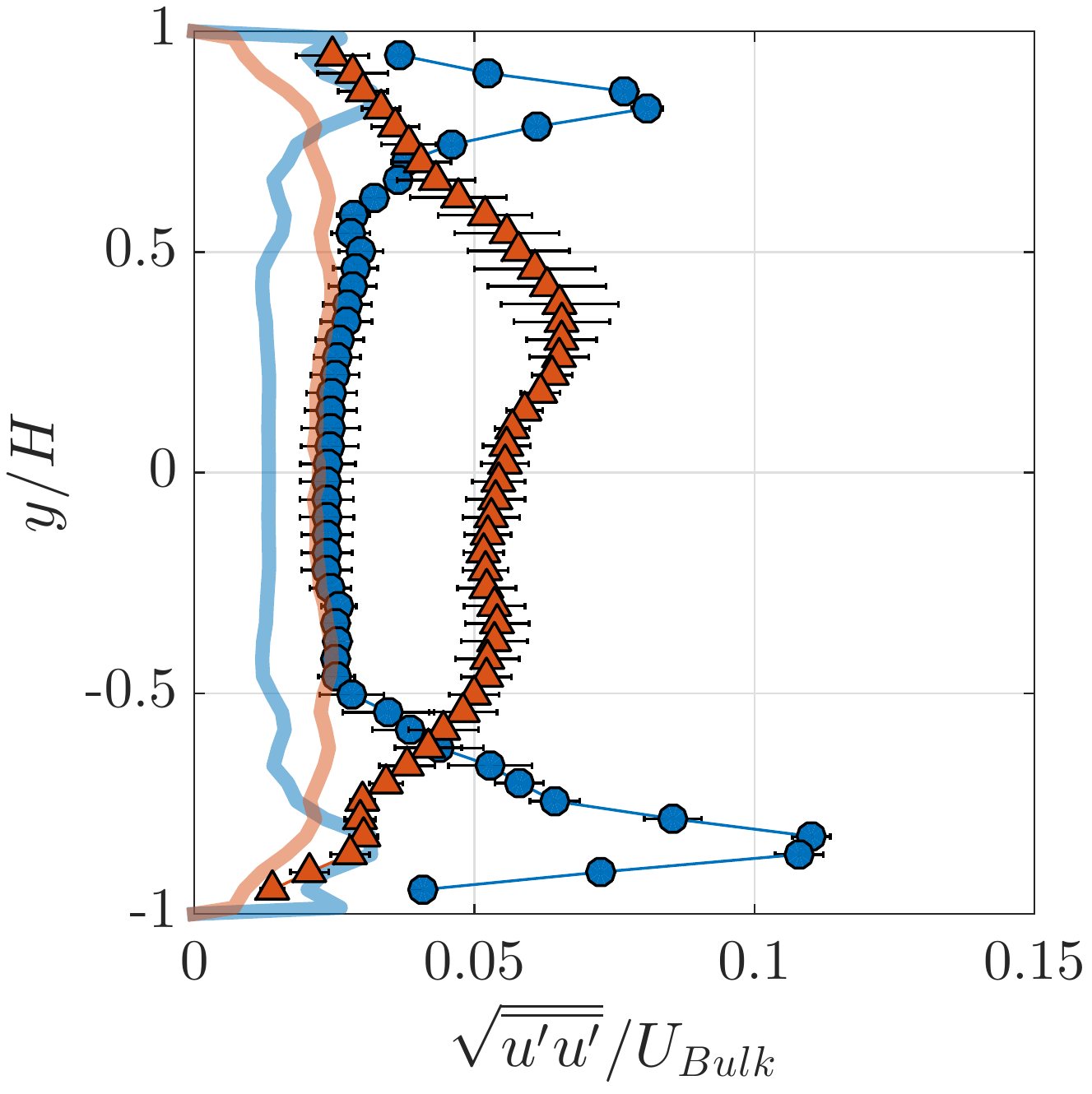}
  \caption{$Re^*$ = 53}
  \label{fig:uu_5p_20lpm}
\end{subfigure}%
\begin{subfigure}{.33\textwidth}
$Q$ = 40 lpm
  \centering
  \includegraphics[height=1\linewidth]{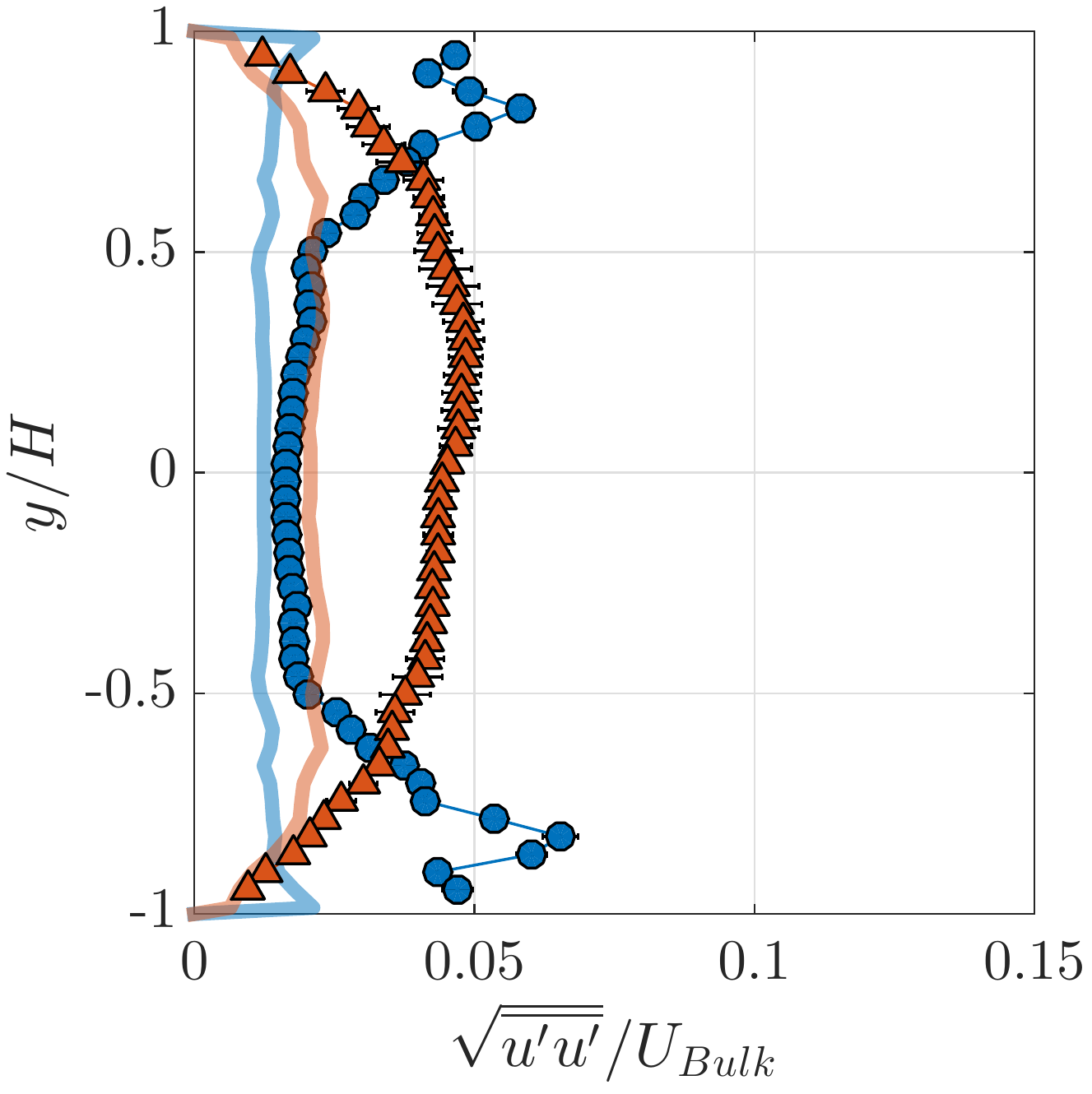}
  \caption{$Re^*$ = 156}
  \label{fig:uu_5p_40lpm}
\end{subfigure}%

\underline{$\phi$ = 10\%}
\begin{subfigure}{.33\textwidth}
  \centering
  \includegraphics[height=1\linewidth]{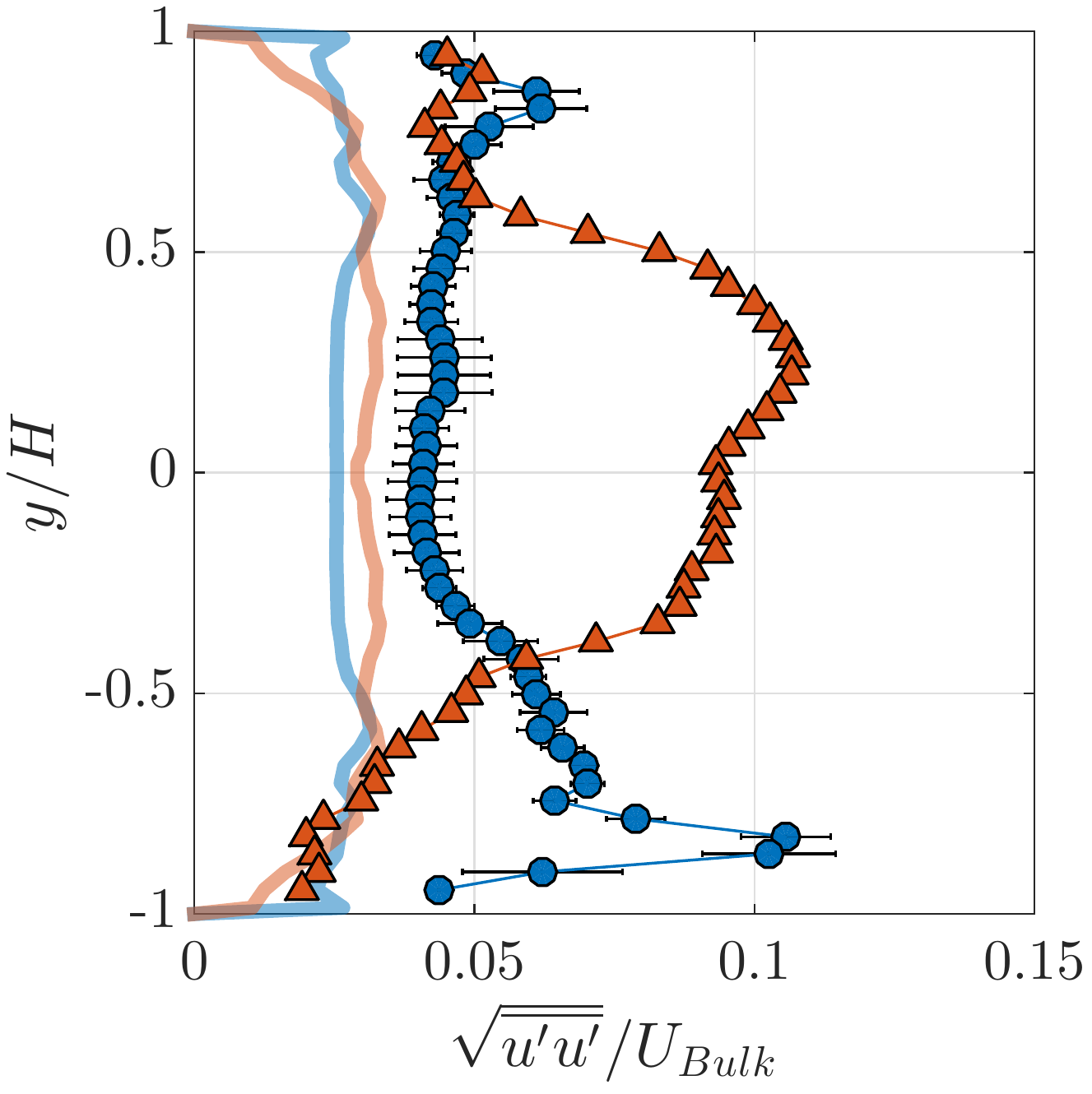}
  \caption{$Re^*$ = 4.2}
  \label{fig:uu_10p_4lpm}
\end{subfigure}%
\begin{subfigure}{.33\textwidth}
  \centering
  \includegraphics[height=1\linewidth]{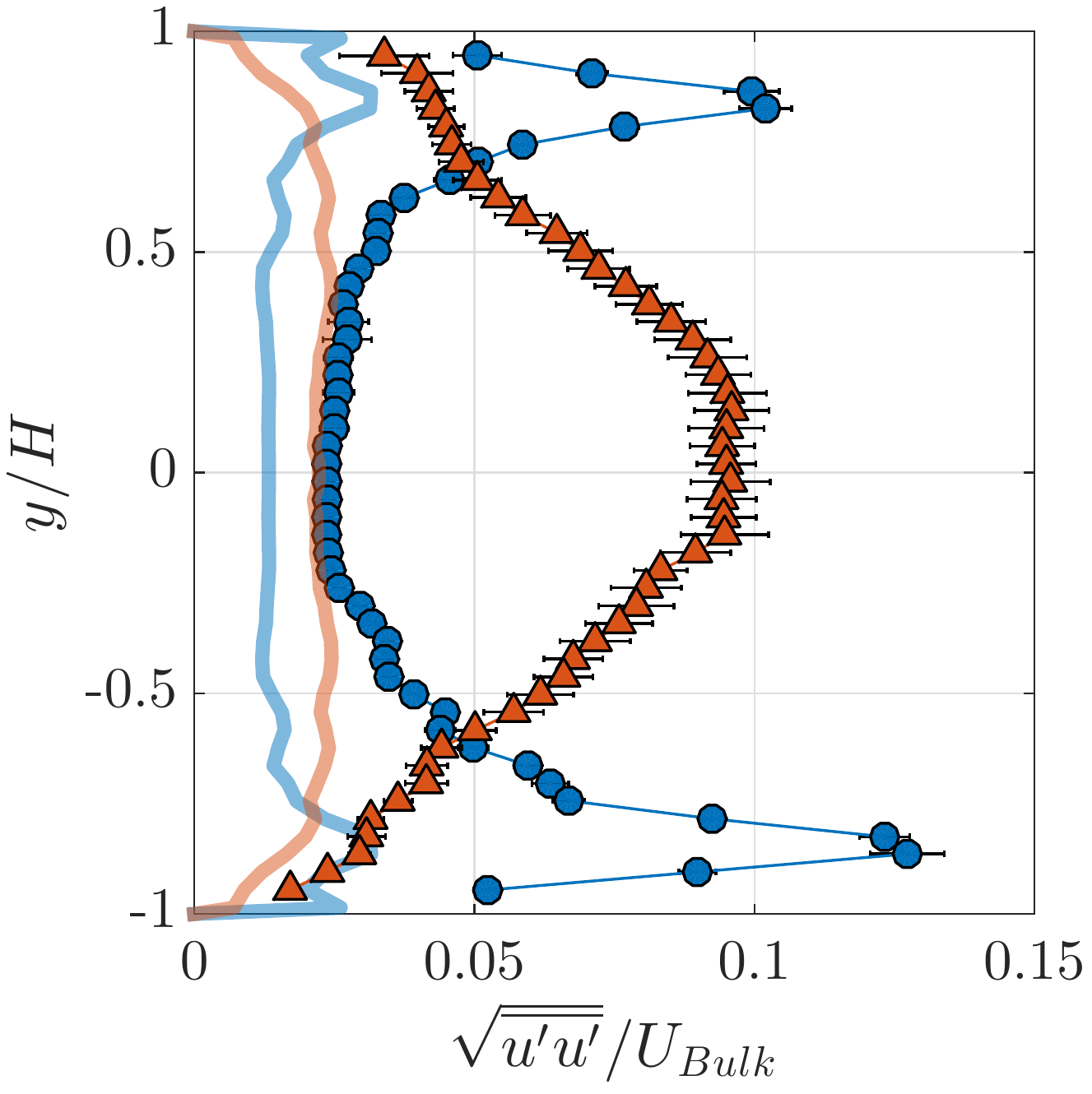}
  \caption{$Re^*$ = 53}
  \label{fig:uu_10p_20lpm}
\end{subfigure}%
\begin{subfigure}{.33\textwidth}
  \centering
  \includegraphics[height=1\linewidth]{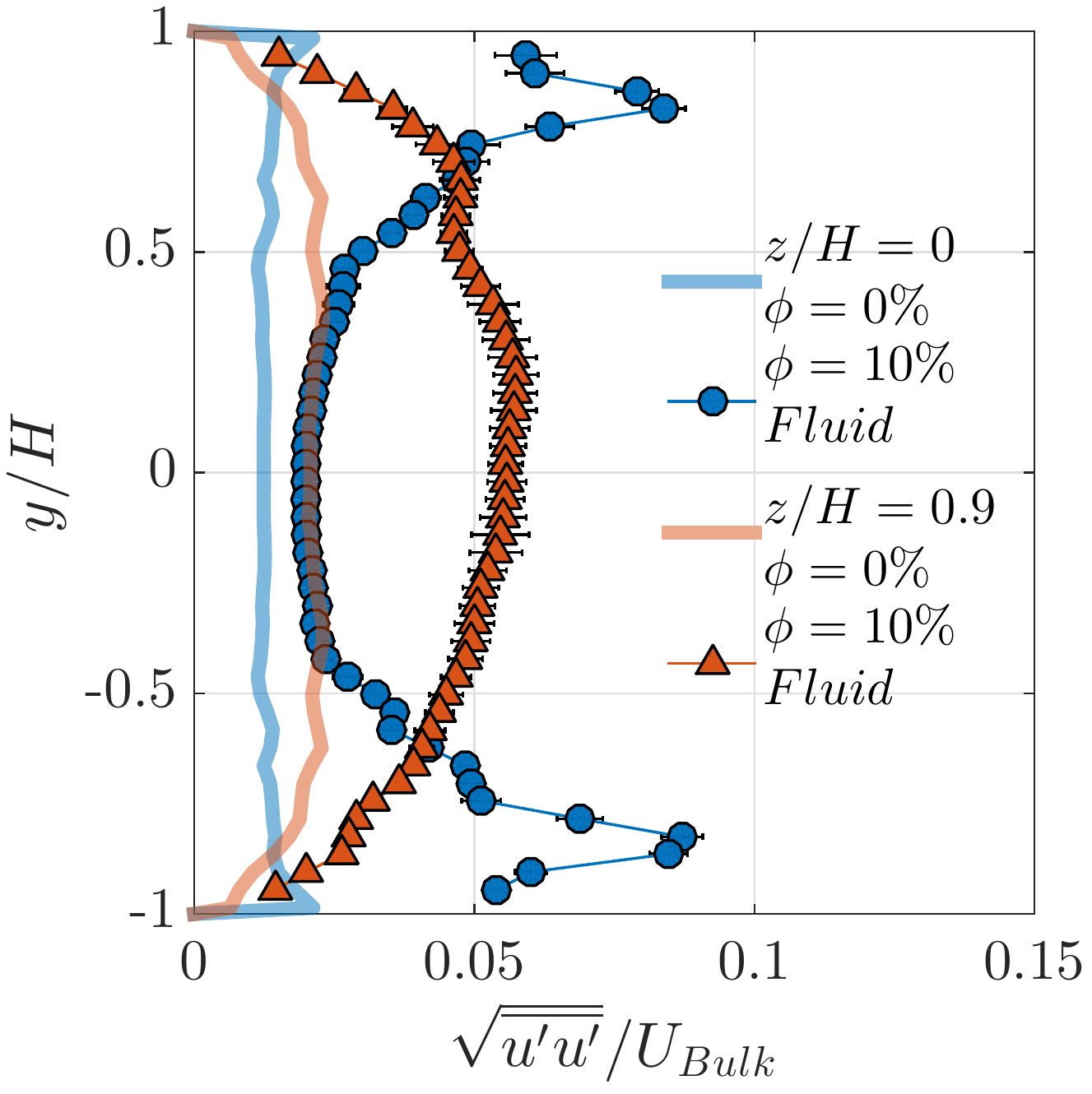}
  \caption{$Re^*$ = 156}
  \label{fig:uu_10p_40lpm}
\end{subfigure}%
\caption{Change of streamwise velocity fluctuations due to introduction of particles at two span-wise planes: $z/H$ = 0 (center-plane) and $z/H$ = 0.9 (plane close to the side-wall). The top row (a--c) corresponds to $\phi$ = 5\% and the bottom row (d--f) corresponds to $\phi$ = 10\%.}
\label{fig:Urms particles}
\end{figure}

\begin{figure}
\centering
\underline{$\phi$ = 5\%}
\begin{subfigure}{.33\textwidth}
$Q$ = 4 lpm
  \centering
  \includegraphics[height=1\linewidth]{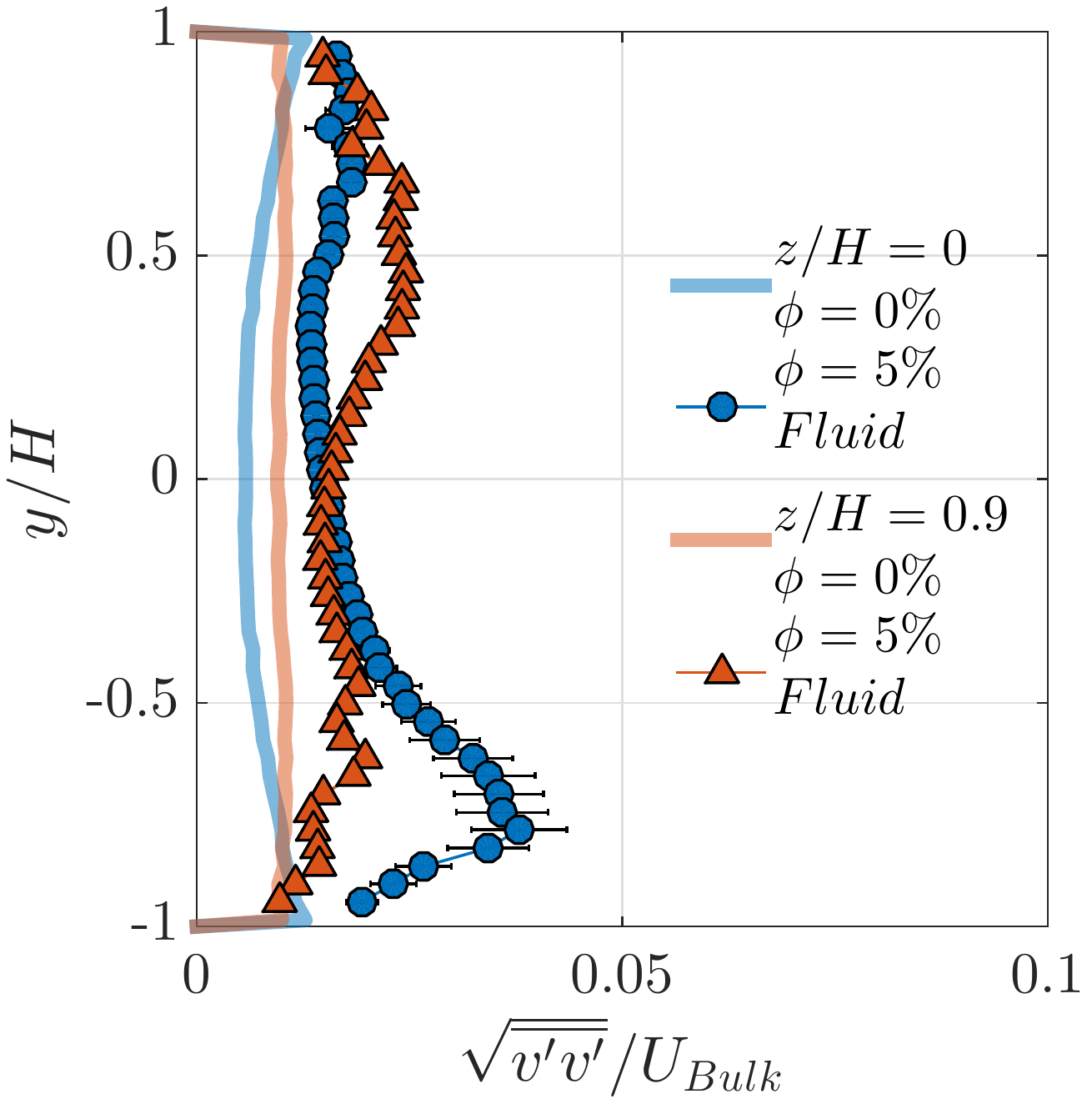}
  \caption{$Re^*$ = 4.2}
  \label{fig:vv_5p_4lpm}
\end{subfigure}%
\begin{subfigure}{.33\textwidth}
$Q$ = 20 lpm
  \centering
  \includegraphics[height=1\linewidth]{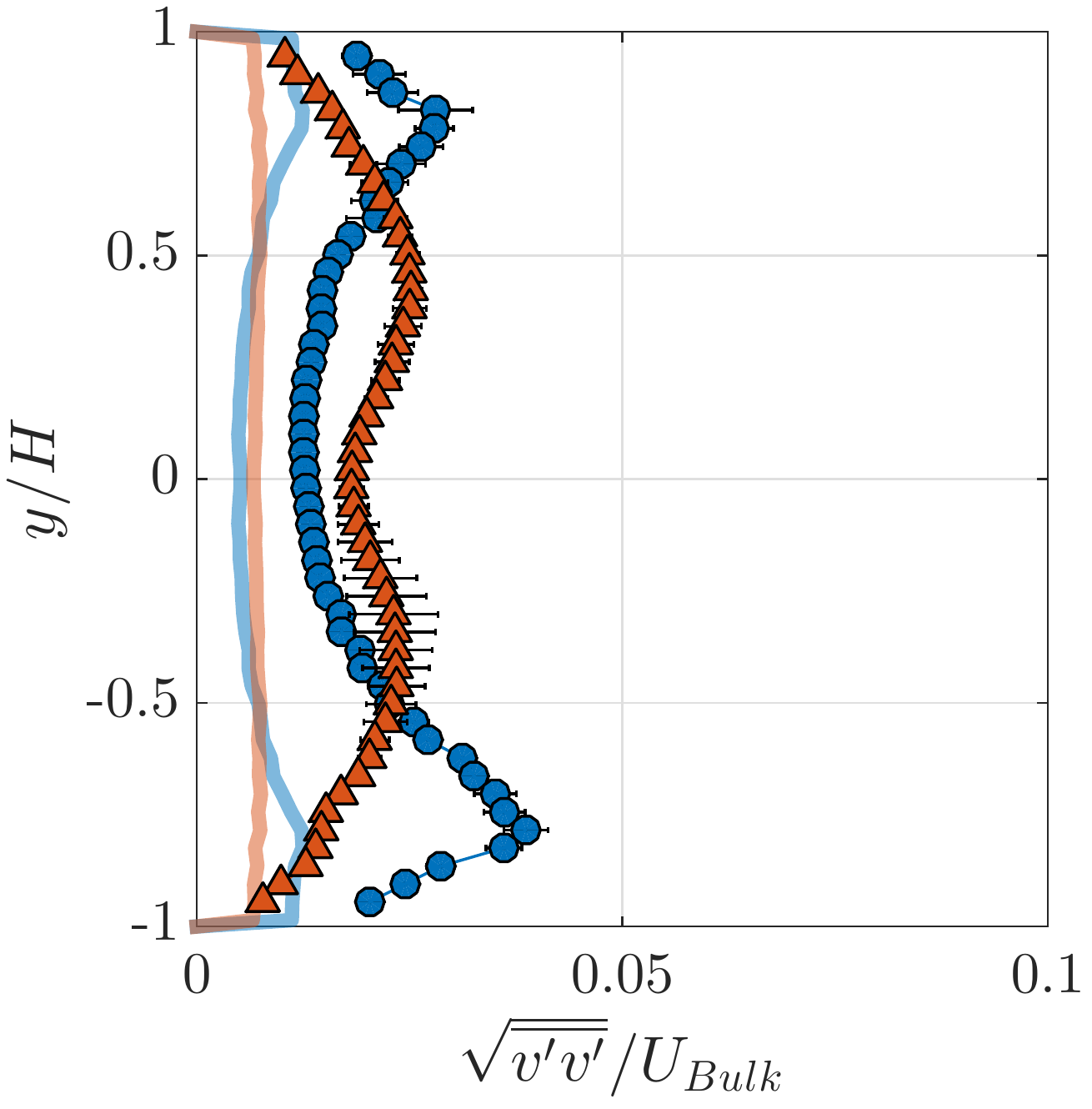}
  \caption{$Re^*$ = 53}
  \label{fig:vv_5p_20lpm}
\end{subfigure}%
\begin{subfigure}{.33\textwidth}
$Q$ = 40 lpm
  \centering
  \includegraphics[height=1\linewidth]{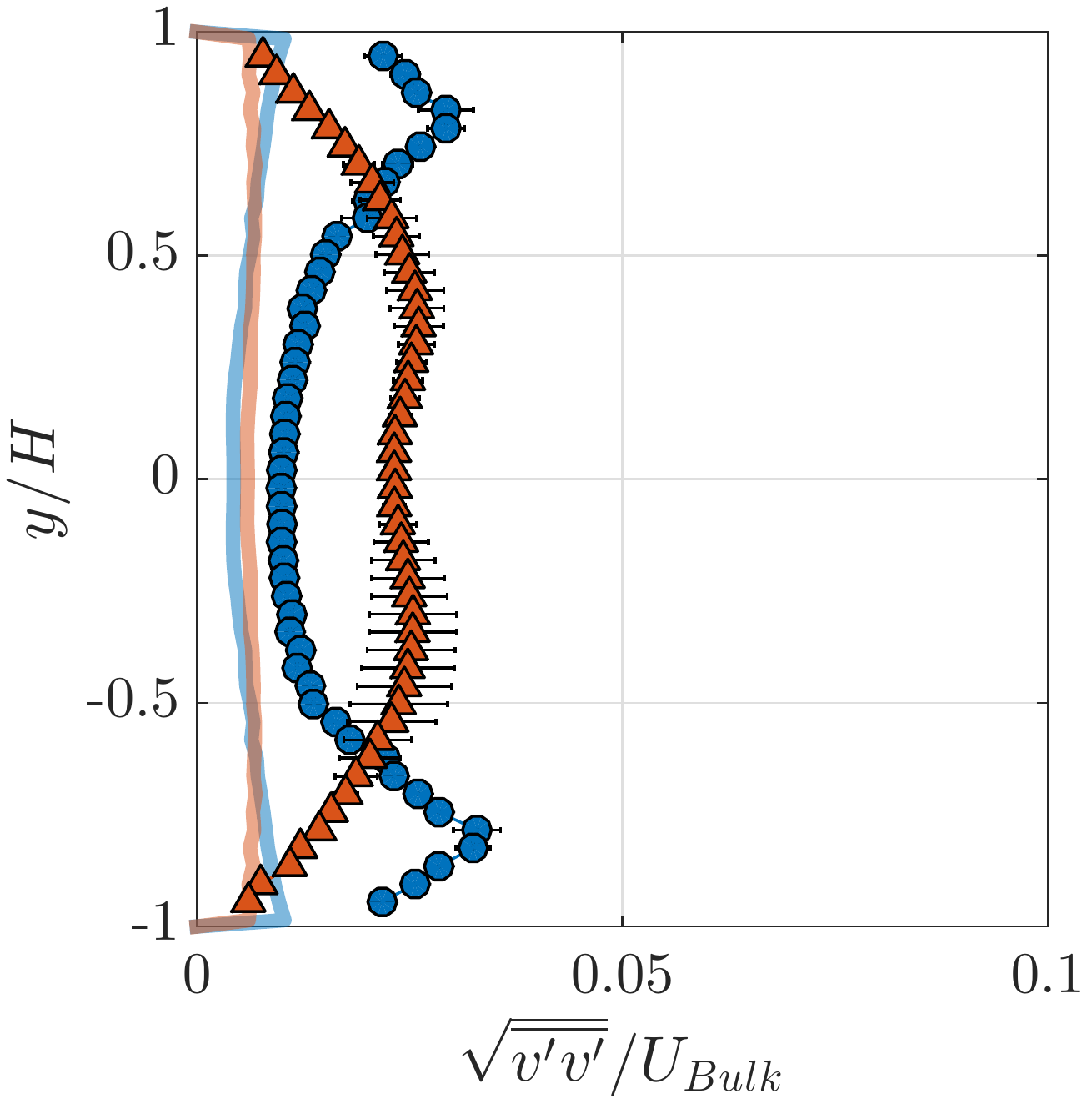}
  \caption{$Re^*$ = 156}
  \label{fig:vv_5p_40lpm}
\end{subfigure}%

\underline{$\phi$ = 10\%}
\begin{subfigure}{.33\textwidth}
  \centering
  \includegraphics[height=1\linewidth]{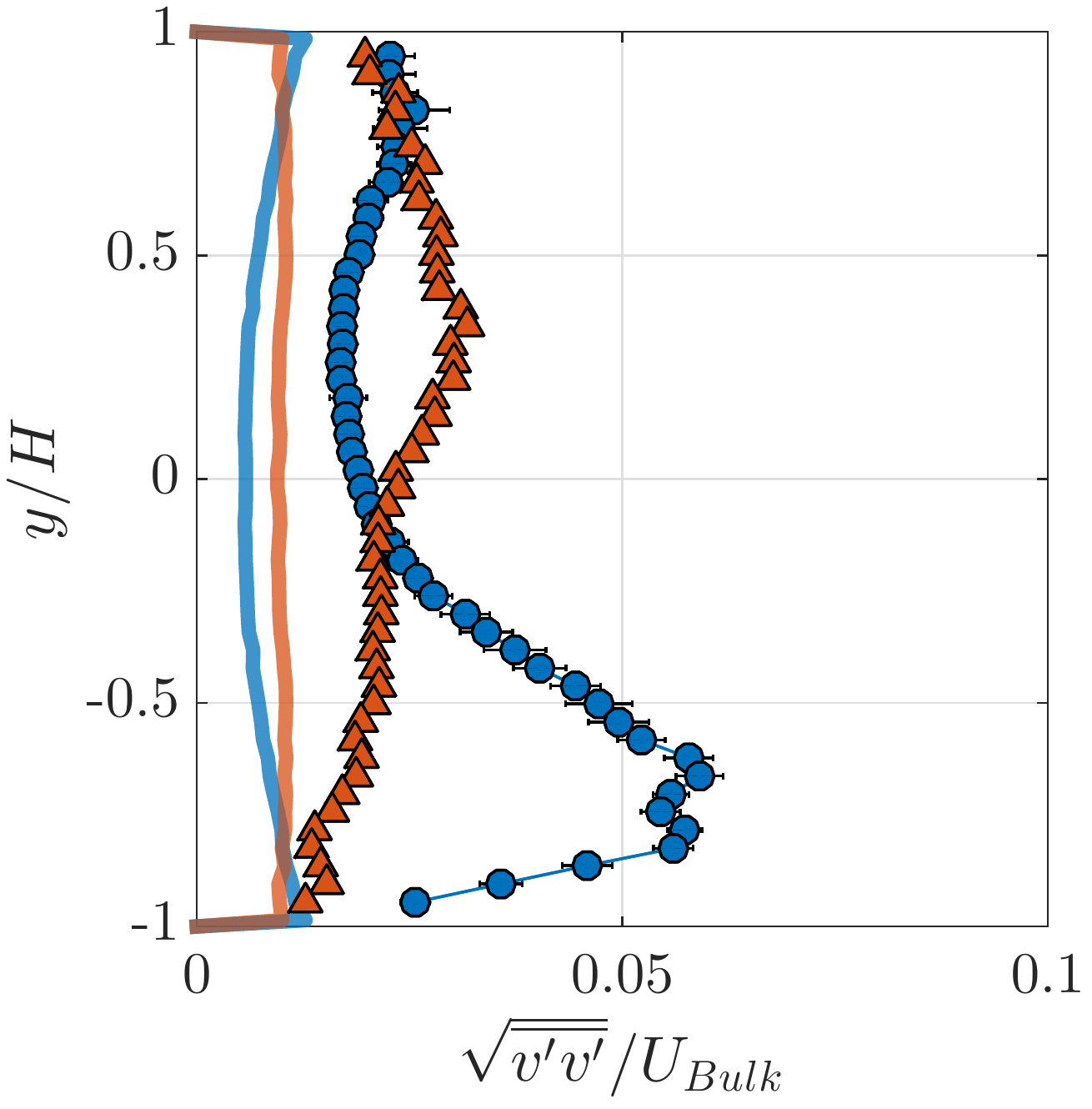}
  \caption{$Re^*$ = 4.2}
  \label{fig:vv_10p_4lpm}
\end{subfigure}%
\begin{subfigure}{.33\textwidth}
  \centering
  \includegraphics[height=1\linewidth]{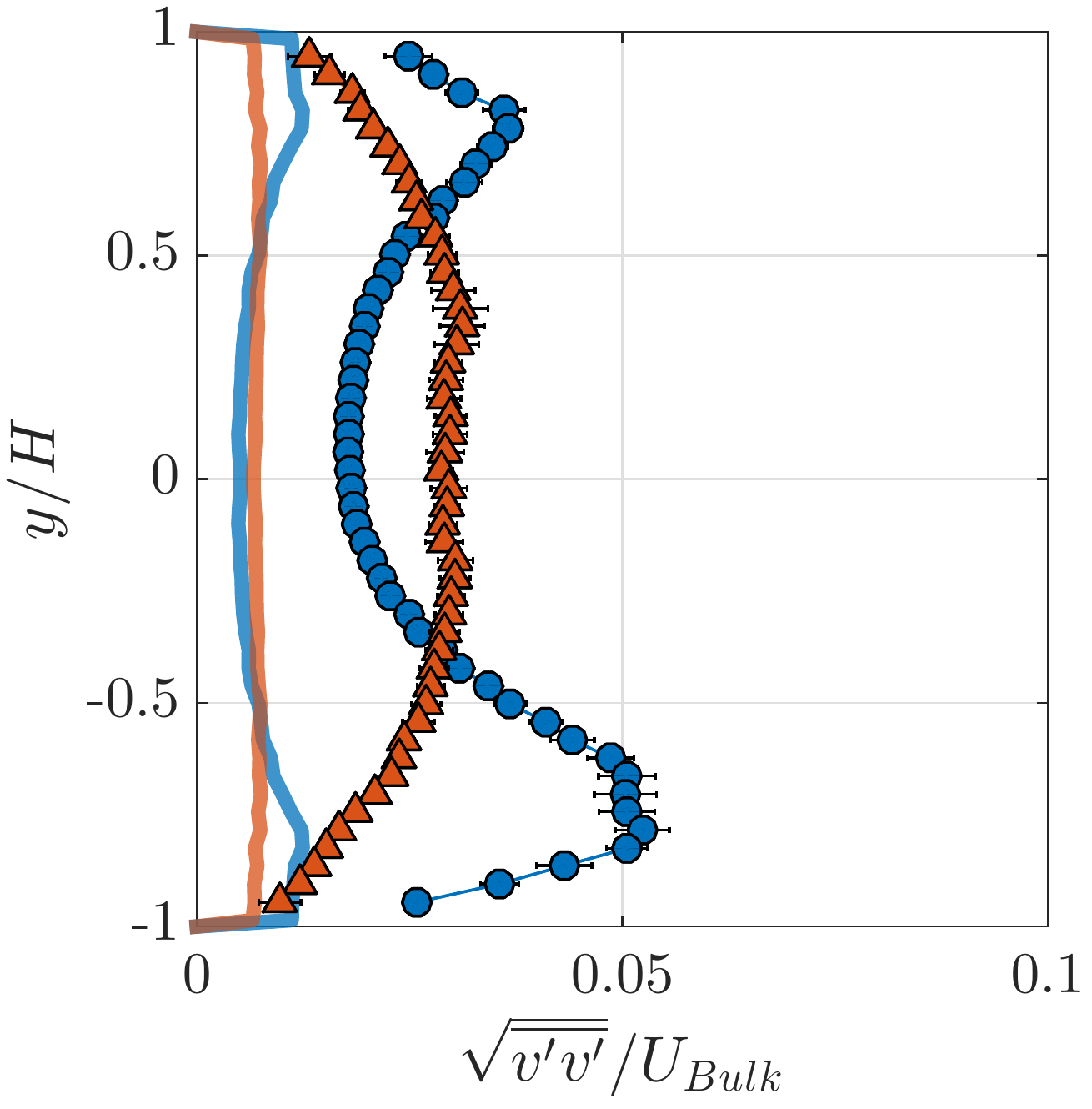}
  \caption{$Re^*$ = 53}
  \label{fig:vv_10p_20lpm}
\end{subfigure}%
\begin{subfigure}{.33\textwidth}
  \centering
  \includegraphics[height=1\linewidth]{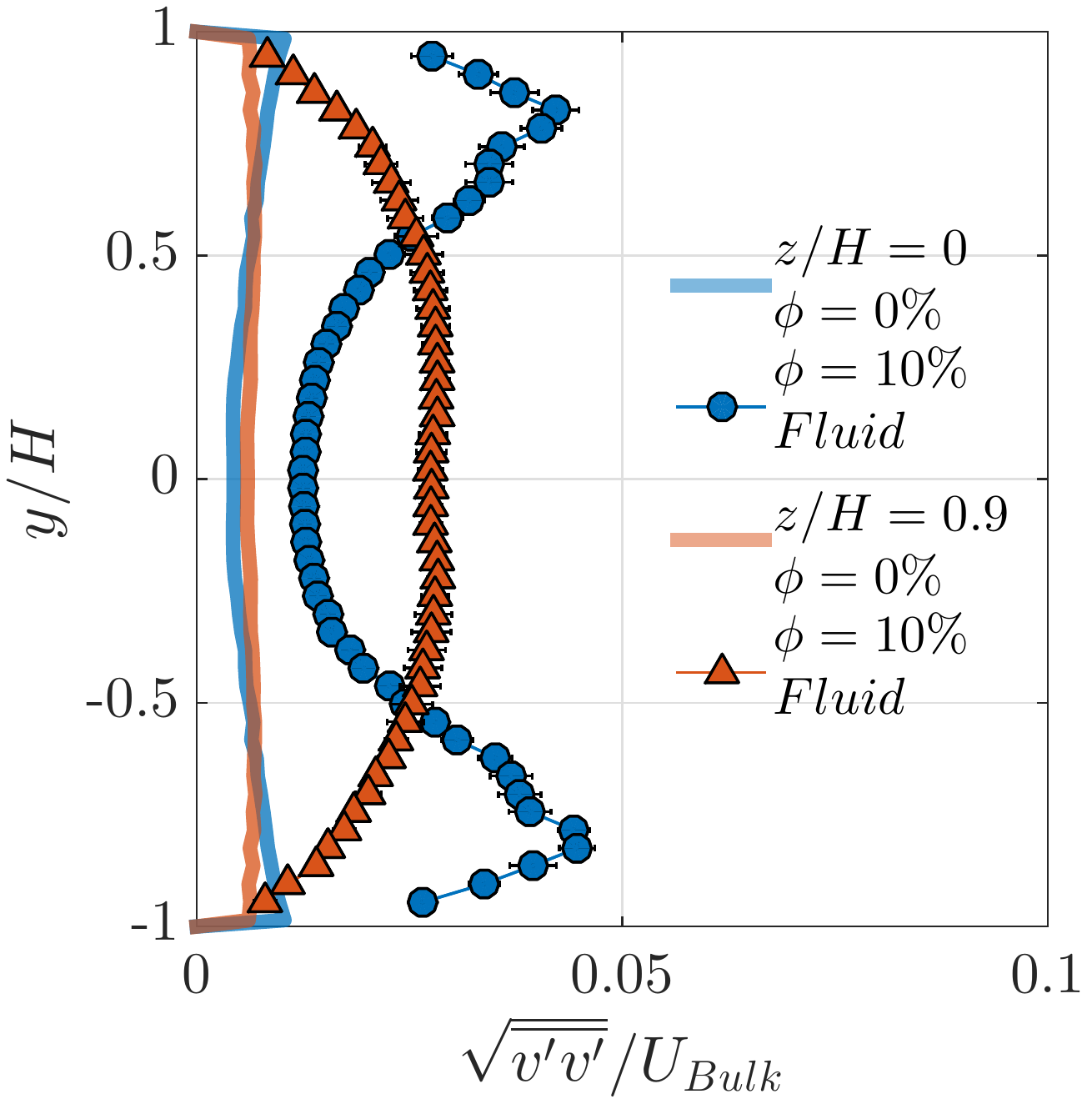}
  \caption{$Re^*$ = 156}
  \label{fig:vv_10p_40lpm}
\end{subfigure}%
\caption{Change of wall-normal velocity fluctuations due to introduction of particles at two span-wise planes: $z/H$ = 0 (center-plane) and $z/H$ = 0.9 (plane close to the side-wall). The top row (a--c) corresponds to $\phi$ = 5\% and the bottom row (d--f) corresponds to $\phi$ = 10\%.}
\label{fig:Vrms particles}
\end{figure}

The presence of particles introduce unsteady disturbances in the surrounding fluid field. The streamwise and wall-normal components of the velocity fluctuations 
associated to these disturbances are displayed in figure \ref{fig:Urms particles} and \ref{fig:Vrms particles}, following  the same scheme as in the last two figures, as also detailed in the captions. Compared to the single phase case, the streamwise and wall-normal fluctuations are higher. For the lowest flow rates, sedimentation causes rather high streamwise fluctuations at the bottom of the centre plane (see $z/H$ = 0 in figures \ref{fig:uu_5p_4lpm} and \ref{fig:uu_10p_4lpm}). 

At higher flow rates, the fluctuation profiles become increasingly symmetric. Considering the duct center plane, streamwise fluctuations reach peak values close to the top and bottom wall, which are also the locations of high particle concentration. Interestingly, the fluctuations are higher for the intermediate $Re^*$ = 53 than at higher $Re^*$ = 156. This is perhaps the result of particles arranging themselves in an intermediate configuration between accumulations at the corners 
observed at low flow rates and the ring-like structure of the high flow rates. In other words, at such $Re^*$, the spread of the particles in the duct is higher as also illustrated in figure \ref{fig:Particle concentration images} and hence for the same $\phi$, the unsteadiness in the fluid field is larger. At higher flow rates, most of the particles cluster near the wall bisectors, the frequency of particles passing in this region is increased and hence, the unsteadiness in the velocities, or departure from the mean velocity, is not so pronounced leading to lower root mean square values in the near-wall planes. Measurements at additional planes will be needed to prove the above qualitative picture; nevertheless, the presence of increased streamwise disturbances in the presence of particles is clearly demonstrated.

The appearance of streamwise fluctuations is also accompanied by wall-normal velocity fluctuations, as depicted in figure \ref{fig:Vrms particles}. Settling of particles at lower flow rates (refer to figures \ref{fig:vv_5p_4lpm} and \ref{fig:vv_10p_4lpm}) leads to higher wall-normal fluctuations near the bottom of the duct while at the highest flow rates, the profiles are symmetric with nearly the same magnitudes for the two highest $Re^*$ (compare figure \ref{fig:vv_5p_20lpm} to \ref{fig:vv_5p_40lpm} and \ref{fig:vv_10p_20lpm} to \ref{fig:vv_10p_40lpm}). 

\begin{figure}
\centering
 \includegraphics[height=0.4\linewidth]{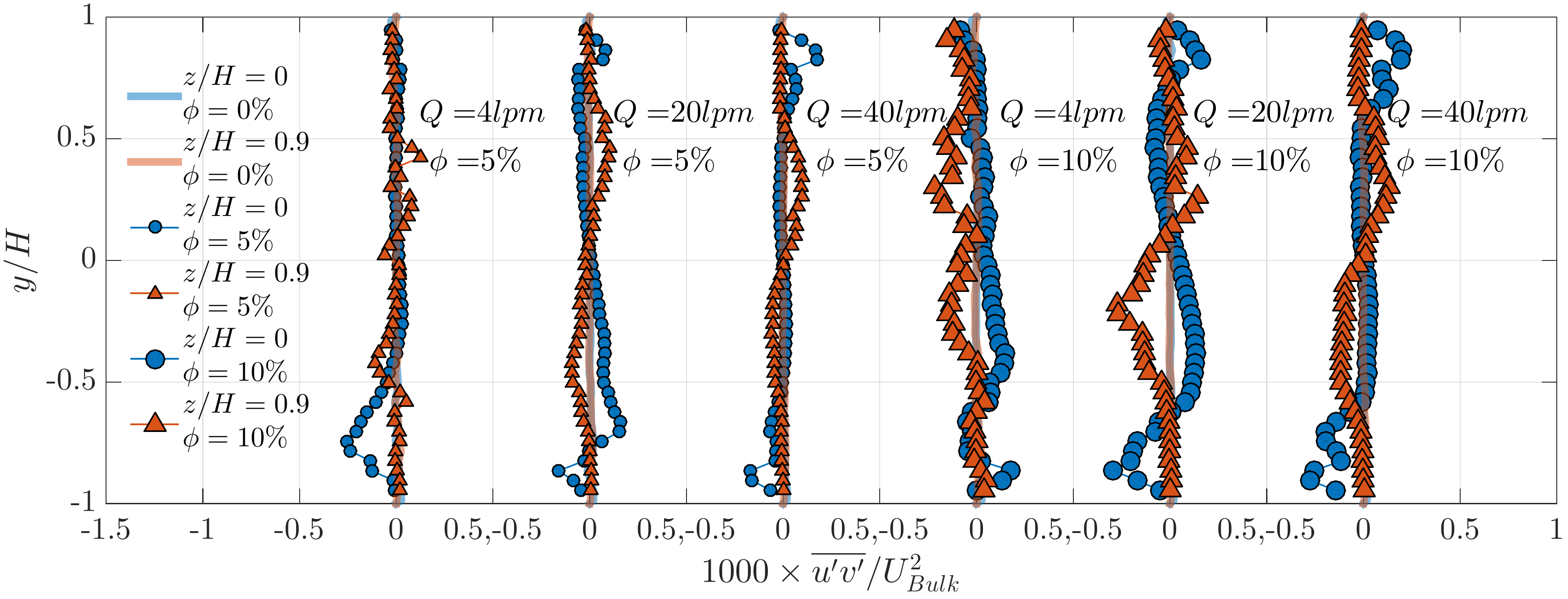}
\caption{Reynolds shear stress due to introduction of particles at two span-wise planes for multiple flow rates (or $Re^*$) and both $\phi$ = 5 and 10\%. The single phase reference case is also shown (solid lines) for comparison. Note that the limits of the x-axis is local to each plot corresponding to a fixed flow rate.}
\label{fig:uv}
\end{figure}

Despite the presence of intense streamwise and wall-normal fluid velocity fluctuations, the correlation between these two components i.e. the Reynolds shear stress $\overline{u'v'}$ does not drastically increase when compared to the corresponding single phase case, as shown in figure \ref{fig:uv}. For reference, the peak values of $\overline{u'v'}/{U^2}_{Bulk}$ in Newtonian turbulence is around of $3-5 \times 10^3$. Thus, the fluctuations generated in the flow are not a signature of the classical turbulence in a duct distinguished by larger Reynolds shear stress, but it is some unsteady flow perturbed by solid particles.

\section{Discussion}

\begin{figure}
\centering
 \includegraphics[height=0.4\linewidth]{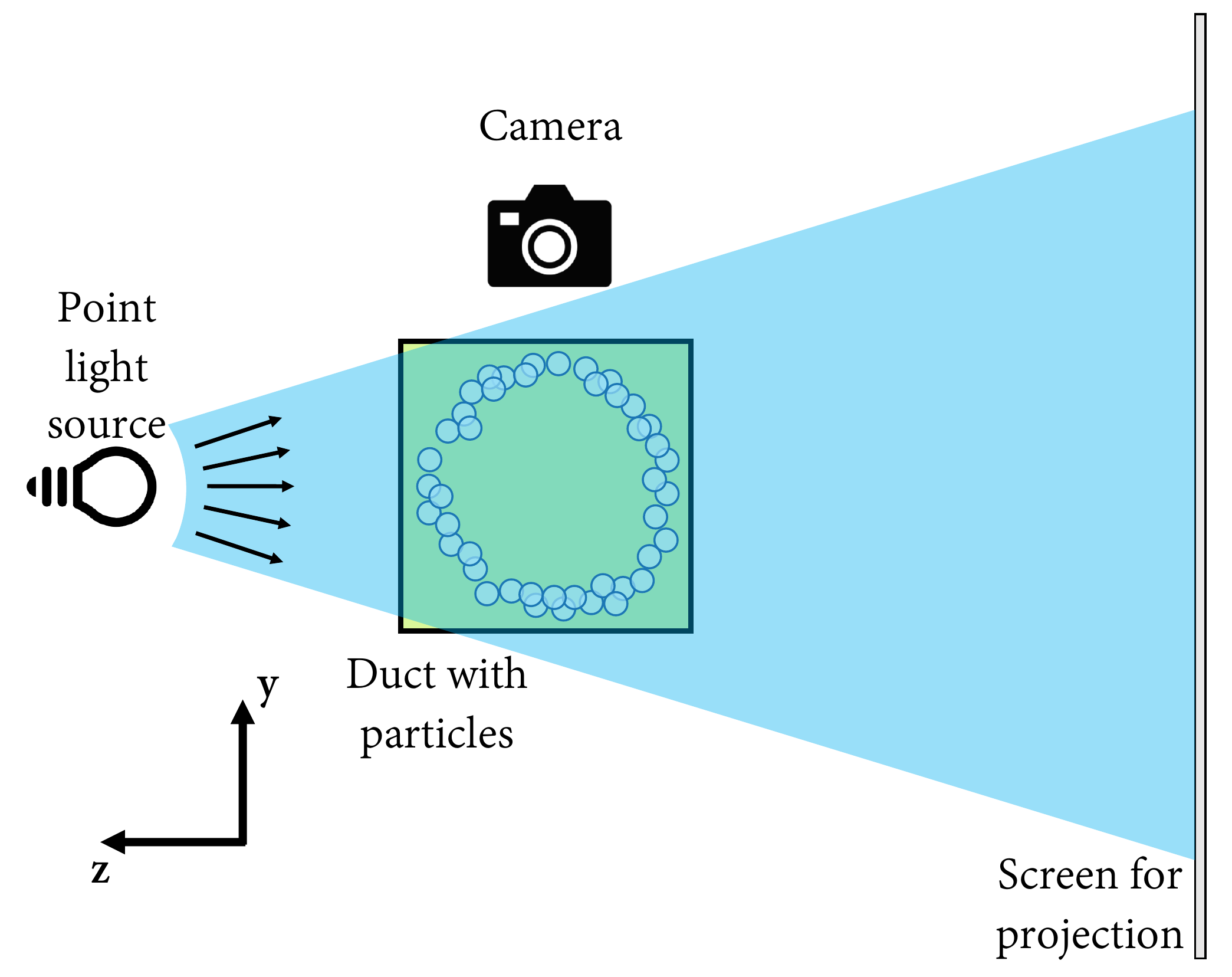}
\caption{Set-up for shadowgraph imaging. The streamwise direction is out of the plane of the paper.}
\label{fig:Shadowgraphy}
\end{figure}

\begin{figure}
\centering
\begin{subfigure}{.50\textwidth}
Entrance $x/H \approx$ 12
  \centering
  \includegraphics[height=0.4\linewidth]{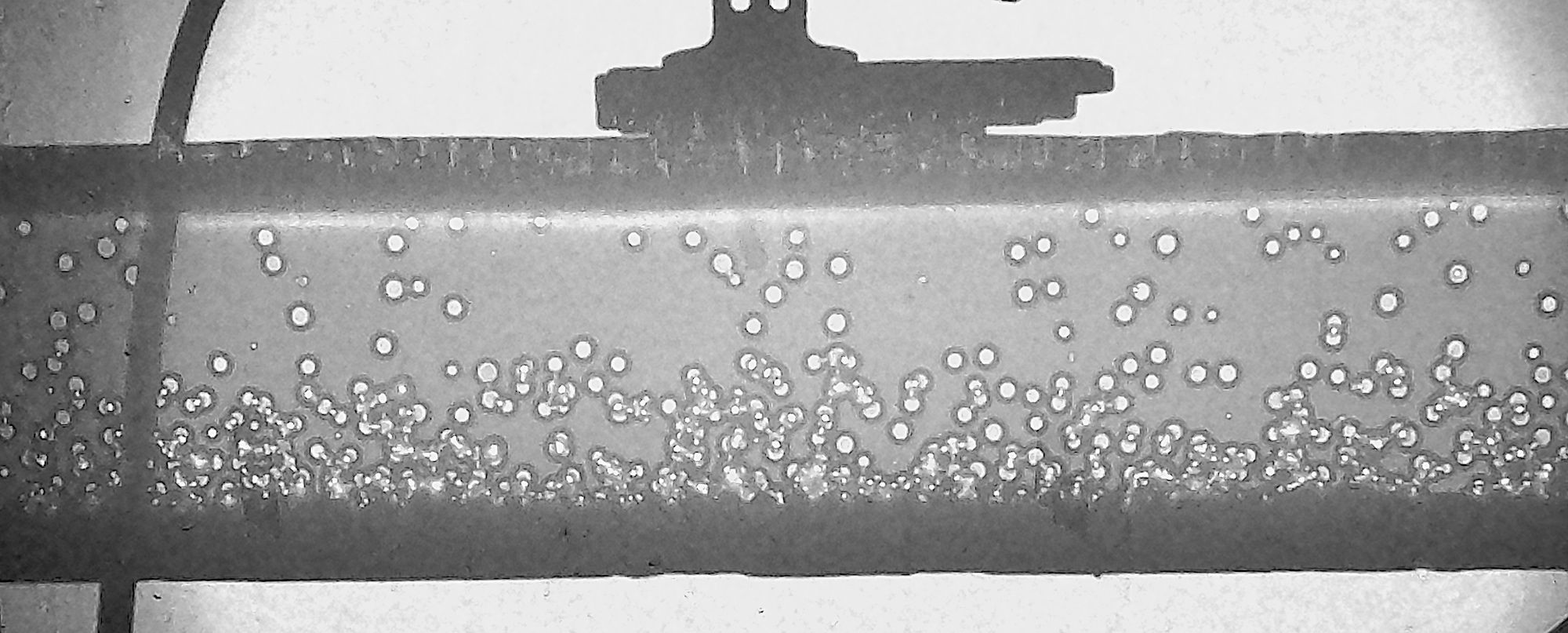}
  \caption{$Q$ = 4 lpm}
  \label{fig:Shadowgraphy_4lpm_entrance}
\end{subfigure}%
\begin{subfigure}{.50\textwidth}
Exit $x/H \approx$ 160
  \centering
  \includegraphics[height=0.4\linewidth]{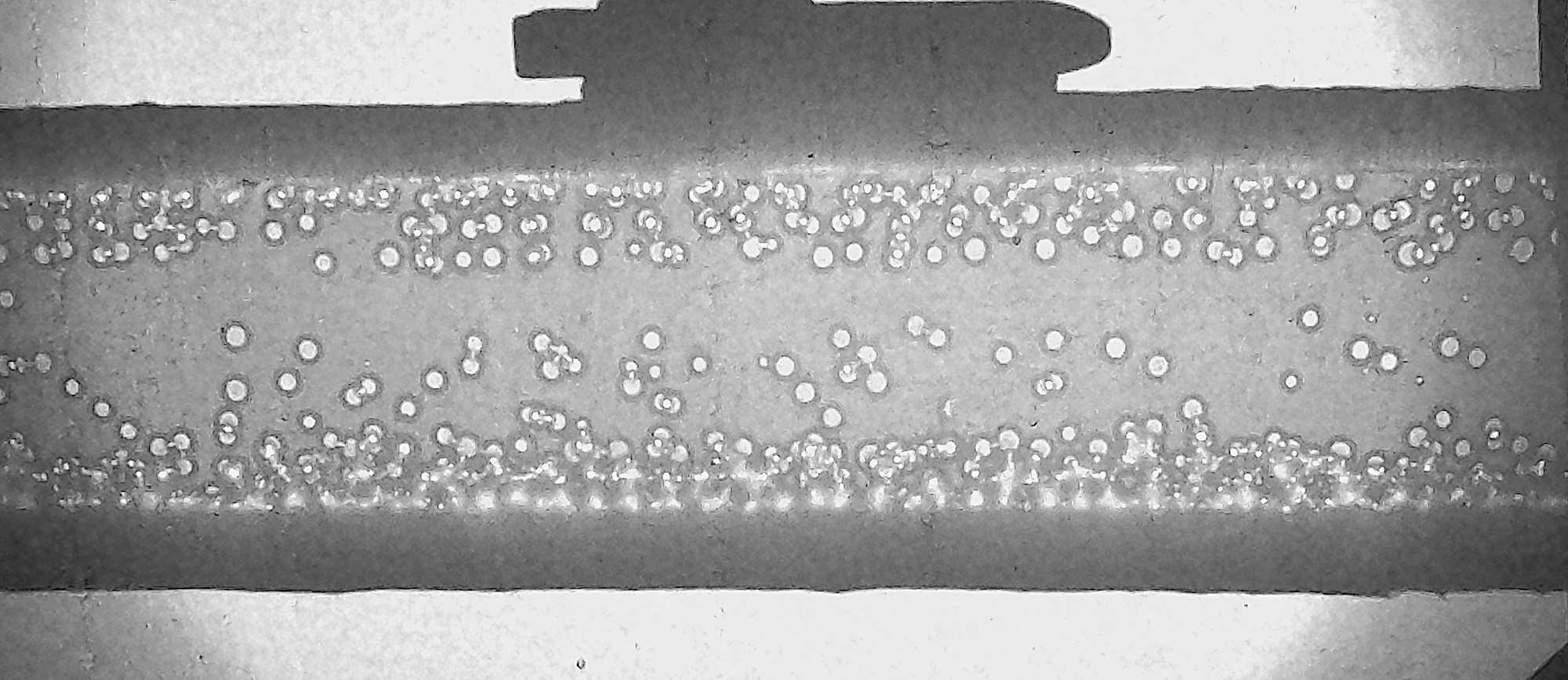}
  \caption{$Q$ = 4 lpm}
  \label{fig:Shadowgraphy_4lpm_exit}
\end{subfigure}
\begin{subfigure}{.50\textwidth}
  \centering
  \includegraphics[height=0.4\linewidth]{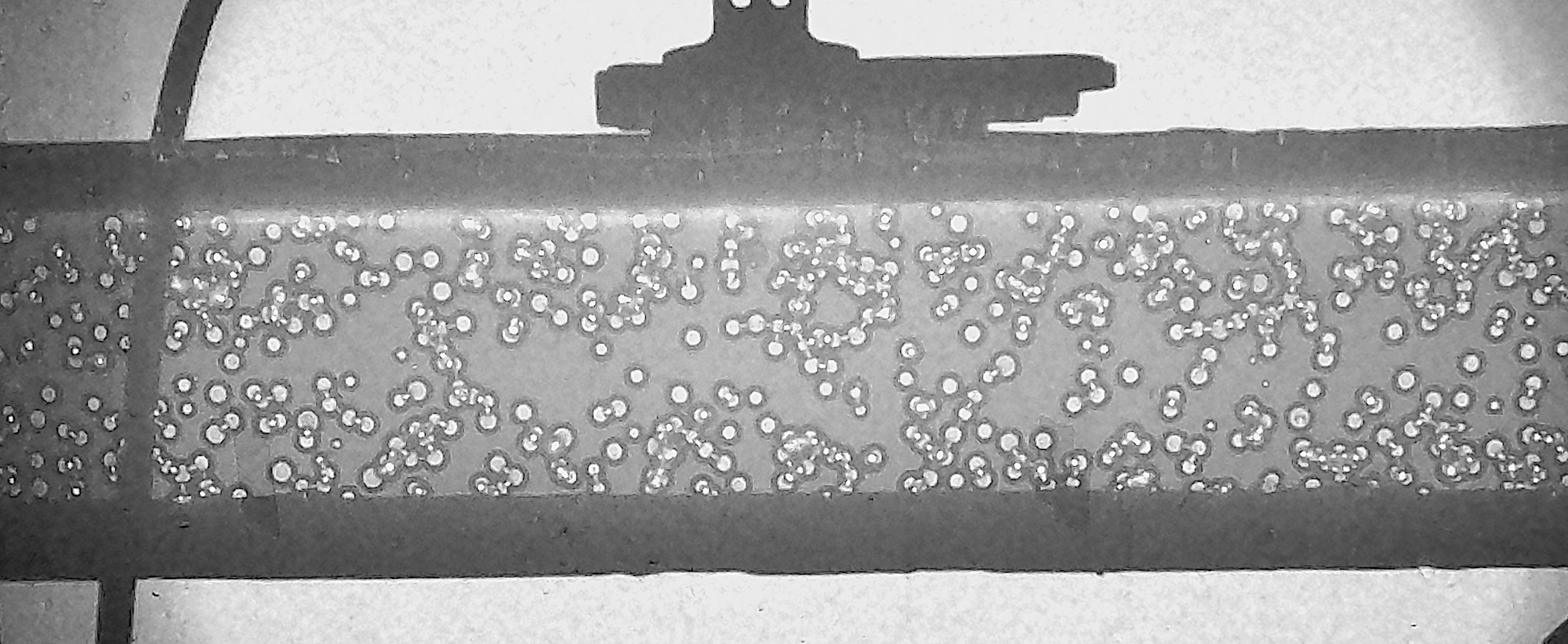}
  \caption{$Q$ = 40 lpm}
  \label{fig:Shadowgraphy_40lpm_entrance}
\end{subfigure}%
\begin{subfigure}{.50\textwidth}
  \centering
  \includegraphics[height=0.4\linewidth]{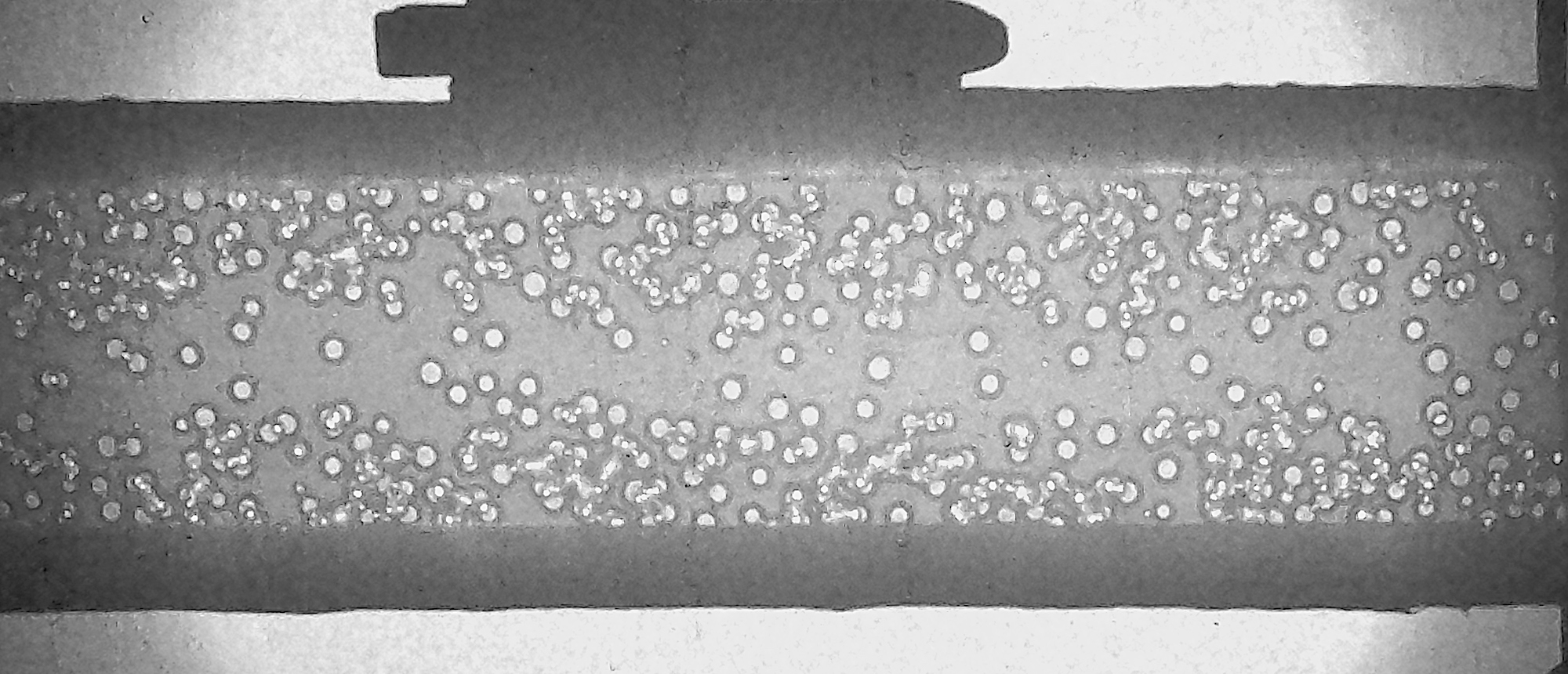}
  \caption{$Q$ = 40 lpm}
  \label{fig:Shadowgraphy_40lpm_exit}
\end{subfigure}
\caption{Visualization of particle distribution at the entrance ($x/H \approx$ 12; a and c) and far downstream ($x/H \approx$ 160; b and d) in the duct at both low ($Q$ = 4 lpm) and high ($Q$ = 40 lpm) flow rates, using simple shadowgraphs. Note: The PIV-PTV measurements were performed at $x/H \approx$ 150.}
\label{fig:Shadowgraphy images}
\end{figure}

In order to shed more light on the overall particle distribution in the duct, the quantitative measurements in figure \ref{fig:Particle concentration profiles} are complemented by flow visualisations obtained with shadowgraphy experiments. The set-up is sketched in figure \ref{fig:Shadowgraphy}, where the duct is illuminated by a point source of light such as a LED. The slight non-uniformity in the refractive index created by the particles inside the suspending medium casts a magnified shadow of the entire volume of a section of the duct on a screen, observable in a dark room. This shadow is captured by a camera as shown in figure \ref{fig:Shadowgraphy images}. 

\begin{figure}
\centering
\begin{subfigure}{.50\textwidth}
  \centering
  \includegraphics[height=0.8\linewidth]{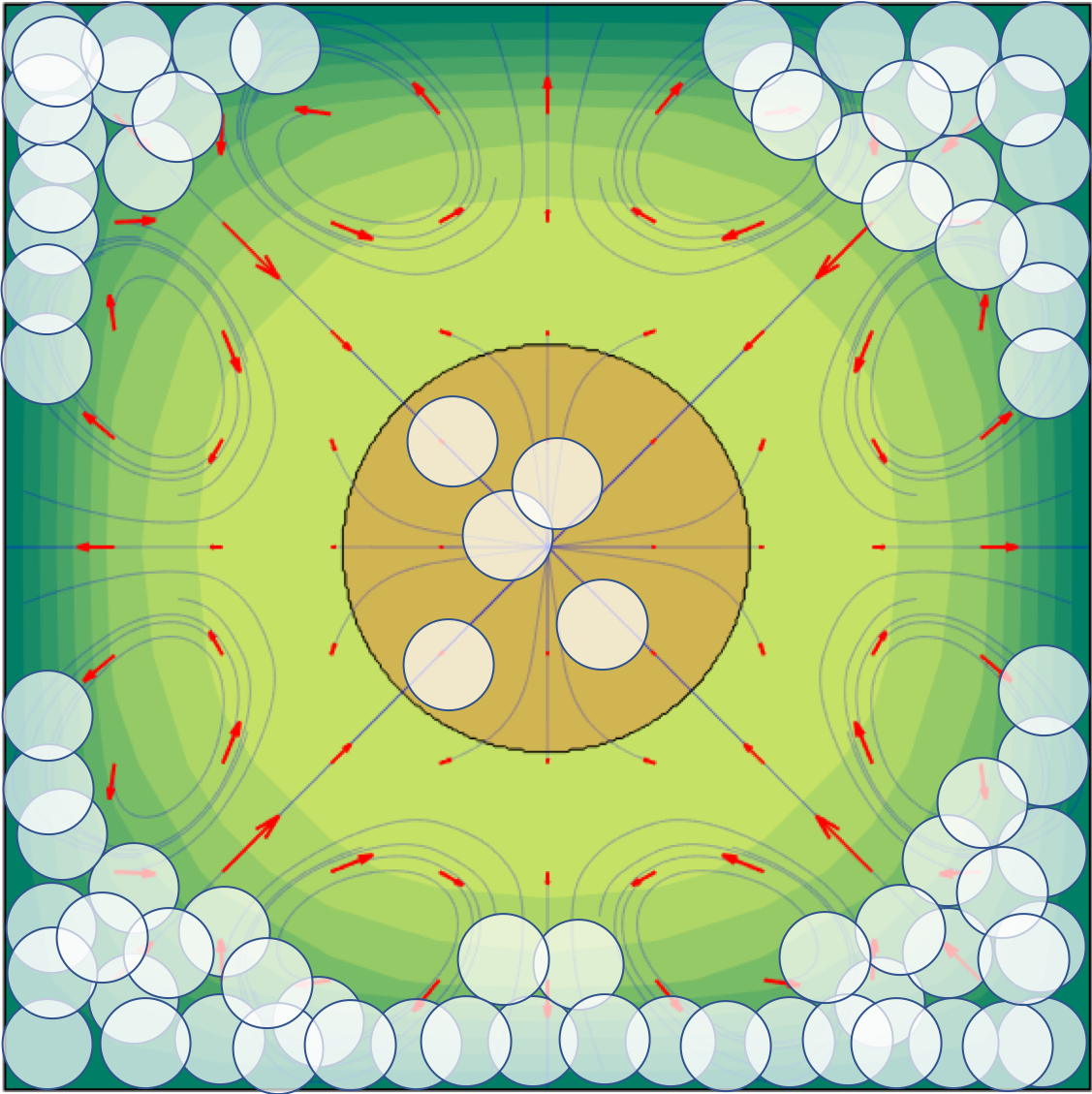}
  \caption{$Q$ = 2 lpm}
  \label{fig:4lpm_particle_distribution}
\end{subfigure}%
\begin{subfigure}{.50\textwidth}
  \centering
  \includegraphics[height=0.8\linewidth]{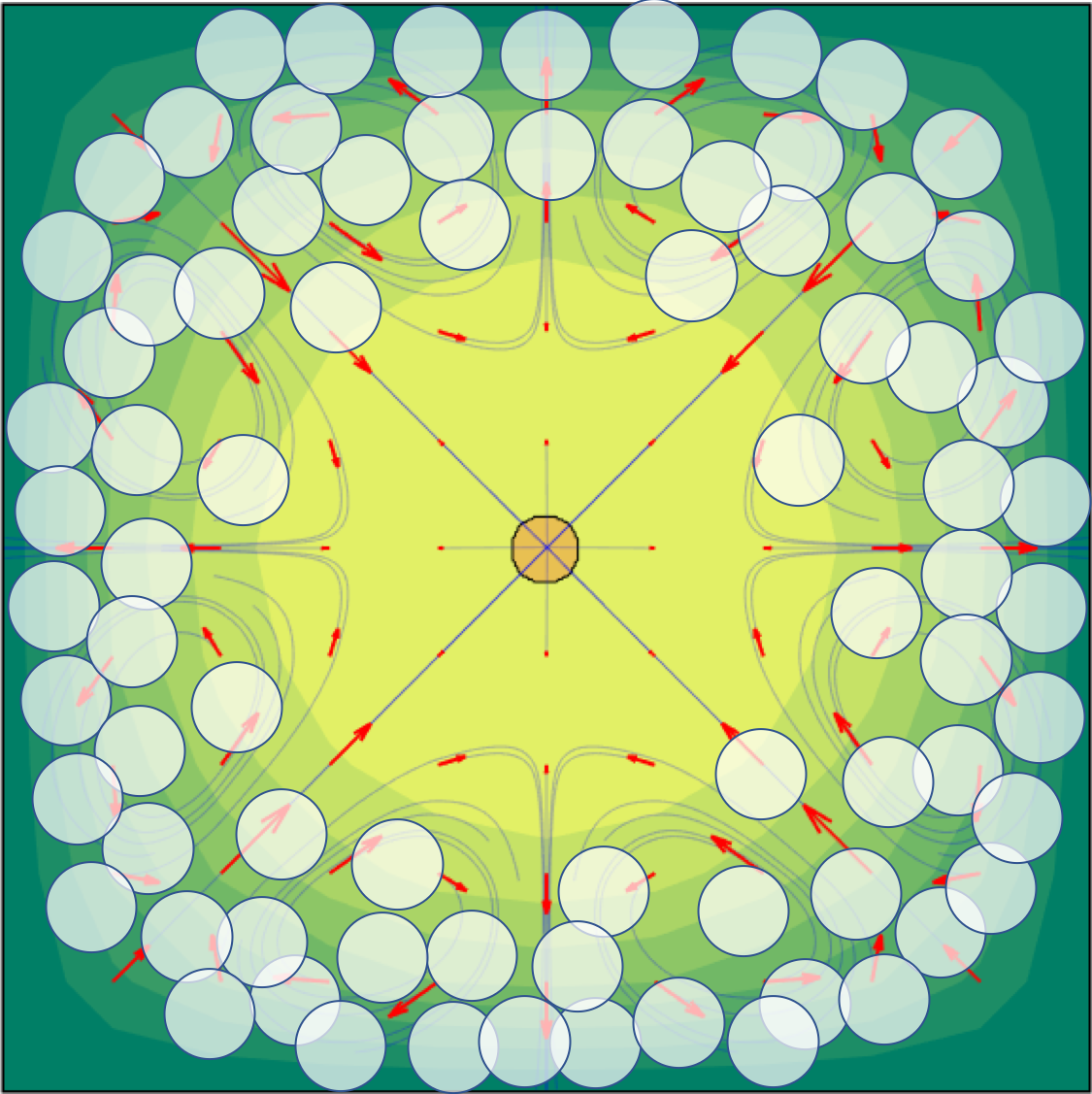}
  \caption{$Q$ = 40 lpm}
  \label{fig:40lpm_particle_distribution}
\end{subfigure}
\caption{Schematic of the particle distribution in the cross-stream section of the duct. The arrows and streamlines representing the normalized secondary flow and the extent of the plug are from the actual experimental measurements for single phase flow at the corresponding flow rates. The contours represent the streamwise velocity normalized by the bulk velocity. Note that the colorbar is the same as in figure \ref{fig:Sec_flow_SP}. The particle distribution is schematically reconstructed from the shadowgraphs and PIV measurements.}
\label{fig: Particle_distribution}
\end{figure}

The qualitative visualization in figures \ref{fig:Shadowgraphy_4lpm_exit} and \ref{fig:Shadowgraphy_40lpm_exit} corresponding to the fully developed flow, along with the quantitative concentration profiles from figure \ref{fig:Particle concentration profiles}, are used to sketch an approximate particle distribution pattern at low and high flow rates, see figures \ref{fig:4lpm_particle_distribution} and \ref{fig:40lpm_particle_distribution}. The increased particle concentration near the bottom wall for low flow rates, due to gravitational forces, is also reproduced in figure \ref{fig:4lpm_particle_distribution}. For consistency, the same number of particles are depicted in both the figures. Note also that visualizations in additional planes besides those reported in figure \ref{fig:Particle concentration profiles} using the PIV laser sheet have contributed to draw the sketches in figures \ref{fig: Particle_distribution}.

In order to explain the observed concentration distribution, it is essential to acknowledge the presence of competing forces of inertial and viscoelastic nature. In a fluid without elasticity, as the particle Reynolds number increases beyond the Stokes regime in a pipe or a duct, the shear-gradient lift forces \citep{ho1974inertial} act to push particles away from the core, whereas the repulsive wall forces act to displace the particles away from the walls, so that an isolated particle reaches a stable equilibrium position somewhere in between the center and the wall, a phenomenon famously known as the Segr\'e-Silberberg effect \citep{segre1961radial}. 
The presence of viscoelasticity disrupts this equilibrium position due to the non-uniform distribution of the first normal stress difference, being maximum at the wall-centre and minimum at the corners and core. This variation of the first normal stress difference across the particle surface causes them to migrate to the corners at low flow rates  \cite[see figure \ref{fig:4lpm_particle_distribution}, and][]{yang2011sheathless, villone2013particle}. Migration towards the core is not as pronounced due to the presence of the non-fluidised core with its infinite viscosity. Hence, the non-zero concentration inside the plug is most likely due to particles that were trapped in the bulk at the inlet to the duct.

With increasing flow rates, the intensity of the secondary flow, induced by the second normal stress differences, increases; as a consequence, the flow sweeps the particles away from the corners, leading to the distribution sketched in figure \ref{fig:40lpm_particle_distribution}. Particles may now follow the helicoidal motion of the secondary flow as reported in the simulations of \cite{villone2013particle} and the experiments of \cite{lim2014lateral}. The secondary flow measured for the single-phase cases is most likely changed by the presence of the particles; however, it is not reported in this study since measurements displayed excessive noise to show a consistent behaviour.
Nonetheless, in a Newtonian suspending medium,  the modifications of the secondary flow due to cross-stream migration of finite-size particles in a square duct
have been shown in numerical studies as in \cite{kazerooni2017inertial}. Previously, \cite{ramachandran2008influence} proved that strong second normal stress differences, and hence secondary flows, can be produced in concentrated suspensions due to an anisotropic particle microstructure. Finally, note that, in our experiments at the higher flow rates, the plug in the core has a negligible size or is perhaps non-existent, and the flow therefore approaches the case of particles in a shear-thinning viscoelastic fluid.

The simple shadowgraphy visualisations also help in assessing the evolution of the particle distribution along the duct streamwise length. Figure \ref{fig:Shadowgraphy_4lpm_entrance} and \ref{fig:Shadowgraphy_40lpm_entrance} show the shadowgraphs of the particle distribution very close to the entrance of the duct (at $x/H \approx$ 12) at low and high flow rates. At the low flow rate, particles seem to enter the duct while being mostly concentrated in the bottom half. This entry profile may either be due to gravitational forces or because of the secondary Dean motion from the bend at the exit of the tank in our setup. 
The installed static mixer meant for cancelling these vortices is, perhaps, less effective at such lower $Re^*$. Nevertheless, as the flow develops, the initial disturbances are dissipated and particles attain an equilibrium concentration as shown in figure \ref{fig:Shadowgraphy_4lpm_exit}, which is captured very close to the location where the PIV measurements are performed. 
Thus, particles migrate to the top against gravitational forces. 
Particles travelling at a higher speed in the core are seen to move without any change in their relative position on account of being locked-up inside the solid plug. On many occasions, clusters of two or more particles are seen to move together inside the plug \cite[which was observed before in the case of particle sedimentation by][]{chaparian2018inline}. The distribution profile just described is attained much earlier than $x/H \approx$ 160 and persists along the remaining length of the duct. Projection from only one direction is insufficient to give full information on the particle distribution and hence, the duct was also illuminated from the bottom to check a second projection. And indeed, a picture very similar to that in figure \ref{fig:Shadowgraphy_4lpm_exit} is obtained, thus confirming the preferential accumulation of particles at the corners. 

Focussing on the highest flow rates, it appears that the concentration profile is more uniform at the entrance (see figure \ref{fig:Shadowgraphy_40lpm_entrance}) and rapidly 
develops to an equilibrium pattern as seen in figure \ref{fig:Shadowgraphy_40lpm_exit}. By visualising fast and slow particles, it appears that very few particles travel near the 
core, which has the highest velocity, or the corners, which has the smallest velocity. Most of the particles travel in the intermediate region between the core and the corners. 
Moreover, the particle motion appears very erratic with instances of rapid change in their velocities. 
This indicates that strong particle-particle and particle-wall interactions are most likely relevant at these concentrations and flow rates. The presence of finite-size particles also causes an increase in the local strain rates,
and the corresponding reduction in the local viscosity for the shear-thinning fluid may intensify the strength of these interactions, which may be an important mechanism for generating the fluid velocity fluctuations shown in figure \ref{fig:Urms particles} and \ref{fig:Vrms particles}.

To conclude, we note that from a visualization perspective, another noteworthy technique can be used, exploiting the property of elastoviscoplastic  flows to come to a rather quick stop after switching off the pump, as mentioned earlier. As a result, the particles get trapped in their \textit{flowing} configuration without much change of the inter-particle distance, thus making it possible to assess their position and hence, the concentration distribution. This can be conveniently done by translating the laser sheet across such a \textit{frozen} flow and capturing still photos which can be later stacked together to reconstruct the full 3D concentration field. Such experiments are not performed here but can be advantageously used for these kind of fluids.

\section{Conclusion}

We have presented an experimental study of the flow of an elastoviscoplastic fluid, Carbopol gel at two different concentrations -- one with a high yield stress HYS and the other with a low yield stress LYS and studied the laminar single-phase flow and the flow in the presence of finite-size spherical particles at relatively high concentrations inside a square duct. 
Pressure drop, mean and fluctuating velocities and concentration fields have been measured using a combination of PIV and PTV; these optical techniques have been  possible due to the refractive-index-matching of the fluid-particle mixture. To facilitate comparative simulations, the rheological properties of the fluids investigated are described in terms of the steady-state flow curves and viscoelastic moduli.

The experimental pressure drop is in agreement, over the range of flow rates considered, with the laminar friction factor based on the semi-empirical Reynolds number $Re^*$ (see equation \ref{eqn:Re}) by \cite{liu1998non}. By making use of the symmetries in a square duct, we have presented the full 3-component velocity field (at moderate spatial resolution) for single phase flow using 2-dimensional PIV in multiple span-wise planes (at high spatial resolution), evidencing the existence of a secondary flow due to the visco-elastic properties of the fluid. Using a low threshold value of the velocity gradient, we have identified the near-circular plug region in the core, whose size reduces with increasing the flow rate or equivalently $Re^*$. At the same $Re^*$, the fluid with a higher yield stress HYS has a higher Bingham number $Bi$ and a larger plug. Also, at the same $Bi$, the ratio of the maximum streamwise velocity to the bulk velocity is larger for a fluid with a higher power law exponent $n$. The strength and size of the streamwise vortices representing the secondary flow increases for reducing plug sizes.

Similar to Newtonian suspending fluids, the friction factor  increases by increasing the particle volume fraction $\phi$ at fixed $Re^*$. This increase seems independent of the rheology of the suspending fluid and matches reasonably well predictions based on the effective viscosity of suspensions of rigid spheres, like the Eilers fit. This matching occurs despite a flow rate dependent non-uniform particle distribution. 
At low flow rates, in the presence of a larger plug and weaker secondary flow, particles migrate inside the fluidised zone, defying gravity, towards the four corners under the influence of the first normal stress difference. Fluid (and particle) velocity reduces in the near-wall planes (where there is a high particle concentration) and increases in the plane of the wall-bisector when compared to the corresponding single-phase flow at the same locations. This increase of the maximum velocity reflects in an increased shear in the core region and results in a smaller plug than for the single-phase flow. At high flow rates, in the presence of a negligibly small plug and a stronger secondary flow, particles arrange along a diffused ring between the core and the corners. Particles are also seen to induce time-dependent fluid velocities leading to uncorrelated fluctuations in the streamwise and wall-normal directions. Visualisation of shadowgraps have, even though qualitatively, confirmed the above concentration profiles. 

The importance and novelty of this experiment, when compared to other studies with particles in non-Newtonian fluids, is the ability to assess the velocity distribution of both the fluid and particle phases along with particle distribution at high bulk concentrations. This will, hopefully, help in future modelling attempts of such systems and shed light on the complex fluid-particle interactions present therein. These results are especially relevant to microfluidic applications which share similar values of the relevant non-dimensional parameters, e.g.\ the Reynolds number $Re^*$ and confinement (or blockage) ratios and where resolved measurements are prohibited by the small length scales. As always, further investigation with different particle sizes, volume fractions and fluid rheologies, perhaps coupled with spatio-temporally resolved measurements, will help to further understand these complex systems.  

\section*{Acknowledgements}

This work was supported by the European Research Council Grant No. ERC-2013-CoG-616186, TRITOS, from the Swedish Research Council (VR), through the Outstanding Young Researcher Award to LB. \r{A}sa Engstr\"{o}m (Rise Bioeconomy AB) is gratefully acknowledged for assistance with the rheological measurements.

\bibliographystyle{jfm}
\bibliography{jfm_template.bib}

\end{document}